\documentclass[9pt,superscriptaddress,amsmath,amssymb,aps,twocolumn]{revtex4-2}
\usepackage{graphicx}
\usepackage{breakurl}
\usepackage[a4paper, top=2cm, bottom=2cm, left=1.5cm, right=1.5cm]{geometry}
\usepackage[colorlinks=true,linkcolor=blue,urlcolor=blue,citecolor=blue,anchorcolor=blue]{hyperref}

\usepackage{xcolor}  

\begin{document}

\title{The structural evolution of temporal hypergraphs through the lens of hyper-cores}

\author{Marco Mancastroppa}
\affiliation{Aix-Marseille Univ, Universit\'e de Toulon, CNRS, CPT, Turing Center for Living Systems, 13009 Marseille, France}
\author{Iacopo Iacopini}
\affiliation{Network Science Institute, Northeastern University London, London, E1W 1LP, United Kingdom}
\author{Giovanni Petri} 
\affiliation{Network Science Institute, Northeastern University London, London, E1W 1LP, United Kingdom}
\affiliation{CENTAI, Corso Inghilterra 3, 10138 Turin, Italy}
\author{Alain Barrat}
\affiliation{Aix-Marseille Univ, Universit\'e de Toulon, CNRS, CPT, Turing Center for Living Systems, 13009 Marseille, France}

\begin{abstract} 
The richness of many complex systems stems from the interactions among their components. The higher-order nature of these interactions, involving many units at once, and their temporal dynamics constitute crucial properties that shape the behaviour of the system itself. An adequate description of these systems is offered by temporal hypergraphs, that integrate these features within the same framework. However, tools for their temporal and topological characterization are still scarce.
Here we develop a series of methods specifically designed to analyse the structural properties of temporal hypergraphs at multiple scales. Leveraging the hyper-core decomposition of hypergraphs, we follow the evolution of the hyper-cores through time, characterizing the hypergraph structure and its temporal dynamics at different topological scales, and quantifying the multi-scale structural stability of the system. We also define two static hypercoreness centrality measures that provide an overall description of the nodes aggregated structural behaviour. We apply the characterization methods to several data sets, establishing connections between structural properties and specific activities within the systems.
Finally, we show how the proposed method can be used as a model-validation tool for synthetic temporal hypergraphs, distinguishing the higher-order structures and dynamics generated by different models from the empirical ones, and thus identifying the essential model mechanisms to reproduce the empirical hypergraph structure and evolution.
Our work opens several research directions, from the understanding of dynamic processes on temporal higher-order networks to the design of new models of time-varying hypergraphs.
\end{abstract}

\maketitle

\section*{Introduction}

Many complex systems composed of interacting elements can be effectively described within the theory of static networks \cite{newman2018networks,Dorogovtsev:2003,barrat2008dynamical}. This powerful framework provides a wide set of techniques and tools to characterize the interactions at different topological scales, through global graph properties (e.g. density), possibly focusing on specific groups of relevant nodes (e.g. $k$-cores) and 
providing various measures of node centralities. Furthermore, this multi-scale characterization helps identify nodes and mesostructures with relevant roles in dynamical processes, since the interaction structure deeply impacts processes unfolding on networks \cite{barrat2008dynamical,castellano2015}. 
Despite the power of network theory, recently several empirical evidences have brought out the limits of this framework, which by definition is restricted to a static description of systems involving only binary interactions.

On the one hand, several systems present time-varying interactions, which follow specific dynamics and temporal patterns \cite{HOLME201297,masuda2016,braha2009time}: for example, human social interactions \cite{karsai2018bursty}, scientific collaborations \cite{petri2018simplicial} and neural systems \cite{HOLME201297,pedreschi2020dynamic}. These systems are represented using \textit{temporal networks}, a generalization of static networks in which nodes interact via links with specific activation and deactivation times \cite{HOLME201297,masuda2016}. Several structural characterization tools for static networks have been generalized to time-varying graphs, showing the non-trivialities emerging from the introduction of the temporal dimension \cite{HOLME201297,masuda2016,braha2009time}: for instance, span-cores can decompose a temporal graph into subgraphs of controlled duration and increasing connectivity \cite{Ciaperoni2020,galimberti2018mining}. Moreover, dynamic processes on temporal networks are also impacted by the network dynamics, especially when the dynamics of and on the network have comparable time scales \cite{HOLME201297,masuda2016,Perra2012,Mancastroppa_2019}.

On the other hand, many complex systems also feature interactions between groups of agents, not reducible to sets of pairs \cite{battiston2020networks,Battiston2021}: this is the case for example of human social interactions \cite{danon2013social}, scientific collaborations \cite{milojevic2014principles} and species interactions in ecosystems \cite{mayfield2017higher}. An adequate description of these systems involves \textit{hypergraphs}, a generalization of networks in which nodes can interact in groups of arbitrary size, i.e., hyperedges \cite{battiston2020networks}. Taking into account such higher-order nature of interactions leads to the definition of new structures and concepts, and the inclusion of higher-order mechanisms in dynamical processes (defined on hypergraphs) leads to the emergence of new phenomena \cite{iacopini2019simplicial,battiston2020networks,Battiston2021,cencetti2023}. Tools to characterize hypergraphs at various scales have only recently been proposed. In particular, the hyper-core decomposition \cite{mancastroppa2023hyper,bianconi2024nature} identifies a doubly nested hierarchy of mesoscopic subhypergraphs, the hyper-cores, composed of nodes progressively more densely connected to each other through interactions of increasing size. This technique provides a global fingerprint of systems described using hypergraphs and identifies structurally central mesostructures that play an important role in higher-order dynamical processes \cite{mancastroppa2023hyper}. This decomposition also comes with an associated centrality measure for nodes, the hypercoreness, which is based on the node structural position at the various interaction orders \cite{mancastroppa2023hyper}.

The increasing attention to the development of frameworks to handle time-varying and non-pairwise structures speaks for the need of using both the temporal and the higher-order nature of interactions to adequately describe and model several complex systems and dynamical processes. 
The integration of these two features has occurred relatively recently within \textit{temporal hypergraphs}, where hyperedges present specific activation times and duration, describing evolving group interactions \cite{battiston2020networks}. Some works focused on defining procedures to construct temporal hypergraphs from data \cite{kirkley2023constructing, sekara2016fundamental}, others on the impact of the hypergraph dynamics on dynamic processes \cite{chowdhary2021simplicial,Neuhauser2021}. Only few attempts have been made to investigate the temporal-topological properties of temporal hypergraphs \cite{sekara2016fundamental,Ceria2023,Cencetti2021,Yao2021,gallo2023higher,iacopini2023temporal}, and a complete structural characterization is still missing. Moreover, synthetic models of temporal hypergraphs have been proposed to identify and replicate the mechanisms that govern the evolution of empirical systems \cite{iacopini2023temporal,petri2018simplicial,di2024percolation,Guo2016,gallo2023higher}, but model-validation tools are still scarce. Therefore, it becomes necessary to develop dedicated multi-scale characterization methods tailored for temporal hypergraphs. These techniques are essential to accurately describe empirical systems, construct and validate synthetic models, and ultimately identify crucial temporal structures for higher-order dynamic processes: how does the higher-order structure evolve at different scales over time? Are there persistent groups of nodes exhibiting dense connections at different interaction orders, or do these configurations change dynamically? Are the most structurally central nodes always the same, or do they undergo changes over time?

Here, we tackle such issues by proposing a multi-scale method for the characterization of temporal hypergraphs at different topological scales. By applying the hyper-core decomposition to successive snapshots of a temporal hypergraph, and by following the evolution of the resulting hierarchical structure, we are able to characterize the structure and its evolution at different scales: macroscopically, following the evolution of the relative sizes of the hyper-cores; mesoscopically, focusing on the dynamics of specific hyper-cores; microscopically, following the position of single nodes in the hyper-core structure over time. 
Measuring the similarity between the hyper-core structure at different times enables the quantification of the structural stability of the system at different topological scales.
We also define two time-aggregated hypercoreness centralities for nodes, based on the node instantaneous hypercoreness and its evolution, which together provide an overall description of its structural behavior.
We apply the proposed approach to several data sets representing systems of diverse nature. This enables us to identify differences and similarities in their structure and evolution, unveiling  temporal patterns, and to establish connections between structural properties and specific activities within the systems. Finally, we illustrate how the proposed method provides a model-validation tool for synthetic models of temporal hypergraphs. 
To this aim, we propose several models of activity-driven temporal hypergraphs 
\cite{Perra2012,petri2018simplicial,mancastroppa2022,le2023modeling} which progressively implement mechanisms for the formation of group interactions of increasingly complexity. We tune these models to mimic the activity patterns of the interaction data sets and show how, following the hyper-core decomposition over time, we are able to distinguish between the hyper-core structures and dynamics generated by the models at different topological scales, providing a quantitative comparison between synthetic models and empirical hypergraphs. 

\section*{Results}

\subsection*{Following 
the hyper-core decomposition of temporal hypergraphs}

Let us consider a time-varying hypergraph $\mathcal{H}$ observed over the time interval $(0,t_{max}]$. We consider a snapshot representation of $\mathcal{H}$ with temporal resolution $\tau$ \cite{kirkley2023constructing}, i.e., the interval $(0,t_{max}]$ is divided into $n=t_{max}/\tau$ time windows of length $\tau$: $\mathcal{H} = \{\mathcal{H}_t \}_{t=1}^{n}$, where 
in each time window $t$ the instantaneous hypergraph 
$\mathcal{H}_t=(\mathcal{V}_t,\mathcal{E}_t)$ is an unweighted static hypergraph formed by the set $\mathcal{V}_t$ of nodes active at least once in $((t-1)\tau,t \tau]$ and by 
the set $\mathcal{E}_t$ of hyperedges active at least once in 
$((t-1)\tau,t \tau]$
(with $N_t=|\mathcal{V}_t|$ and $E_t=|\mathcal{E}_t|$). 
A hyperedge $e=\{i_1,i_2,...,i_m\} \in \mathcal{E}_t$ represents a group interaction between nodes $i_k \in \mathcal{V}_t$ $\forall k=1,...,m$: it consists in a set of $m$ nodes, with $m \in [2,M_t]$, where $M_t = \max_{e \in \mathcal{E}_t} |e|$. We denote with $\Psi_t(m)$ the hyperedge size distribution in the time-window $t$ \footnote{We consider only interactions of size $m \geq 2$ and neglect the presence of singletons, i.e. hyperedges of size $m=1$, since here we focus on the characterization of how the elements of the system interact with each other. Moreover, the singletons are immediately pruned in the hyper-core decomposition.}.

We propose to characterize the structural evolution of the temporal hypergraph $\mathcal{H}$ by applying the hyper-core decomposition procedure to each snapshot $\mathcal{H}_{t}$ \cite{mancastroppa2023hyper}. The hyper-core decomposition decomposes static hypergraphs 
into series of subhypergraphs of increasing connectivity, ensured by hyperedges of increasing sizes. Specifically, 
the  $(k,m)$-hyper-core of the snapshot  
$\mathcal{H}_t=(\mathcal{V}_t,\mathcal{E}_t)$ is defined as 
the maximum subhypergraph that contains all the nodes $i \in \mathcal{V}_t$ involved in at least $k$ distinct hyperedges of size at least $m$ within the subhypergraph itself (see Methods and \cite{mancastroppa2023hyper}). 

The set of nodes belonging to the $(k,m)$-core but not to the $(k+1,m)$-core forms the $(k,m)$-shell. Each node $i$ in the temporal hypergraph can thus be assigned a time-varying $m$-shell index $C_m(i,t)$, which defines the maximum $k$ such that $i$ belongs to the $(k,m)$-hyper-core but not to the $(k+1,m)$-hyper-core at time $t$. This leads to the definition of 
the hypercoreness $R(i,t)$ of node $i$ in $\mathcal{H}_t$ by
\cite{mancastroppa2023hyper}:
\begin{equation}
R(i,t)=\sum_{m=2}^{M_t} C_m(i,t)/k_{max}^m(t) \ ,
\label{eq:hypercoreness}
\end{equation} 
where $k_{max}^m(t)$ is the maximum connectivity at order $m$ for the snapshot $t$, such that the $(k_{max}^m(t),m)$-core is not empty, but the $(k_{max}^m(t)+1,m)$-core is empty. 
$R(i,t) \in [0,M_t-1]$
summarizes the centrality properties of $i$ with respect to the hyper-core decomposition at time $t$ by taking into account its relative
depth in the $(k,m)$-core structure at all interaction orders \cite{mancastroppa2023hyper}
\footnote{It is possible to define a whole
family of hypercoreness centralities \protect\cite{mancastroppa2023hyper}
by arbitrarily weighing the different hyperedge sizes $m$ in Eq.
\protect\eqref{eq:hypercoreness}. Here we consider the simplest
``size-independent'' hypercoreness in which all sizes contribute equally}. 

By considering the hyper-core decomposition of the successive snapshots forming the temporal hypergraph, we can thus follow the temporal evolution of its higher-order hierarchical structure, and obtain a characterization of the higher-order dynamics at several scales, as we now discuss.

\textbf{Macroscopic scale.} The fraction of nodes within the $(k,m)$-hyper-cores, $n_{(k,m)}$, as a function of $k$ and $m$ constitutes the {\em filling profile} of the hyper-cores, and
provides information on the distribution of nodes in the various cores and shells. 
Following its evolution across successive snapshots yields information on how the overall system's cohesiveness changes over time.
The filling profile can indeed detect changes in the underlying higher-order hierarchical structure, since different distributions of nodes in the hyper-cores reflect different configurations of interactions in the nested hierarchy \cite{mancastroppa2023hyper}: for instance, a smooth decay of $n_{(k,m)}$ with $k$ and $m$ suggests the presence of nodes progressively more densely connected with each other through interactions of larger sizes (homogeneously populated shells), while the alternation of plateaus and abrupt drops reveals the presence of a non-trivial structure, with nodes poorly or densely connected with each other, without intermediate behaviours (unevenly filled shells). 
Thus, the similarity between the hyper-cores filling profiles of two different snapshots, $n_{(k,m)}(t)$ and $n_{(k,m)}(t')$, provides a quantitative estimate of the stability of the macroscopic hyper-core structure over time. 
While several similarity measures can be defined between the filling profiles of two hypergraphs, we consider here the root-mean-square deviation similarity, defined as follows for the filling profiles $a_{(k,m)}$ and $b_{(k,m)}$ of two static hypergraphs $\mathcal{A}$ and $\mathcal{B}$ with respective maximum connectivities $k_{max}^m(\mathcal{A})$ and $k_{max}^m(\mathcal{B})$ $\forall m$, and 
respective maximum hyperedge sizes $M_{\mathcal{A}}$ and $M_{\mathcal{B}}$:
\begin{equation}
\Sigma(\mathcal{A},\mathcal{B}) = 1- \sqrt{\frac{\sum\limits_{k=1}^{\overline{K}} \sum\limits_{m=2}^{\overline{M}} \left(a_{(k,m)} - b_{(k,m)}\right)^2}{\overline{K} \, (\overline{M}-1)-1}},
\label{eq:Sigma}
\end{equation}
with $\overline{K}= \max\limits_m\{ \max\{k_{max}^m(\mathcal{A}),k_{max}^m(\mathcal{B})\} \}$ and $\overline{M}=\max\{M_{\mathcal{A}},M_{\mathcal{B}}\}$
(in this way $\Sigma \in [0,1]$)
\footnote{The maximum similarity $\Sigma=1$ is obtained when the two hyper-cores filling profiles are identical $a_{(k,m)}=b_{(k,m)} \, \forall k \in [1,\overline{K}], \, \forall m \in [2,\overline{M}]$; the minimum similarity $\Sigma=0$ is obtained when the hypergraphs feature the two maximally different configurations, $a_{(1,2)}=1$, $a_{(k,m)}=0$ otherwise, and $b_{(k,m)}=1 \, \forall k \in [1,\overline{K}], \, \forall m \in [2,\overline{M}]$, i.e. in one case the $(1,2)$-core contains the entire population while all the other hyper-cores are empty, and in the other case all the hyper-cores are maximally filled with the entire population.} 
\footnote{This measure can be applied to any couple of hypergraphs with different populations, numbers of hyperedges, 
distributions of hyperdegrees $P(D_m^{\mathcal{H}})$ $\forall m \in [2,M]$ 
and distributions of interactions size $\Psi(m)$. In general, systems with similar $P(D_m^{\mathcal{H}})$ and $\Psi(m)$ feature a higher similarity compared to those with different distributions.}. 
The temporal similarity matrix $\Sigma(t,t') = \Sigma(\mathcal{H}_t, \mathcal{H}_{t'})$ provides then a way to explore the existence of various temporal patterns in the hyper-core decomposition of the system at different times, and to unveil the presence of stable periods, recurrences or sudden changes
\cite{masuda2019detecting,sugishita2020recurrence,pedreschi2020dynamic,braha2009time} \footnote{Note that different similarity measures could be considered to build such similarity matrix.}.

\textbf{Mesoscopic scale.} By following the hyper-core decomposition over time, it is moreover possible to study the  
temporal stability and changes occurring in subhypergraphs with specific structural roles. To this aim, we can consider a given set of shells or cores, and compare their sets of nodes $A$ in two different snapshots $t$ and $t'$ through the 
Jaccard similarity $J(t,t')=|A_t \cap A_{t'}|/|A_t  \cup A_{t'}|$.
The matrix $J(t,t')$ quantifies the stability over time of the set of nodes forming the cores under scrutiny.
In particular, we will here focus on the set of the most central hyper-cores of each snapshot, i.e. the $(k_{max}^m,m)$-hyper-cores $\forall m$. We can then determine whether these cores are stable, involving always the same nodes across snapshots, or whether their composition evolves, due to changes of connectivity of individual nodes: this can happen even when the macroscopic structure remains similar 
(as found in temporal networks where the most connected nodes can vary with time \cite{braha2006centrality}, or a core-periphery structure can be stable even when the composition of the core strongly fluctuates \cite{pedreschi2020dynamic}).

Moreover, empirical data include sometimes meta-data
(see Methods) describing properties or attributes of the nodes or hyperedges, and dividing them into classes based on their specific function or context. For instance, data describing social interactions can be enriched by information on the individuals involved (e.g., to which class they belong in a school environment, to which department or which role they have in a work environment). Such information makes it possible to study whether different groups or classes of nodes have different higher-order structural properties, and whether specific hyper-cores are preferentially composed by specific nodes or specific types of hyperedges.
For instance, one can identify the most represented class in each hyper-core at each time, and follow over time which types of nodes or hyperedges are dominant in the most central hyper-cores.

\textbf{Microscopic scale.} At the node level, the hypercoreness  $R(i,t)$ gives an instantaneous measure of the centrality of a node in each snapshot. It is thus possible, for each node of interest, to follow its trajectory in the hyper-core structure through the evolution of its hypercoreness. More precisely, in order to make the hypercoreness values comparable across different snapshots, we consider the temporal evolution of the relative position of each node $i$ in the hypercoreness ranking:
\begin{equation}
r(i,t) = \frac{R(i,t)}{\max\limits_{j \in \mathcal{V}_t} \{R(j,t)\}}.
\end{equation}
The evolution of $r(i,t)$ with $t$ indeed 
reflects the movements that node $i$ undergoes within the hierarchical structure, potentially navigating towards more central or more superficial cores. 

The set of all $R(i,t)$ moreover provides an instantaneous node hierarchy within the time window $t$. Such a hierarchy might  fluctuate from one snapshot to the next \cite{braha2006centrality}, and the Pearson correlation coefficient $\varrho(t,t')=\varrho(R(i,t),R(i,t'))$ of the nodes hypercoreness values between two time snapshots $t$ and $t'$ provides information on the stability of the node ranking over time, i.e., on how the nodes change their respective structural positions over time. Just as $\Sigma(t,t')$ for the global scale and $J(t,t')$ for intermediate scales, this measure can unveil correlation patterns at various time-scales: for example, a high and constant $\varrho(t,t')$ indicates that nodes tend to keep their relative structural positions over time, while constantly low values correspond to an unstable situation with nodes continuously changing place in the hierarchy. 

Note that, as not all nodes are active in each snapshot, 
we can compute $\varrho(t,t')$ in two ways: (i) $\rho^*(t,t')$
takes into account only the nodes that are active in both $t$ and $t'$, while (ii) $\rho(t,t')$ is computed considering all nodes active in at least one of them (setting the hypercoreness
of inactive nodes to $0$).
The difference between $\rho(t,t+1)$ and $\rho^*(t,t+1)$ provides information on the structural properties of nodes just after entering the system or right before leaving it:
$\rho \lesssim \rho^*$ indicates that nodes have mainly low hypercoreness when joining/leaving the system, while $\rho \ll \rho^*$ indicates that nodes joining/leaving the system tend to be central.

\subsection*{Time-aggregated hypercoreness centralities}

The hypercoreness centrality of nodes in static hypergraphs has been shown to provide information on their importance for dynamic processes involving higher-order interactions unfolding on such hypergraphs \cite{mancastroppa2023hyper}. Many processes however unfold on time-varying hypergraphs 
\cite{chowdhary2021simplicial,Neuhauser2021}, hence a 
time-aggregated ranking of nodes summarizing the evolution of their instantaneous coreness could prove useful. 

We first define \textbf{the snapshot activity $a_w(i) \in [0,n]$,} 
given by the number of time windows in which node $i$ is active, and
\textbf{the average number of interactions when 
active $\overline{h}(i)=D(i)/a_w(i)$}, where $D(i)$ is the total number of hyperedges in which $i$ is involved in the temporal hypergraph. 
We then introduce two time-aggregated centrality measures that summarize the positions of the nodes in the hyper-core structure over time:
\begin{itemize}
\item \textbf{the aggregated hypercoreness $W$}:
\begin{equation}
W(i)=\sum_{t=1}^{n} \frac{R(i,t)}{\max\limits_{j \in \mathcal{V}_t}
\{R(j,t)\}}=\sum_{t=1}^{n} r(i,t) ,
\end{equation}
takes into account how deep $i$ is in the hyper-core structure at the various interaction orders in each time window, and simply aggregates this information over time. 
\item \textbf{the activity-averaged hypercoreness $\overline{W}$}:
\begin{equation}
\overline{W}(i)=\sum_{t=1}^{n} \frac{r(i,t)}{a_w(i)} = \frac{W(i)}{a_w(i)} ,
\label{eq:over_W}
\end{equation}
averages $W$ over the activity of the nodes.
\end{itemize}
$W$ and $\overline{W}$ provide complementary information.
Indeed, a high $W$ can be obtained either for 
a node $i$ that is very active (high $a_w(i)$) but not very central (small $r(i,t)$) or for a node $j$ that is 
not very active (low $a_w(j)$) but central when active 
(high $r(j,t)$). These two situations are distinguished when taking into account also $\overline{W}$, as $\overline{W}(i)$ will then be 
small while $\overline{W}(j)$ will be large.
Together, the time-aggregated hypercoreness measures $W(i)$ and $\overline{W}(i)$ thus provide a two-dimensional picture taking into account both the activity of nodes and the evolution of their relative centralities over time.\\

The approach outlined above is general and can be applied to  empirical data of higher-order interactions evolving over time describing a variety of systems. In the following, we showcase its interest using: a data set of scientific collaborations \cite{APS,journals}, several data sets of physical proximity interactions between individuals in various environments \cite{SP,genois2015,Genois2018,ISELLA2011166,cattuto2013,barrat2011,Mastrandrea2015,Toth2015,Sapiezynski2019} (a hospital \cite{cattuto2013}, a conference \cite{Genois2018}, three schools \cite{barrat2011,Mastrandrea2015,Toth2015}, a university \cite{Sapiezynski2019} and a workplace \cite{genois2015,Genois2018}), and a data set of email communications \cite{emailEUdata,benson2018,Benson_site}. These data sets present different statistical, topological and temporal properties (e.g., interaction size distribution, temporal patterns due to system-specific activities). In the main text we specifically analyse data describing scientific collaborations, and proximity data collected in a hospital and in a university. Results for the other data sets are reported in the Supplementary Material (SM). In addition, we show ---in the case of the university data--- how the analysis of the hyper-core structure over time can contribute to the validation of models of time-varying hypergraphs.

\subsection*{Dynamics of the higher-order structure of scientific collaborations}
The scientific collaborations data set of the American Physical Society (APS) provides the list of papers published in APS journals from 1893 to 2021 \cite{APS,journals}. From this data set, we build a temporal hypergraph (see Methods) in which each node corresponds to an author and each hyperedge represents a paper connecting its co-authors. Each hyperedge is endowed with a label indicating the journal in which the paper was published and its publication date.
We consider a 5-years temporal resolution, i.e., each temporal snapshot is formed by all papers published in a 5-years time window
(see SM for a different temporal resolution), and we consider the period 1962-2001 (earlier years having only much smaller numbers of nodes and hyperedges).

\begin{figure*}
    \centering
    \includegraphics[width=0.9\textwidth]{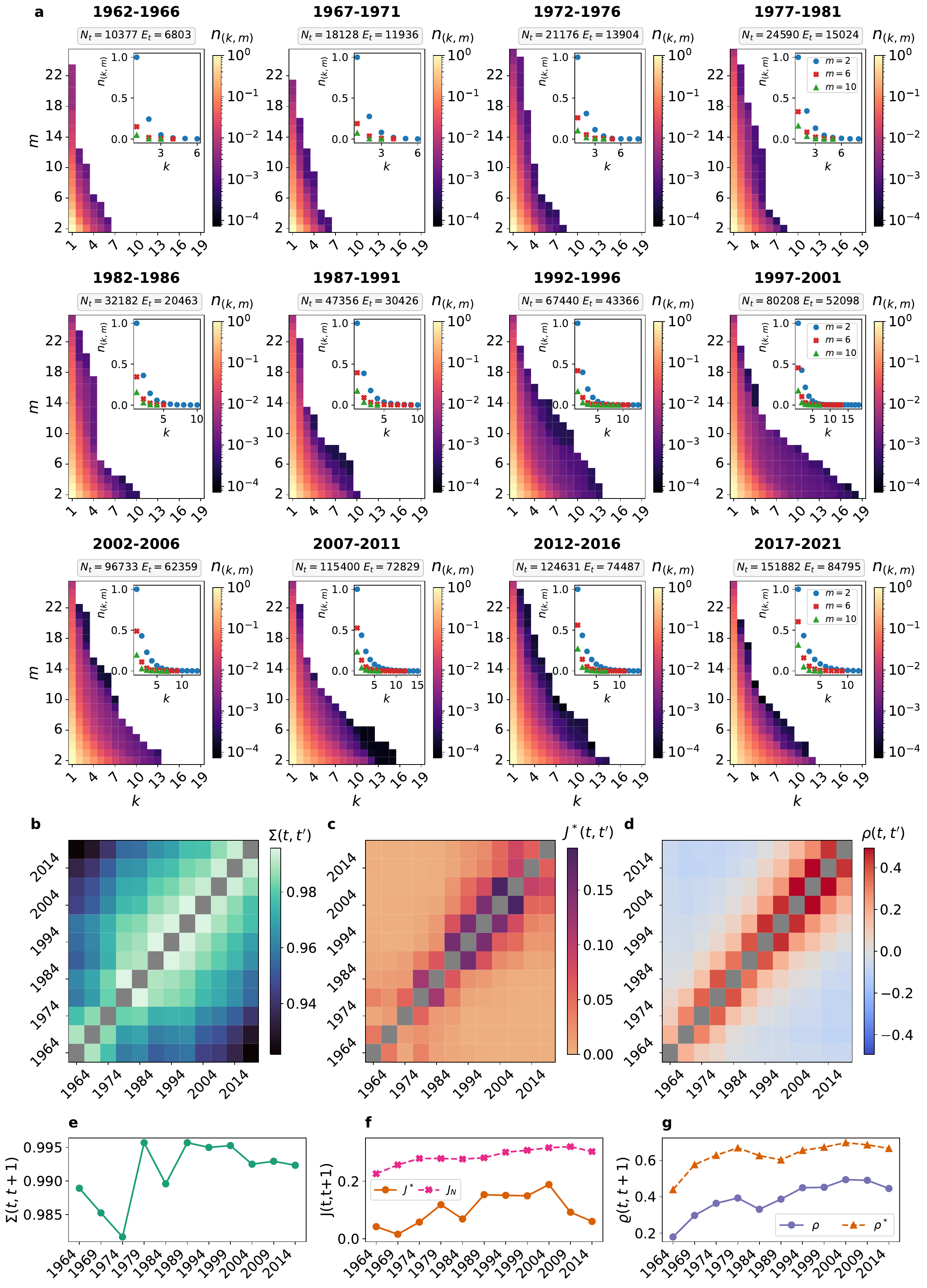}
    \caption{\textbf{Evolution of the hyper-core structure in APS scientific collaborations.} 
    \textbf{a:} fraction of nodes $n_{(k,m)}$ in the $(k,m)$-core as a function of $k$ and $m$ for each 5-years time window.  The numbers of active nodes $N_t$ and hyperedges $E_t$ are also reported and the insets show $n_{(k,m)}$ as a function of $k$ for $m=2$, $m=6$ and $m=10$. 
    \textbf{b:} root-mean-square deviation similarity $\Sigma(t,t')$ between $n_{(k,m)}(t)$ and $n_{(k,m)}(t')$ (grey diagonal: $\Sigma(t,t)=1$).
    \textbf{c:} Jaccard similarity $J^*(t,t')$ between the sets of nodes belonging to the most central hyper-cores, i.e. to the $(k_{max}^m,m)$-cores $\forall m$, at time $t$ and $t'$ (grey diagonal: $J^*(t,t)=1$). 
    \textbf{d:} Pearson correlation coefficient $\rho(t,t')$ between the nodes hypercoreness at times $t$ and $t'$, considering all the nodes that are active in at least one of the snapshots (grey diagonal: $\rho(t,t)=1$). 
    \textbf{e:} similarity $\Sigma(t,t+1)$ vs. $t$.  
    \textbf{f:} temporal evolution of $J^*(t,t+1)$ and Jaccard similarity $J_N(t,t+1)$ between the entire population in two consecutive time windows.
    \textbf{g:} temporal evolution of the correlation between the nodes hypercoreness in consecutive snapshots, considering all the nodes that are active in at least one of the snapshots, $\rho(t,t+1)$, or only those active in both, $\rho^*(t,t+1)$. 
    }
    \label{fig:figure1}
\end{figure*}

\begin{figure}
    \centering
    \includegraphics[width=0.7\columnwidth]{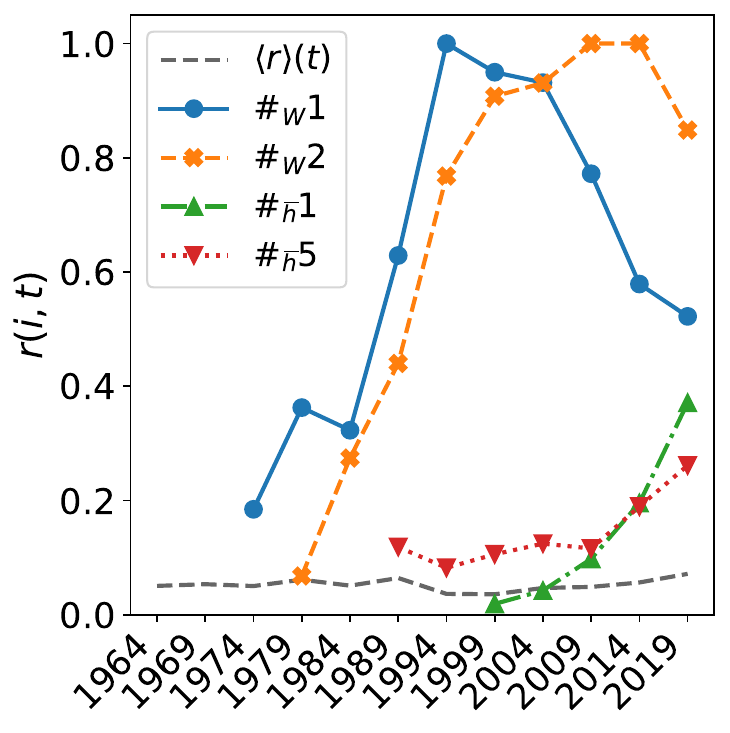}
    \caption{\textbf{Hypercoreness evolution for selected nodes in the APS scientific collaborations.} We show the temporal evolution of the hypercoreness $r(i,t)$ for four authors and the mean $\langle r \rangle (t)$ value (average on active nodes): we show the authors I.Y. Lee ($\#_W1$) and R.V.F. Janssens ($\#_W2$), who occupy respectively the first and second position in the ranking produced by the aggregated hypercoreness $W$ over the period 1942-2021, and the authors Guang-Can Guo ($\#_{\overline{h}}1$) and Loren N. Pfeiffer ($\#_{\overline{h}}5$), who occupy respectively the first and fifth position in the ranking produced by the average number of interactions per active windows $\bar{h}$ over the period 1942-2021. 
    }
    \label{fig:figure2}
\end{figure}

Figure  \ref{fig:figure1}\textbf{a} shows the evolution of
the global hyper-cores structure as given by the filling profiles, which do not simply expand in a monotonous fashion as the numbers of nodes and hyperedges increase over the years. Initially the system presents only $(k,m)$-hyper-cores with low connectivity $k$, especially for large hyperedge sizes $m$; then, the filling profile undergoes an expansion towards higher $k$ and higher $m$ values. At first, $k_{max}^m$ increases for high interaction orders $m$ and only later at low orders. Furthermore, the increase in $k_{max}^m$ is non-monotonic with respect to time, especially for low $m$: $k_{max}^m$ for $m \gtrsim 2$ grows up to a maximum in the 1997-2001 snapshot, and then decreases and stabilizes in the following years (as we will discuss below, this behavior can be traced back to a specific scientific community and its collaboration dynamics). Thus, the cohesiveness of the scientific community first increased through connected large size collaborations, then an increase in cohesiveness occurred at all orders until 1997-2001. The cohesiveness of the community then relaxed to a lower but stationary level in the last 20 years.

Although the size of the interactions and the density of collaborations change over time, the overall structure of the filling profiles remains similar instead (Fig. \ref{fig:figure1}\textbf{a}). In fact, the hyper-cores always present a rapid and progressive emptying of the cores as $k$ and $m$ increase: superficial shells (low $k$) are densely populated, and shells become gradually less populated with increasing $k$ and $m$. The root-mean-square deviation similarity $\Sigma(t,t')$ between the hyper-cores filling profiles at time $t$ and $t'$ presents very high values for all pairs $(t,t')$ 
(Fig. \ref{fig:figure1}\textbf{b}), indicating a stable structure: the similarity is particularly high between consecutive snapshots (Fig. \ref{fig:figure1}\textbf{e}), and decreases monotonically  when $|t'-t|$ increases. 

We investigate the mesostructural level through the similarity $J^*(t,t')$ between the sets of nodes belonging to the most central cores, i.e. to the $(k_{max}^m,m)$-hyper-cores $\forall m$ at different times. Figure \ref{fig:figure1}\textbf{c},\textbf{f}
shows that the stability of the central cores is low, even between adjacent time windows. This is not only due to the fact that the set of authors change over time, as $J^*$ is much lower than the Jaccard coefficient $J_N$ between the sets of authors in different time windows. $J^*(t,t')$ moreover decreases to $0$ as soon as the time difference $|t'-t|$ exceeds 2-4 time windows, indicating a completely different composition of the central hyper-cores. Note that a tendency to increase the stability of the central cores can be seen until $\approx 2010$ (Fig. \ref{fig:figure1}\textbf{c},\textbf{f}), although it decreases again afterwards.
Overall the $J^*$ values remain low, indicating that the nodes sitting in the most central hyper-cores change over time. 

\begin{figure*}
    \centering
    \includegraphics[width=\textwidth]{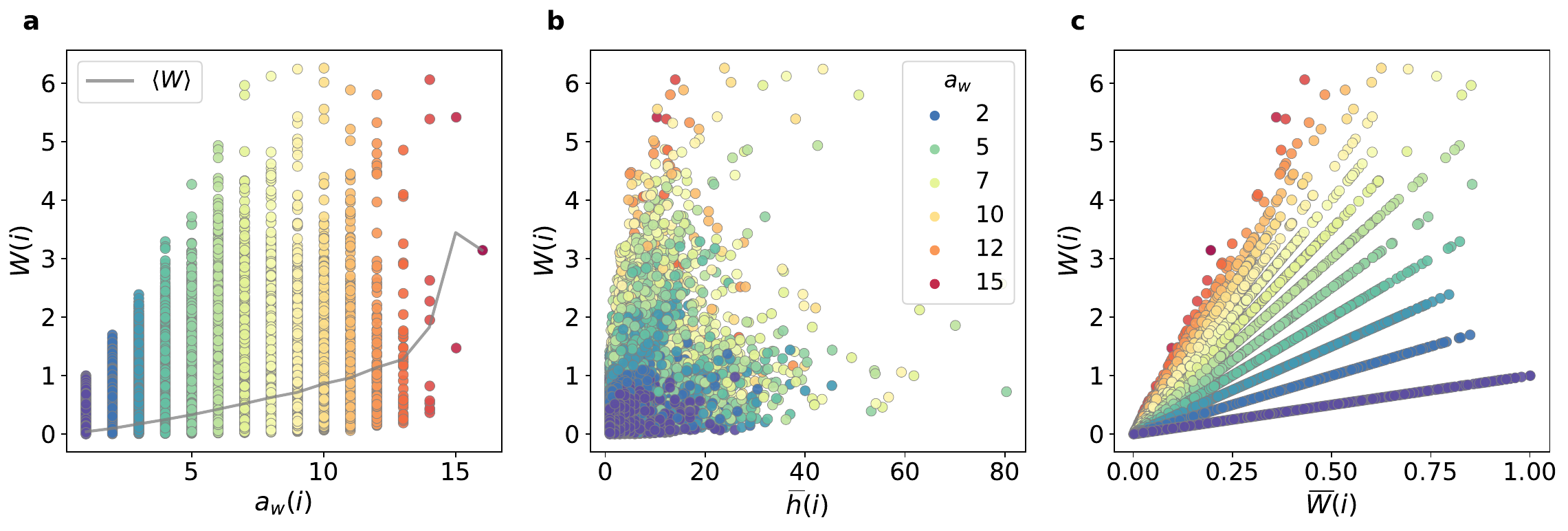}
    \caption{\textbf{Time-aggregated hypercoreness in APS scientific collaborations 1942-2021.} \textbf{a:} scatter plot of the aggregated hypercoreness $W(i)$ as a function of the snapshot activity $a_w(i)$ for all nodes $i$, and average aggregated hypercoreness $\langle W \rangle$ as a function of $a_w$. 
    \textbf{b:} aggregated hypercoreness $W(i)$ vs. average number of interactions per active window $\overline{h}(i)$ for all nodes $i$. 
    \textbf{c:} aggregated hypercoreness $W(i)$ as a function of the activity-averaged hypercoreness $\overline{W}(i)$. 
    In all panels the points are colored according to the activity $a_w$ of the corresponding node.
    }
    \label{fig:figure3}
\end{figure*}

We further explore this instability using the correlation $\rho(t,t')$ of nodes hypercoreness across different time windows, as shown in Fig. \ref{fig:figure1}\textbf{d},\textbf{g}. A positive correlation is observed between the hypercoreness values of nodes in successive snapshots, but the correlation $\rho(t,t+1)$ computed using all nodes active at least once in $(t,t+1)$ is lower than 
$\rho^*(t,t+1)$, which takes into account only nodes active in both snapshots (Fig. \ref{fig:figure1}\textbf{g}). As discussed above, this indicates that some nodes with high centrality leave the system, and/or nodes enter the system and gain immediately a central position. As the temporal distance $|t'-t|$ increases, the correlation $\rho(t,t')$ progressively decreases. 
Moreover, the correlations tend to increase with $t$: 
$\rho(t,t+1)$ increases with $t$ and the decrease of  
$\rho(t,t')$ with $t-t'$ becomes slower  
(Fig. \ref{fig:figure1}\textbf{d},\textbf{g}), indicating an increased stability in centrality rankings as time evolves.

The correlation between hypercoreness values decays to zero in approximately 3-5 time windows and then reaches negative values: this suggests a progressive inversion of the rankings over time, with nodes successively increasing and decreasing their hypercoreness and rankings, as driven by the unfolding of their academic careers. 
Figure \ref{fig:figure2} indeed gives some examples of the evolution of individual nodes' relative hypercoreness $r(i,t)$, which are a reflection of the academic trajectories of the corresponding scientists. Some nodes have a bell-shaped hypercoreness profile, entering the system with a low centrality, progressively moving towards the more central cores and then back to lower ranks.
This can describe the academic trajectory of a young researcher, who enters into the scientific community, becomes central and then progressively leaves the community due to retirement or a change in the topic/journals reference of their research. Other nodes present instead a rather stable ranking, and, for individuals having entered the system more recently, only the upward trend of increasing centrality is observed.

To characterize the nodes' overall behaviours, we moreover compute their time-aggregated centrality measures, and show the results in Fig.~\ref{fig:figure3}. On average, the aggregated hypercoreness
$\langle W \rangle$ increases with the activity snapshot $a_w$ (Fig. \ref{fig:figure3}\textbf{a}), but a large variability in the values of $W$ is observed at given $a_w$.
Some nodes can be very active but display a low centrality, while nodes with moderate activity can reach large values of $W$.
The average number of interactions per active window $\overline{h}$ 
is also only weakly correlated with $W$, and the nodes with highest $W$ do not coincide with those with largest $\overline{h}$ 
(see Fig. \ref{fig:figure3}\textbf{b}).
Finally, the aggregated and activity-averaged hypercoreness, $W$ and $\overline{W}$, also do not produce the same ranking (see Fig. \ref{fig:figure3}\textbf{c}). 
Some nodes are not often active (low $a_w$) with medium-low $W$ but high $\overline{W}$: these authors appear in few windows but within very connected communities, therefore are very central on average when active but their low $a_w$ make them less relevant in aggregated terms. Other nodes are very active (high $a_w$) with medium-high $W$ but relatively low $\overline{W}$: such authors are often active either with a low centrality or with non-monotonous hypercoreness profile (see Fig. \ref{fig:figure2}). Overall, the combined information of $W$ and $\overline{W}$ provide a more complete description of nodes structural behavior on the whole time span than when considering only one of these centralities.

\begin{figure*}
    \centering
    \includegraphics[width=\textwidth]{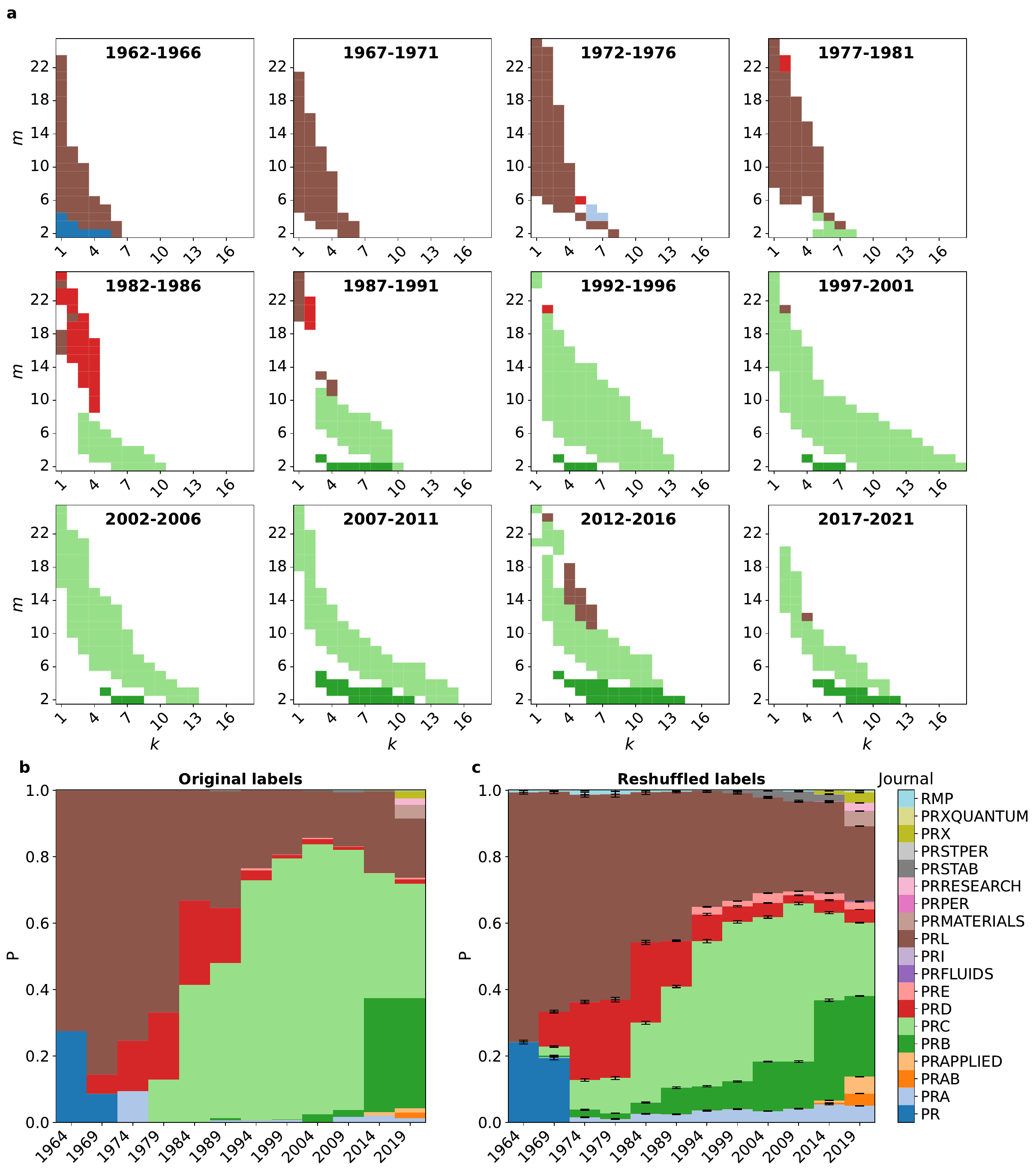}
    \caption{\textbf{Prevalent APS scientific communities in hyper-cores.} 
    \textbf{a:} temporal evolution over 5-years time windows of the prevalent journal within each $(k,m)$-hyper-core of the APS data set, defined as the most frequent hyperedge label in each core (we consider a journal dominant only if its frequency is larger than 0.5; white indicates hyper-cores which are empty or where a dominant journal cannot be defined).
    \textbf{b:} relative frequency $P$ of the various journals within the most central hyper-cores, i.e. $(k_{max}^m,m)$-cores $\forall m$, and its temporal evolution. 
    \textbf{c:} same as \textbf{b} for the randomized data. We average the relative frequency over 50 randomized realizations of the hypergraph (see Methods). The error bars give the standard errors.
    }
    \label{fig:figure4}
\end{figure*}

We finally leverage the fact that each hyperedge representing a scientific article is labelled by the journal it was published in to examine the importance of the various APS journals in the hyper-core structure. 
The APS journals can be interdisciplinary (e.g. PRL) or specialized in a specific research field (e.g. PRC for nuclear physics, PRD for high-energy physics, PRB for condensed matter physics), thus representing a specific research area \cite{journals} (see SM). 

For each $(k,m)$-core we consider all the hyperedges it contains and their labels, and we identify the dominant journal (namely, whose frequency exceeds $0.5$; if no journal is represented by more than half of the hyperedges, we consider that no journal dominates) \footnote{In the aggregation procedure to create each snapshot, some hyperedges can fully overlap (i.e., the same group of authors can publish more than one article). Although we do not consider weighted hyperedges, in such a case we assign a multiple label composed of the set of journals in which these articles with the same co-authors were published.}. 
Figure \ref{fig:figure4}\textbf{a} shows the resulting evolution of the hyper-cores dominant journal. 
Initially, PR and PRL dominate within all the hypercores, since they were the only available journals together with RMP (not shown in the figure, see SM). Then, the more superficial cores present a mixed composition, while the most central ones are first dominated by PRL in the period 1962-1981; subsequently in 1982-1986, central cores are mostly formed by the high-energy physics community (PRD) for large collaboration sizes, while at low order the nuclear physics area dominates (PRC). 
Starting from 1992, PRC dominates the most central hyper-cores at all orders: the non-monotonic behavior observed in the core structure, with the maximum connectivity in 1997-2001, is predominantly due to interactions within the nuclear physics area. This could be due to several discoveries in the field occurring in the preceding years (e.g. the discovery of the W and Z bosons \cite{WZ} or the discovery of top quarks \cite{PRL1,PRL2}), which boosted collaborations in the community, favouring and increasing cohesion. After this phase the nuclear physics area remains overall dominant. Moreover, this non-monotonic behavior can also be identified in the hyper-core decomposition of the hypergraphs obtained by considering only the papers published in PRC (see SM). 
Recently, the condensed matter physics community (PRB) is also expanding its contribution to the central cores at low interaction orders. 
The relative contribution of the scientific communities to the set of the most central cores is summarized in Fig. \ref{fig:figure4}\textbf{b}: PRL is the dominant journal in the first time windows, while the share of PRC increases rapidly starting in the 80s; 
the share of PRB becomes also important from 2012-2016 and in 2017-2021 new journals start gaining relevance (e.g. PRX). 

As the number of scientists and articles in various fields are neither homogeneous nor constant, we check whether such patterns are simply due to the relative abundance of authors and articles in the different journals. To this aim, we build a randomized version of the temporal hypergraph, which preserves in each time window the hypergraph structure and the total number of interactions of each order for each label, but destroys any correlation between the nodes and the label of the hyperedges in which they participate (see the reshuffling procedure in Methods). We consider 50 randomized realizations and for each hyper-core we estimate the average frequency of each label. The patterns of topic dominance in the most central cores is significantly different in the reshuffled version compared to the empirical case (see Fig. \ref{fig:figure4}\textbf{b},\textbf{c} and SM). For example, in the reshuffled case PRA, PRB and PRE are significantly more represented in the central cores, while PRC is instead less represented than in the original data.

It is also possible to consider a different time resolution for building the temporal hypergraph, to investigate e.g. the dynamics at shorter time-scales, or to focus on one specific scientific community by considering the hypergraph formed by articles published in one specific journal. We refer to the SM for some results in such directions.

\subsection*{Higher-order structure dynamics of interactions in a hospital}

We now consider a data set of face-to-face interactions in a hospital (LH10), collected within the SocioPatterns collaboration \cite{SP,cattuto2013}. The original data set has a temporal resolution of $20$ seconds. We first build a temporal hypergraph in which each node corresponds to an individual and each hyperedge represents a group interaction, defined with a temporal resolution of 5 minutes (see Methods and \cite{iacopini2019simplicial}). 
As the data set covers a period of 96 hours, we first study differences in the daily aggregated hypergraph structures, i.e., we aggregate the temporal hypergraph over 
24-hours time windows (thus here we obtain $n=4$ time windows). 
Each node is assigned with a label indicating its social role: \texttt{Med} for doctors, \texttt{Param} for nurses, \texttt{Admin} for administrative staff, and \texttt{Patient} for patients.

\begin{figure*}
    \centering
    \includegraphics[width=\textwidth]{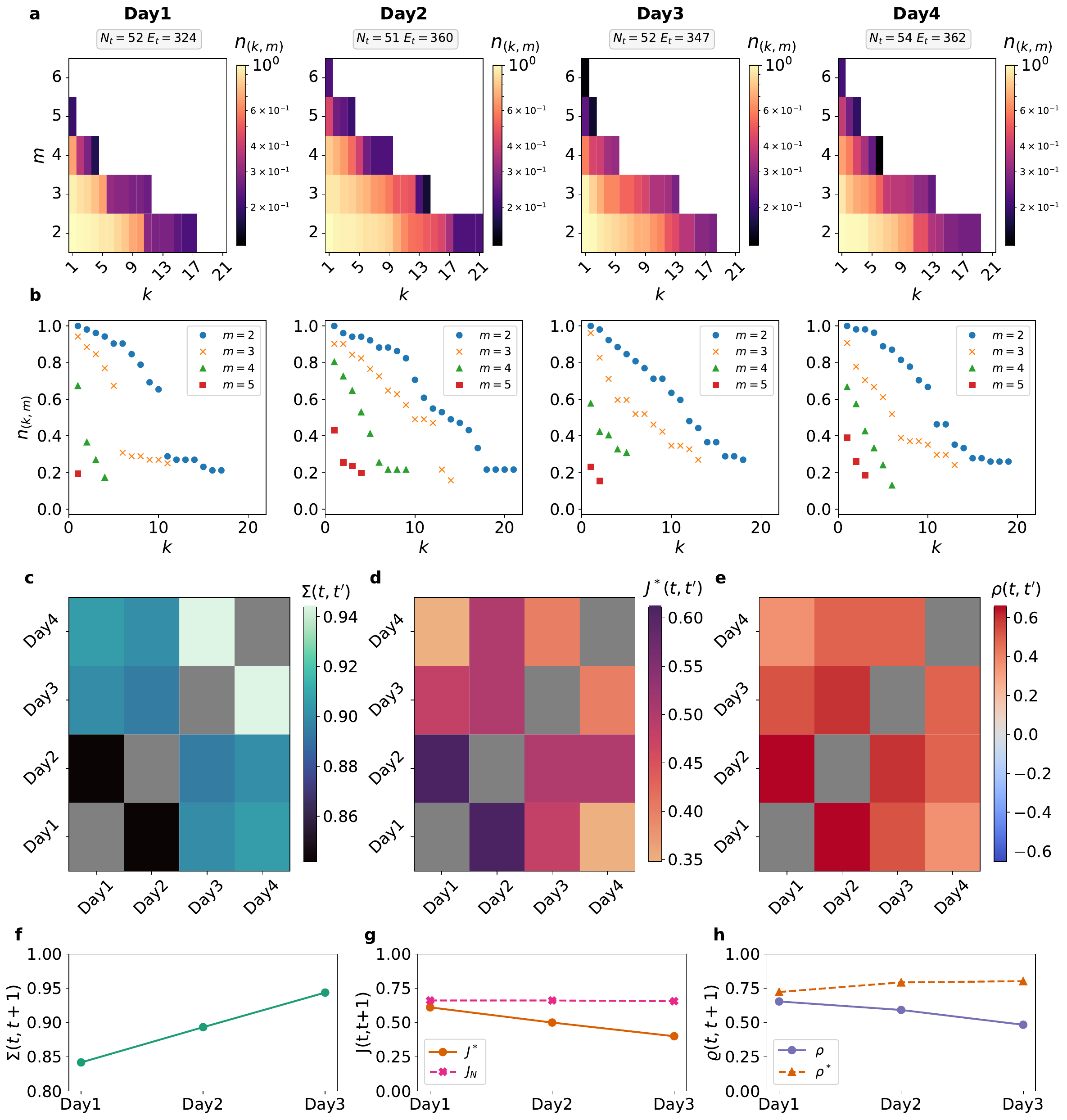}
    \caption{\textbf{Hyper-core structure evolution in daily interactions within a hospital (LH10)}. 
    \textbf{a:} relative population $n_{(k,m)}$ of the $(k,m)$-core as a function of $k$ and $m$ for each time window. The number of active nodes $N_t$ and hyperedges $E_t$ is reported for each snapshot.  
    \textbf{b:} $n_{(k,m)}$ as a function of $k$ for fixed values of $m$. 
    \textbf{c:} root-mean-square deviation similarity $\Sigma(t,t')$ between $n_{(k,m)}(t)$ and $n_{(k,m)}(t')$  -- the grey diagonal corresponds to $\Sigma(t,t)=1$;  
    \textbf{d:} Jaccard similarity $J^*(t,t')$ between the sets of nodes belonging to the most central hyper-cores, i.e. the $(k_{max}^m,m)$-cores $\forall m$, at time $t$ and $t'$ -- the grey diagonal corresponds to $J^*(t,t)=1$. 
    \textbf{e:} Pearson correlation coefficient $\rho(t,t')$ between the nodes hypercoreness at time $t$ and $t'$, considering all the nodes that are active in at least one of the snapshots -- the grey diagonal corresponds to $\rho(t,t)=1$.
    \textbf{f:} similarity $\Sigma(t,t+1)$ as a function of $t$.
    \textbf{g:} temporal evolution of both the similarity $J^*(t,t+1)$ and the Jaccard similarity $J_N(t,t+1)$ between the entire population in consecutive time windows. 
    \textbf{h:} temporal evolution of the correlation between the nodes hypercoreness in consecutive snapshots, considering all the nodes that are active in at least one of the snapshots, $\rho(t,t+1)$, or that are active in both, $\rho^*(t,t+1)$.
    }
    \label{fig:figure5}
\end{figure*}

\begin{figure}
    \centering
    \includegraphics[width=0.7\columnwidth]{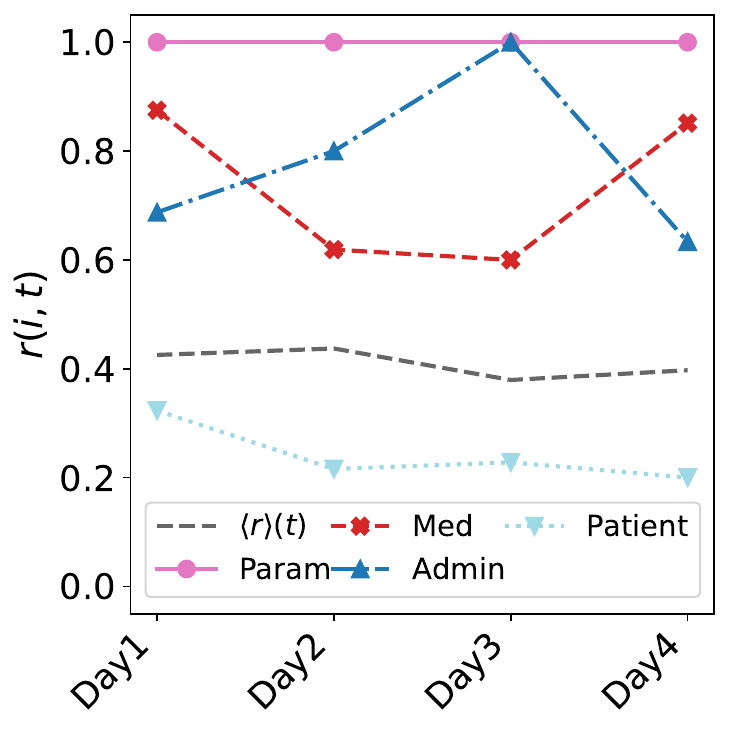}
    \caption{\textbf{Hypercoreness evolution in the temporal hypergraph of daily interactions within a hospital (LH10).} We show the temporal evolution of the hypercoreness $r(i,t)$ for four agents with different social role: a paramedic (\texttt{id=1210}), a medic (\texttt{id=1144}), a member of the administrative staff (\texttt{id=1098}) and a patient (\texttt{id=1383}). The dashed line shows the mean $\langle r \rangle (t)$ (averaged only on active nodes). Note how the patient's hypercoreness is always lower than average while the paramedic's hypercoreness is always maximal.}
    \label{fig:figure6}
\end{figure}

The maximum size of interactions $M_t$ and the maximum connectivity values $k_{max}^m(t)$ $\forall m$, i.e. the cohesiveness of the system, are rather stable over different days (Fig. \ref{fig:figure5}\textbf{a},\textbf{b}). However, nodes are differently distributed within the cores. On the first day, the population of the $(k,m)$-cores features sharp drops when $k$ increases, followed by plateaus: these correspond to densely populated shells at small $k$ followed by almost empty shells. 
In other days, the structure instead presents a more progressive emptying of the cores as $k$ increases, hence shells are populated more homogeneously (even if some jumps and plateaus of reduced sizes are still present). 
The root-mean-square deviation similarity $\Sigma(t,t')$ between the hyper-cores filling profiles at time $t$ and $t'$ still presents high values for all pairs $(t,t')$ (see Fig. \ref{fig:figure5}\textbf{c}), however the similarity is lower than the one observed for the APS data set. Moreover, the similarity $\Sigma$ between consecutive snapshots increases over time (Fig. \ref{fig:figure5}\textbf{f}).

Mesoscopically the system is quite stable (see Fig. \ref{fig:figure5}\textbf{d},\textbf{g}): the similarity $J^*(t,t')$ between the nodes in the most central cores at time $t$ and $t'$ presents medium-high values, and $J^*(t,t')$ slightly decreases when increasing $|t'-t|$ and in consecutive time windows it still assumes values close to the similarity of the entire population $J_N$, even if decreasing over time. The composition of the most central cores is thus quite stable, therefore in general the nodes maintain the same position in the core structure. This is confirmed by the correlation $\rho(t,t')$ in the nodes hypercoreness between two snapshots (see Fig. \ref{fig:figure5}\textbf{e},\textbf{h}). The correlation $\rho(t,t')$ presents high values. As we will explore further below, this stability in the composition of central cores and in the behavior of the nodes is due to the difference in the roles played by the different individuals in the hospital, which limits the mobility of the nodes in the hyper-core structure.

Note that, even if the position of the nodes in the hyper-core structure is fairly stable over time, the evolution of the hypercoreness $r(i,t)$ for single nodes can show different trajectories. This is evident when disaggregating by social role, as for the examples in Fig. \ref{fig:figure6}: the nodes can present a stable dynamic with a constant position in the core structure, as shown by the patient and the paramedic cases, or a non-monotonic dynamic, with movements from more central cores towards more superficial ones and vice-versa, as for the doctor and the administrative staff member.

\begin{figure*}
    \centering
    \includegraphics[width=\textwidth]{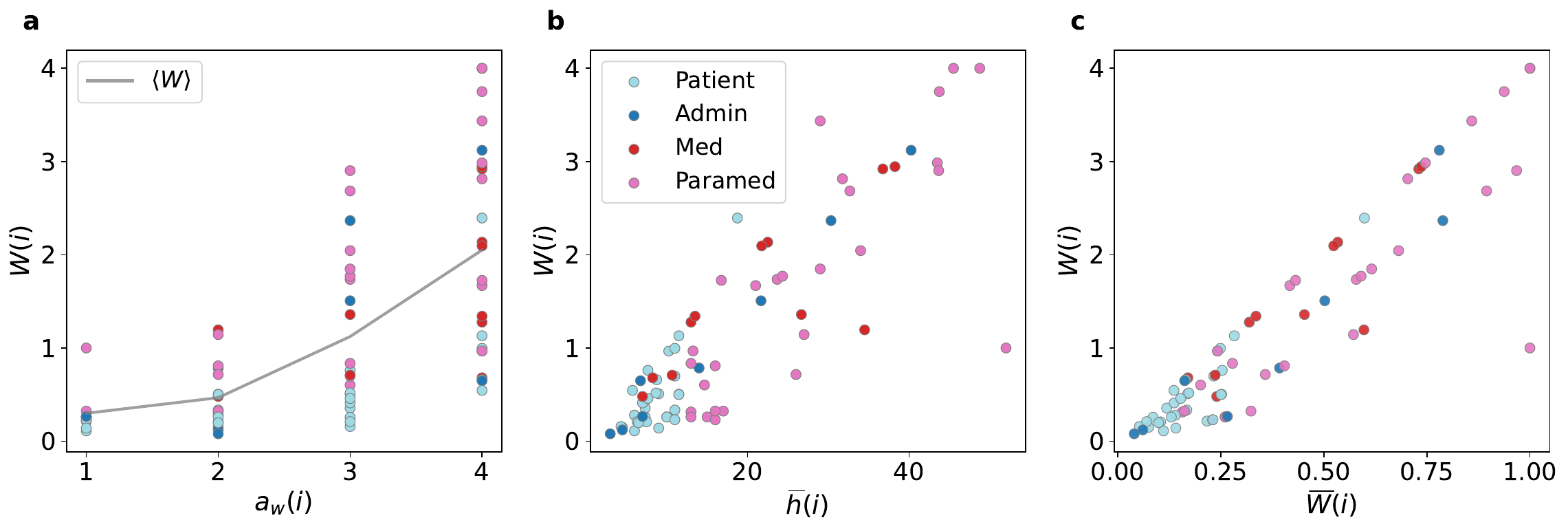}
    \caption{\textbf{Time-aggregated hypercoreness in a hospital (LH10).}  
    \textbf{a:} scatter plot of the aggregated hypercoreness $W(i)$ as a function of the snapshot activity $a_w(i)$ for all nodes $i$, and averaged aggregated hypercoreness $\langle W \rangle$ as a function of $a_w$.
    \textbf{b:} aggregated hypercoreness $W(i)$ vs. average number of interactions per active window $\overline{h}(i)$ for all nodes $i$.  
    \textbf{c:} aggregated hypercoreness $W(i)$ as a function of the activity-averaged hypercoreness $\overline{W}(i)$. 
    In all panels points are colored according to the node's social role.
    }
    \label{fig:figure7}
\end{figure*}

These different behaviours are summarized by the time-aggregated centrality measures. In general the aggregated hypercoreness $W$ increases with the snapshot activity $a_w$ (see Fig. \ref{fig:figure7}\textbf{a}), however nodes with the same $a_w$ can have very different $W$. Analogously, $W$ and the average number of interactions when active $\overline{h}$ are positively correlated, but there are outliers, which produce different top positions in the corresponding rankings (see Fig. \ref{fig:figure7}\textbf{b}). By taking the structure into account, the aggregated hypercoreness can thus provide a different and more detailed information than the activity or the average number of interactions. The aggregated and activity-averaged hypercoreness show that the nodes that are globally relevant are also relevant, on average, when active (see Fig. \ref{fig:figure7}\textbf{c}). Nevertheless, the produced rankings are still different since some nodes are relevant when active (high $\overline{W}$), but not globally (low $W$). By combining the two time-aggregated hypercoreness measures we obtain information on the different overall behaviors of the nodes (see Fig. \ref{fig:figure6}).

\begin{figure*}
    \centering
    \includegraphics[width=\textwidth]{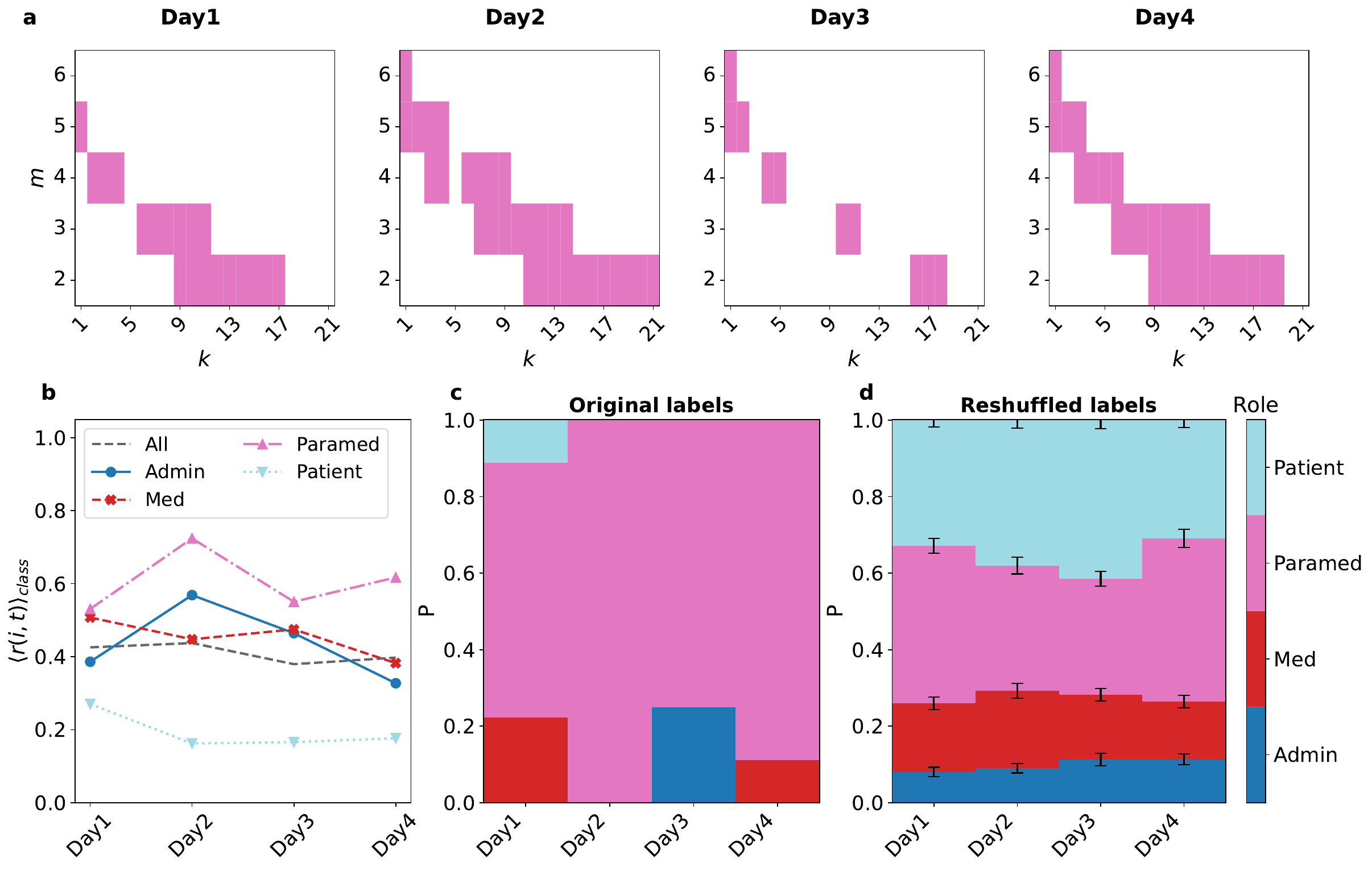}
    \caption{\textbf{Prevalent social role in hyper-cores of a hospital (LH10).} 
    \textbf{a:} temporal evolution over 24-hours time windows of the prevalent social role in each $(k,m)$-hyper-core of the LH10 data set, defined as the most frequent label in the core: we use a color code for identifying social roles and we consider a role dominant only if its frequency is larger than 0.5. In white are indicated hyper-cores which are empty or where no dominant role can be identified. 
    \textbf{b:} temporal evolution of the hypercoreness $r(i,t)$ averaged over all nodes (dashed black line) and averaged over each distinct class. 
    \textbf{c:} temporal evolution of the relative frequency $P$ of the various social roles within the top 15\% positions of the nodes ranking given by the hypercoreness $r(i,t)$.
    \textbf{d:} same as \textbf{b}, but in this case we consider the relative frequency $P$ averaged over 50 randomized realizations of the hypergraph (see Methods). In this case, we also show error bars corresponding to the standard errors. 
    }
    \label{fig:figure8}
\end{figure*}

We finally expose strong differences in the temporal and structural properties of specific roles in the hospital. Figure \ref{fig:figure7} shows that the activity $a_w$ is quite independent of the social role; however, the patients have a homogeneous behavior occupying always the lower positions in all the rankings produced by the other time-aggregated centrality measures; on the contrary the nurses, doctors and administrative staff present a more heterogeneous behaviour, presenting a wide range of centrality values. Nurses constitute the most structurally and temporally relevant group according to all the time-aggregated centrality measures, always occupying the top positions of the rankings (see Fig. \ref{fig:figure7}).

The nurses have a key role also mesoscopically: in each $(k,m)$-hyper-core indeed, we identify the dominant social role when possible by checking whether more than half of the nodes of a core belong to one category. In the superficial cores it is not possible to identify a dominant role, however in the most central cores the nurses dominate in all time windows and at all interaction orders (see Fig. \ref{fig:figure8}\textbf{a}). Nurses thus constitute the most densely connected social group at all the orders of interaction, thus the interactions structure in the most central cores is attributable to their activities. 

The dominant role of nurses is further highlighted microscopically by considering the evolution of the average hypercoreness $r(i,t)$ within each specific class (see Fig. \ref{fig:figure8}\textbf{b}). All roles present a quite stable average hypercoreness: patients and nurses present a hypercoreness notably lower and higher than the average, respectively, while doctors and administrative staff are close to the average behavior. Moreover, if we consider the instantaneous ranking produced by the hypercoreness and estimate the frequency of each role, we find that nurses always dominate the top positions (see Fig. \ref{fig:figure8}\textbf{c}). This pattern is not due to a difference in numbers of nodes or hyperedges, as we check by comparing the results with a reshuffled data set in Fig. \ref{fig:figure8}\textbf{d}: we generate 50 random realizations of the hypergraph, which completely preserve in each time window the structure of the hypergraph and the total number of nodes with each label, but destroys correlations between the labels of interacting nodes (see the reshuffling procedure in Methods). 
The frequencies of the different social roles in the top positions of the hypercoreness ranking, averaged over all the realizations, shows strong differences compared to the original case.\\

While we have here focused on the changes occurring between different days, it is possible to consider a different temporal resolution to focus e.g. on specific activities in the system occurring at a different time scales: in the SM we consider as an example the evolution occurring within a single day with 2-hours time windows. 

In the SM we also apply the proposed analysis to data sets describing interactions between individuals in different contexts (see Methods).  
In some contexts, the composition of the hyper-cores present
a strong structural variability and instability: this corresponds e.g. to conferences or workplaces where different days can bring very different patterns of connections.
A more stable structure is obtained in others, with high stability of the cores composition, e.g. systems in which patterns of interactions are repeated over time due to role and activities constraints, such as in schools and hospitals (see SM).  
Such differences in the results highlight and confirm the interest of following the hyper-core decomposition over time as a characterization tool for temporal hypergraphs.

\subsection*{A validation tool for time-varying hypergraph models}
\label{sez:models}

We now illustrate how the hyper-core decomposition can also help  with the validation of synthetic models of temporal hypergraphs. More precisely, it can serve as a tool to quantitatively validate whether a model reproduces given hierarchical structures and structural dynamics of interest, such as those of an empirical temporal hypergraph, at several topological and temporal scales. 
To showcase the potential of this as a tool, we consider several models of temporal hypergraphs of increasing complexity, and tune them to reproduce the activity patterns of a data set. We then apply the previously described approach to each model and to the original data set, identifying differences among the models, and ultimately investigating which model ingredients make it possible to generate a non-trivial hierarchical structure that resembles the one found in the data.
 
\begin{figure*}
    \centering
    \includegraphics[width=0.9\textwidth]{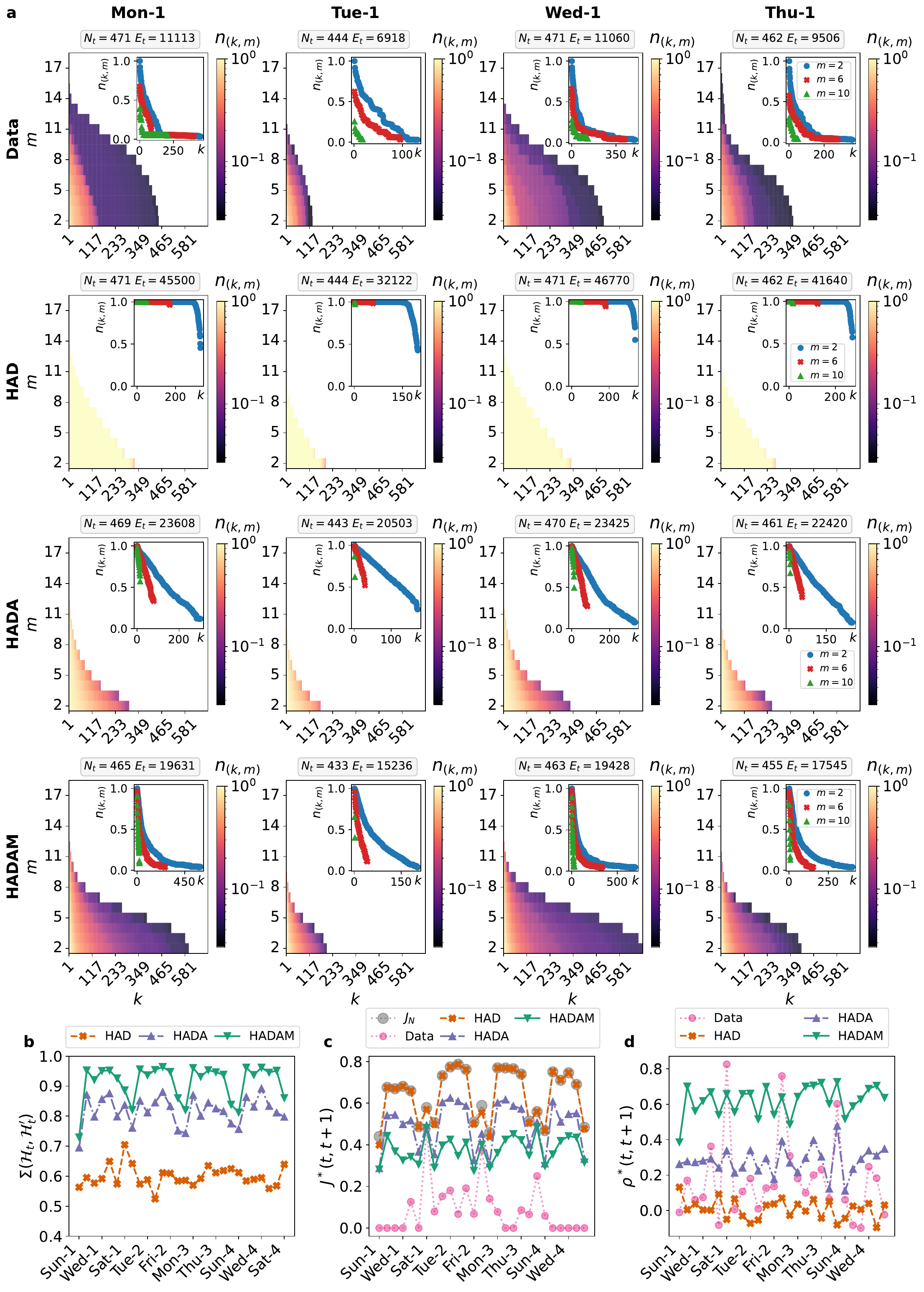}
    \caption{\textbf{Hyper-cores structure in time-varying hypergraphs models.} We consider the CopNS data set as well as the HAD, HADA and HADAM models adjusted to the CopNS node activities and hyperedge size distributions, and aggregated over 1-day time windows. 
    \textbf{a:} relative population $n_{(k,m)}$ of the $(k,m)$-core as a function of $k$ and $m$ from Monday to Thursday of the first week; the number of active nodes $N_t$ and hyperedges $E_t$ are also reported. The insets show $n_{(k,m)}$ as a function of $k$ for fixed values of $m$. The first row corresponds to the empirical data; the second, third and fourth rows correspond to the hypergraphs generated respectively with the HAD, the HADA and the HADAM models.  
    \textbf{b:} similarity $\Sigma$ between the hyper-cores filling profiles of the empirical hypergraph $\mathcal{H}_t$ and each of the synthetic models $\mathcal{H}_t'$ in the same time window $t$.  
    \textbf{c:} similarity $J^*(t,t+1)$ between the most central hyper-cores, i.e. $(k_{max}^m,m)$-cores $\forall m$, in two consecutive snapshots, and Jaccard similarity $J_N(t,t+1)$ between the entire population of the data set in  consecutive time windows.  
    \textbf{d:} Pearson correlation coefficient $\rho^*(t,t+1)$ between the nodes hypercoreness in two consecutive snapshots, considering all the nodes that appear in both time snapshots. In panels \textbf{c}-\textbf{d} we consider both the data set and the corresponding synthetic models.
    }
    \label{fig:figure9}
\end{figure*}

For simplicity, we consider models within the class of 
activity-driven (AD) networks: these models are based on simple mechanisms for the formation of interactions \cite{Perra2012},
and can be refined to include increasing complex realistic features 
and tuned to reproduce many properties of empirical data sets
 \cite{Karsai2014,Ubaldi2016,attr1,attr2,le2023modeling}. 
We consider here several generalizations taking into account higher-order interactions, in a similar spirit as \cite{petri2018simplicial,di2024percolation}.
In each model, we consider a population of $N$ nodes: each node is assigned at each time $t$ an activity parameter $a_t(i)$, which represents the node  propensity to generate interactions and sets its activation rate (Poissonian activation dynamics). When a node is active, it generates a hyperedge of size $m$, drawn from a distribution $\Psi_t(m)$ (which potentially depends on the time step $t$). The remaining $(m-1)$ nodes are selected in the population with mechanisms depending on the specific AD model. 
We consider the following models:
\begin{itemize}
\item \textbf{Higher-order activity-driven model (HAD).} This model is the hypergraph generalization of the standard AD network \cite{Perra2012} and of the simplicial activity-driven model (SAD) \cite{petri2018simplicial}. Each active node creating an hyperedge of size $m$ chooses the $m-1$ nodes to interact with uniformly at random from the whole population. This model takes into account only the heterogeneity of the agents behaviour, through their activities, and the one of the size of the groups. Interactions are instantaneous and there is no memory between successive time steps.

\item \textbf{HAD model with attractiveness (HADA).} This model corresponds to the hypergraph generalization of the AD network with attractiveness \cite{attr1,attr2,mancastroppa2022,le2023modeling}. Each node is also assigned with an attractiveness parameter $b_t(i)$, which defines the intensity with which the node attracts active interactions. Each active node, to create an interaction of size $m$, selects the $m-1$ other nodes in the population randomly with probability proportional to their attractiveness $b$. The interactions are instantaneous and there is no memory. We consider $b_t(i)=a_t(i)$ $\forall i$ at each time, i.e. the most (less) active nodes are also the most (less) attractive ones, as observed in empirical systems \cite{attr1,attr2}. 

\item \textbf{HAD model with memory (HADAM).} This model is the HADA with the introduction of an additional memory mechanism, similar to that proposed in the AD networks with memory \cite{Karsai2014,Ubaldi2016}. For each active node $i$, we denote by $l_t(i)$ the number of other nodes with which it has already interacted in previous time steps. The active node $i$, to create an interaction of size $m$, selects the $m-1$ other nodes (i) with probability $p_t(i)=1/(1+l_t(i))$, among those not yet encountered, (ii) with probability $(1-p_t(i))$ among those already met. These nodes are selected: in the former case, with probability proportional to their attractiveness $b(j)$; in the latter case, with probability proportional to their attractiveness $b(j)$ and to the number of times they have already met with the active node $w_{ij}$. 
\end{itemize}

Each model can be fed by empirical data in the following manner.
Given an empirical temporal 
hypergraph $\mathcal{H}$ and its snapshot representation $\{\mathcal{H}_t\}_{t=1}^{n}$, for each model we consider the same population size as the empirical hypergraph; moreover, we use the empirically observed hyperedges size distribution $\Psi_t(m)$ at each time step, and we tune the activities $a_t(i)$  so that the total number of interactions at each time, $n_t^{tot}$, and the total number of interactions in which each node is involved, $n_t(i)$, replicate the empirical ones (see Methods for more details on the hypergraphs generation).
Here specifically, we consider a data set of human interactions in a University, collected within the Copenhagen Network Study (CopNS) \cite{Sapiezynski2019} (in the SM we also apply the same analysis to the hospital data set). We build a temporal hypergraph from the data by considering each individual as a node, and each hyperedge as a group interaction with a temporal resolution of 5 minutes (see Methods). The data set covers a period of 4 weeks.

Once we have generated the three synthetic temporal hypergraphs, we aggregate both data and models on 1-day time windows. We then apply the hyper-core decomposition to each time window and compare the resulting structures and their temporal evolution at this time scale. We mainly focus here on the first working days of the first week of the data, and we show in the SM that similar temporal and structural patterns are obtained also for other days and weeks.

The original data set presents a non-trivial filling of the cores, with significant differences over time (see Fig. \ref{fig:figure9}\textbf{a}): on Monday the $(k,m)$-cores present a rapid emptying for all orders when $k$ increases, with a rapid drop in the population (densely populated shells), followed by an extended plateau (empty shells); a similar structure is obtained on Wednesday and Thursday, but with some differences in the drops widths, in the plateaus extensions and in the maximum connectivity values; on Tuesday instead, the structure is very different, the maximum connectivity values are much lower and the plateau observed in the other time windows is almost absent. These filling profiles suggest the presence of a rich hierarchical structure in the hypergraph that changes over time.

The HAD model, despite replicating the activities and hyperedge distribution sizes of the data, has a very different hyper-core decomposition, which does not display any hierarchical structure
(Fig. \ref{fig:figure9}\textbf{a}): all $(k,m)$-cores are equally populated by the whole population until $k \sim k_{max}^m$, then $n_{(k,m)}$ quickly collapses to zero; all the shells are empty apart for those with $k \sim k_{max}^m$ which contain the entire population. The model thus does not replicate the empirical hierarchical structure nor its evolution, neither mesoscopically, since all cores coincide with the entire population, nor microscopically, since all the nodes have the same position in the core structure. This is expected due to the interaction mechanism of the model ---which generates a completely mean-field structure. 

By contrast, the temporal hypergraph obtained from the HADA model does present a hierarchical structure: the population of the $(k,m)$-cores decreases progressively and smoothly with $k$ at all orders $m$, indicating the presence of uniformly populated shells. The system presents a hierarchy both mesoscopically, since there are groups of nodes more densely connected, and microscopically, since the nodes are distributed on the various shells. The model partially replicates the changes in the maximum connectivity, but it does not completely reproduce the empirical hierarchical structure, as the shapes of $n_{(k,m)}$ vs. $k$ are rather different from the empirical ones (insets of Fig. \ref{fig:figure9}\textbf{a}).

\begin{figure*}
    \centering
    \includegraphics[width=\textwidth]{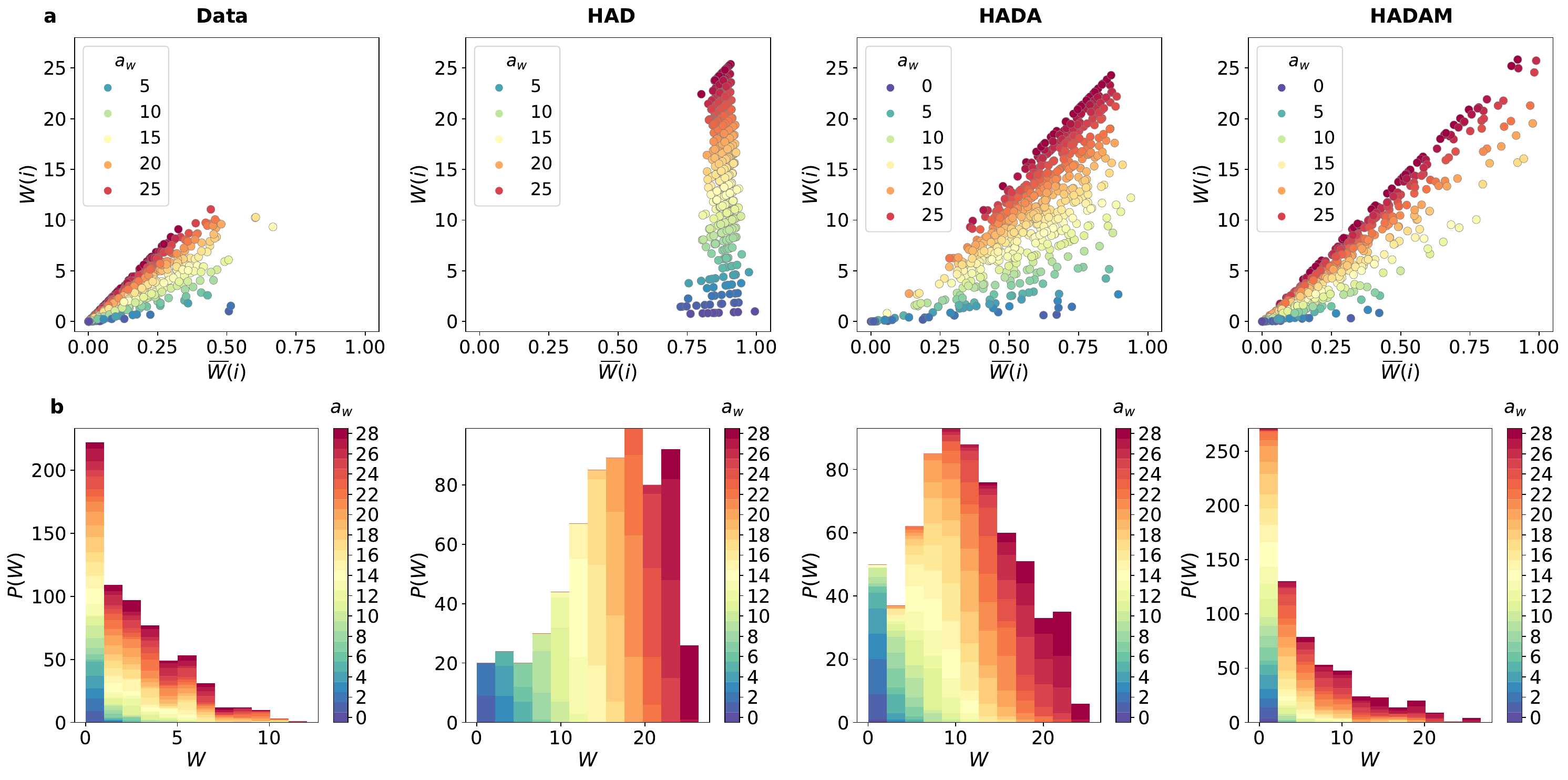}
    \caption{\textbf{Time-aggregated hypercoreness in time-varying hypergraphs models.} We consider the CopNS data set with 1-day time windows over four weeks, as well as the three synthetic models. 
    \textbf{a:} scatter plots of the aggregated hypercoreness $W$ as a function of the activity-averaged hypercoreness $\overline{W}$ for each node: the points are colored according to the snapshot activity $a_w$ of the corresponding node. 
    \textbf{b:} histograms giving the number of nodes $P(W)$ with aggregated hypercoreness $W$: within each bar we distinguish the relative frequency of nodes belonging to each class $a_w$, through stacked bars. In all panels, we consider both the empirical hypergraphs (first column) and the corresponding synthetic temporal hypergraphs (second column - HAD, third column - HADA, and fourth column - HADAM).
    }
    \label{fig:figure10}
\end{figure*}

Finally, the synthetic hypergraphs generated using the HADAM model present a rich hierarchical structure
that reproduces quite well the empirical one and its evolution, both in the maximum connectivity and in the filling profiles. Indeed, the memory effect drives the creation of interactions between nodes that have already met several times in the past, thus favoring non-trivial patterns with densely connected groups of nodes. Some quantitative differences with the empirical structure are still observed, such as a more progressive emptying of the cores with $k$, and slightly different $k_{max}^m$ values.

Figure \ref{fig:figure9}\textbf{b} provides a quantitative comparison of the hyper-core structures generated by each model with the empirical one, through the root-mean-square deviation similarity $\Sigma$ between the respective hyper-cores filling profiles in each time window. As expected from the above considerations, the hyper-core structure of the HADAM model is the most similar to the empirical one with $\Sigma \sim 0.95$, followed by the HADA model ($\Sigma \sim 0.80$), and by the HAD model ($\Sigma \sim 0.60$). Similar results are also obtained with other similarity measures (see SM).

At the mesoscopic scale, the empirical data present a strong instability in the most central cores (see Fig. \ref{fig:figure9}\textbf{c}), with a very low similarity $J^*(t,t+1)$ between consecutive snapshots. The HAD model, on the contrary, presents a very high stability in the deepest cores, reproducing the empirical similarity $J_N$ of the entire population, as expected since the whole population composes the most central cores (see Fig. \ref{fig:figure9}\textbf{a}). The HADA and the HADAM models yield a lower stability of the central cores: the variations in activity and memory effects are enough to generate changes in the mesoscopic hierarchical structure and similarities closer to the empirical case, even if still higher. At the microscopic level, the empirical data set alternates phases with low and high hypercoreness correlations in consecutive snapshots $\rho^*(t,t+1)$, (see Fig. \ref{fig:figure9}\textbf{d}): during the weekdays the structural position of nodes change a lot across days (low $\rho^*$), because of varying activities, while during the weekends it is quite stable (high $\rho^*$). On the contrary, the three models present approximately constant correlation values: the HAD model trivially does not present any correlation $\rho^* \sim 0$, since the model does not generate any hierarchy of nodes in any time window; the HADA model instead presents higher correlations $\rho^* \sim 0.30$, as the system generates a hierarchical structure with high-activity nodes being the most central over time; finally, the HADAM model presents the highest correlations $\rho^* \sim 0.60$, since the memory forces the creation of correlations in nodes behavior over time and could be balanced only by strong changes in nodes activity. 

These results are further confirmed by comparing the entire similarity matrices of the models with the ones of the empirical hypergraph at different scales (see SM, for the matrices $\Sigma(t,t')$, $J^*(t,t')$, $\rho(t,t')$ and $\rho^*(t,t')$): the HADAM model better reproduces the evolution and temporal stability of the empirical system at all the temporal and structural scales, while the HADA and HAD models feature larger differences, with the HAD model leading to the widest discrepancy (see SM).

We finally compare in Fig. \ref{fig:figure10} the behaviour of the time-aggregated centralities measures in the data and models.
The original data set presents a wide variability. In fact, even if the aggregated hypercoreness $W$ and the activity-averaged hypercoreness $\overline{W}$ are positively correlated, there are nodes very central on average when active (high $\overline{W}$) but globally not relevant (low $W$) and vice-versa. This suggests different node hypercoreness trajectories and node movements across the core structure (see SM). The system also presents a heterogeneous distribution of the aggregated hypercoreness $W$, $P(W)$, which provides a clear ranking of nodes. Moreover, nodes with the same snapshot activity $a_w$ can present very different structural behaviors, indeed the activity is unevenly distributed in the $W$ classes: the nodes with relevant structural role (high $W$) are frequently active (high $a_w$), but nodes poorly structurally relevant (low $W$) can have very different activity values.

In the HAD model all nodes have approximately the same activity-averaged hypercoreness $\overline{W}$ but different values of the aggregated one $W$ (see Fig. \ref{fig:figure10}): the HAD model does not produce any hypercoreness hierarchy of nodes in any time window, therefore on average when a node is active it has the same centrality as the others $\overline{W}$. The aggregated hypercoreness $W$ differentiate among the nodes only through their temporal persistence in the system, i.e. through $a_w$. The distribution of $W$ appears homogeneous and peaked.

The HADA model creates a hierarchy of nodes both in terms of $W$ and $\overline{W}$ (see Fig. \ref{fig:figure10}): in this case, the most globally central nodes are also relevant on average when active, while nodes that are less central globally can feature different behaviours when active, either being very central or not. The distribution $P(W)$ appears homogeneous and peaked, with a gradual increase in the activity $a_w$ of nodes more relevant. Even if it features a hypercoreness hierarchy, the model does not reproduce the empirical distribution of the aggregated hypercoreness $P(W)$,
and yields a stronger correlation between $W$ and $a_w$ than in the empirical data.

The HADAM model yields a hierarchy both in terms of $W$ and $\overline{W}$ (see Fig. \ref{fig:figure10}), replicating quite well the empirical patterns, even if there are nodes with time-aggregated hypercoreness values, $W$ and $\overline{W}$, higher than those empirically observed. The distribution $P(W)$ is heterogeneous, with few nodes with very high $W$, and also the heterogeneity in nodes structural and temporal behaviours is well reproduced, since the distribution of $a_w$ in the $W$ classes well replicate the empirical case. \\

Overall, these results show how the hyper-core decomposition allows to validate the hypergraph models structurally and temporally at different scales. The three temporal models are generated starting from the same amount of information extracted from the empirical data set and are tuned to replicate the same statistical and temporal properties. The HAD model fails to produce and replicate the hierarchical structure at any of the scales considered, as the model generates a mean-field structure without hierarchy. The introduction of attractiveness in the HADA model generates a hierarchical structure that however still strongly differs  from the empirical one, as the model generates a more progressive core-periphery structure. The memory effect introduced in the HADAM model makes it possible to obtain a hierarchical structure that resembles quite well the empirical one at all scales, except for a stronger correlation between the nodes hypercoreness rankings. Note that analogous results can be obtained also considering other data sets (see the SM).

\section*{Discussion} 

Recently, there has been a recognition of the importance of going beyond pairwise and static representations for complex systems~\cite{HOLME201297, battiston2020networks}.
In this article, we have put forward a method for the structural and dynamic characterization of temporal hypergraphs, which represent time-varying systems involving higher-order interactions.
The approach is based on decomposing the hypergraph into hyper-cores over time, and it provides a multi-scale characterization: macroscopically, it follows the higher-order hierarchical structure over time, monitoring the stability of the overall hyper-core structure; mesoscopically, it follows the evolution of specific hyper-cores, observing whether stable groups of nodes are densely connected to each other or whether they change over time; microscopically, it follows the structural behavior of single nodes, monitoring their movements across the hierarchical structure, towards more superficial or more central hyper-cores.
The approach provides several similarity measures that quantitatively estimate the higher-order structural stability of the system at different topological scales, also identifying temporal patterns in the structure evolution. 
We moreover introduced two time-aggregated centrality measures of nodes, by aggregating the instantaneous hypercoreness or by averaging it over the node's activity. These last measures provide additional information on the behavior of the nodes, as opposed to other centrality measures that do not account for higher-order structural properties. 

We applied the method to a wide range of data sets describing different systems, characterizing each of them and identifying similarities and differences: for example, stronger instability characterizes systems where the nature of the interactions favors variability in the interaction patterns, such as scientific collaborations, conferences, universities and workplaces; a more stable structure is observed instead in systems with patterns of repeated interactions due tho functional roles, such as schools and hospitals. We also linked structural properties of nodes to specific roles and activities in the systems, thus identifying relevant functions and their evolution over time. 

The proposed method represents also an effective model-validation tool, since it allows to quantitatively estimate whether a synthetic temporal hypergraph can replicate the structure of an empirical hypergraph and its evolution at different topological scales, and to compare several candidate models. 
In this direction, we proposed several models of activity-driven hypergraphs with increasing complexity in the mechanisms that drive the hyperedges formation and we estimated their structural-temporal differences and similarities with respect to the empirical systems. We have shown that models taking into account solely the node activities and the hyperedges size distribution over time cannot reproduce the empirical higher-order structure and its evolution. By contrast, introducing attractiveness and memory, while keeping the model simple, yields non-trivial hyper-core structures and to obtain a behaviour closer to the one empirically measured.

Our work opens several research directions and future perspectives. It lays the foundations for the development of new characterization techniques for time-varying hypergraphs \cite{battiston2020networks}: for example, it represents a first step for the definition of a core decomposition of temporal hypergraphs, which is a highly challenging task 
because of the difficulties in defining a procedure taking into account both non-dyadic interactions and the temporal dimension to generalize e.g. the span-core decomposition of temporal networks \cite{galimberti2018mining,Ciaperoni2020}. Our work also provides  insights for the understanding of higher-order dynamic processes on temporal hypergraphs, since hyper-cores play an important role in dynamic processes \cite{mancastroppa2023hyper}: understanding how the multi-scale evolution of the underlying hypergraph affects dynamic processes is of great interest, in order to fully assess the coupling between the dynamics of and on the hypergraph. This is crucial also for the planning of adaptive measures and interventions, e.g. to maximize or prevent the spread of information on a time-varying hypergraph.
Finally, our approach provides tools to guide the design of new models for temporal hypergraphs capable of reproducing higher-order structural properties of empirical systems at different topological scales. Here we have proposed examples of activity-driven hypergraphs featuring different interesting properties \cite{Perra2012,petri2018simplicial,mancastroppa2022}, however more complex models could be devised \cite{iacopini2023temporal,di2024percolation,Guo2016,gallo2023higher,le2023modeling}, 
for example introducing correlations between the activity of nodes and the size of hyperedges of which they are member, or considering memory and attractiveness mechanisms involving groups of nodes.

\section*{Methods}

\subsection*{Hyper-core decomposition}
Let us consider an unweighted static hypergraph $\mathcal{H}_t=(\mathcal{V}_t,\mathcal{E}_t)$, composed by the set of its nodes $\mathcal{V}_t$ and by the set of its hyperedges $\mathcal{E}_t$. A hyperedge $e=\{i_1,i_2,...,i_m\} \in \mathcal{E}_t$ consists in a set of $m$ nodes $i_k \in \mathcal{V}_t$ $\forall k=1,...,m$, with $m \in [2,M_t]$, where $M_t = \max_{e \in \mathcal{E}_t} |e|$. 

The hyper-core decomposition is a procedure that decomposes the hypergraph $\mathcal{H}_t$ into $(k,m)$-hyper-cores, i.e., a double hierarchy of nested subhypergraphs of increasing connectivity, provided by hyperedges of increasing size. Specifically, the $(k,m)$-hyper-core of $\mathcal{H}_t$, denoted as $\mathcal{F}_t^{(k,m)}=(\mathcal{A}_t^{(k,m)},\mathcal{S}_t^{(k,m)})$, is defined as the maximum subhypergraph that contains all the nodes $i \in \mathcal{V}_t$ involved in at least $k$ distinct hyperedges of size at least $m$ within the subhypergraph itself. It contains all the hyperedges that are subsets of interactions in the original hypergraph $\mathcal{H}_t$, of size at least $m$ and that contain only nodes of $\mathcal{A}_t^{(k,m)}$. Therefore, $\mathcal{A}_t^{(k,m)}=\{i \in \mathcal{V}_t \, \text{s.t.} \, D_m^{\mathcal{F}_t^{(k,m)}}(i) \geq k\}$ and $\mathcal{S}_t^{(k,m)}=\{e \cap \mathcal{A}_t^{(k,m)} \, \text{s.t.} \, e \in \mathcal{E}_t \wedge |e \cap \mathcal{A}_t^{(k,m)}|\geq m\}$, where $D_m^{\mathcal{F}_t^{(k,m)}}(i)$ is the number of distinct interactions of size at least $m$ in which the node $i$ is involved in $\mathcal{F}_t^{(k,m)}$. Note that the $(k,m)$-hyper-core includes the $(k,m+1)$- and $(k+1,m)$-hyper-cores, producing a doubly nested hierarchical structure which, by increasing $k$ and $m$, progressively identifies groups of nodes more densely connected with each other through interactions of increasing order \cite{mancastroppa2023hyper}. The $(k,m)$-hyper-core is obtained by removing progressively and iteratively all the nodes with $D_m<k$ and all the hyperedges of size smaller than $m$ \cite{mancastroppa2023hyper}. 

\subsection*{Data description and preprocessing}

We consider data sets covering a wide range of interaction systems and present different statistical, topological and temporal properties (see SM). 

\paragraph*{Scientific collaborations.} The American Physical Society (APS) scientific collaborations data set \cite{APS,journals} consists in all the APS publications from 1893 to 2021: for each paper the date of publication, the journal and the list of authors are indicated. 

We initially addressed some issues appearing in the data:
({\it i}) information is missing for some papers, for example on the author list: in these cases we removed the corresponding entries from the data set;
({\it ii}) the same author \textit{"Name Surname"} can appear with the full extended name, as \textit{"N. Surname"}, \textit{"N Surname"} or \textit{"Na. Surname"}; analogously with middle names \textit{"Name Second Surname"} or \textit{"Name-Second Surname"}. To minimize the impact of these inconsistencies, we:
({\it a}) identified all entries with the same \textit{"Surname"};
({\it b}) reassigned the papers associated to dotted names to the corresponding extended name, carrying out the reassignment only in case of uniqueness. Some dotted names do not have or have several extended correspondences, making a unique reassignment impossible: in these cases we consider the contracted name as if it were a unique additional author. See the SM for further details on the size of the various issues. The performed approach reduces the problems related to author identification, but does not completely eliminate the issue: it is still possible that two authors have the same name, therefore the publications are attributed as if they were a single individual. Moreover, in the presence of large collaborations, not all authors are listed \cite{Tomasello2017}. Such issues cannot be eliminated through preprocessing of the data without additional information sources to perform a cross-source analysis \cite{Tomasello2017}. However, even without such additional information, the preprocessed data set gives a good enough picture of the scientific interactions as our purpose is here demonstrative and we do not seek to give precise ranking indications concerning scientists, nor follow in detail some careers. 

We thus use the data to build a hypergraph in which each node is an author, a hyperedge represents a paper connecting the co-authors, and it is assigned with a label indicating the corresponding journal. Since we focus on the pattern of collaborations between authors, rather than on the absolute scientific production, we do not take into consideration papers with a single author. We obtain a temporal hypergraph with 1-day resolution, and we focus on 1942-2021. We consider 5-years adjacent time windows and aggregate the temporal hypergraph within each of them, obtaining a sequence of unweighted static hypergraphs. Each static hypergraph is composed of all the nodes and hyperedges active at least once in the considered time window. The same group of authors can have co-authored several papers in the same time window producing fully overlapping hyperedges: in this case we consider only one hyperedge (unweighted hypergraph) and we assign a multiple label to it, including all the journals in which the same group of authors published.

\paragraph*{Physical proximity.}
We consider several data sets of human face-to-face interactions obtained through RFID wearable proximity sensors, made publicly available by the SocioPatterns collaboration \cite{SP,Genois2018,ISELLA2011166} and by the Contacts among Utah's School-age Population project \cite{Toth2015}. These data sets describe interactions between individuals in several settings and cover different time periods: a workplace (InVS15 \cite{genois2015,Genois2018} - 2 weeks), a conference (SFHH \cite{Genois2018} - 2 days), a hospital (LH10 \cite{cattuto2013} - 4 days), two primary-schools (LyonSchool \cite{barrat2011}, Utah\_elem \cite{Toth2015} - 2 days) and a high-school (Thiers13 \cite{Mastrandrea2015} - 1 week). The data consist in each case in lists of time-resolved pairwise interactions between individuals (nodes), i.e., temporal networks with a time resolution of 20 seconds. To identify group interactions and transform such temporal networks into temporal hypergraphs, we carried out the following procedure \cite{Iacopini2022,mancastroppa2023hyper}: 
({\it i}) pairwise interactions are aggregated over 5-minutes time intervals; 
({\it ii}) cliques, i.e. fully connected clusters, are identified in each time step;
({\it iii}) in each time interval the maximum cliques, i.e. cliques not fully contained in another clique, are identified and promoted to hyperedges. 
This procedure generates temporal hypergraphs with 5-minutes resolution. Some data sets have moreover node labels providing information on single nodes properties, e.g. class of each student for LyonSchool, Thiers13, Utah\_elem, social role for LH10 and working department for InVS15. 

We also consider time-resolved data describing physical proximity events between students in a University, collected through the Bluetooth signal of cellphones during 4 weeks within the Copenhagen Network Study \cite{Sapiezynski2019,iacopini2023temporal} (CopNS). The data set provides pairwise interactions between individuals (nodes) with a temporal resolution of 5 minutes and with information on the signal intensity: we perform the preprocessing procedures described in \cite{iacopini2023temporal}, obtaining a temporal hypergraph with 5-minutes resolution. 

\paragraph*{Email.}
Finally, we consider a data set describing email communications within an European institution (email-EU \cite{emailEUdata,benson2018,Benson_site} - 17 months). This data set is publicly available as a temporal hypergraph: each node represents a user, each hyperedge corresponds to an email and involves both the recipients and the sender of the message. The sending time is provided for each hyperedge with 1-second resolution and the information on the directionality of the email is discarded. 

\subsection*{Labels reshuffling procedures}
We implement two reshuffling procedures, one for systems with hyperedge labels (e.g. APS), and one for those with node labels (e.g. LH10). \\

\textbf{Hyperedge labels reshuffling.} We consider a temporal hypergraph $\mathcal{H}=\{\mathcal{H}_{t}\}_{t=1}^{t=n}$, in which each hyperedge $e$ is assigned with one or multiple labels. We obtain a reshuffled realization of the temporal hypergraph $\mathcal{H}'$ in the following way: for each static snapshot $\mathcal{H}_{t}$, we randomly select two hyperedges $e$ and $f$ of the same size $m$ and, if they have different labels $l_e$ and $l_f$, we perform a label swap so that $e$ will have the new label $l_e'=l_f$ and $f$ will have the new label $l_f'=l_e$. In the case of hyperedges $e$ with multiple labels $[l_e^1,l_e^2,...,l_e^n,...,l_e^q]$, one of the labels is randomly selected $l_e^n$, and the label swap is performed only with it. The procedure is repeated $10^5$ times for each size $m \in [2,M_t]$ and for each static snapshot $\mathcal{H}_{t}$ (if the number of hyperedges of size $m$ is at least 4 and at least two different labels are available). The described procedure preserves in each temporal snapshot the hypergraph structure, the overall number of hyperedges with each label at each order of interaction, while it  destroys the correlations between the nodes and the labels of the hyperedges in which they are involved. \\

\textbf{Node labels reshuffling.} We consider a temporal hypergraph $\mathcal{H}=\{\mathcal{H}_{t}\}_{t=1}^{t=n}$, in which each node $i$ is assigned with a label $l_i$. We obtain a reshuffled realization $\mathcal{H}'$ of the temporal hypergraph in the following way: for each temporal snapshot $\mathcal{H}_{t}$, we randomly select two nodes $i$ and $j$ and, if they have different labels $l_i$ and $l_j$, we perform a label swap so that $i$ will have new label $l_i'=l_j$ and $j$ will have new label $l_j'=l_i$. The procedure is repeated $10^4$ times for each temporal snapshot. The described procedure preserves the hypergraph structure and the overall number of nodes with a specific label in each temporal snapshot, but it destroys the correlations between the labels of interacting nodes.

\subsection*{Temporal hypergraphs models}
We generate different synthetic temporal hypergraphs starting from the properties of the empirical hypergraph we want to model. Let us consider an empirical temporal hypergraph $\mathcal{H}$ observed over the time interval $(0,t_{max}]$. We consider $n=t_{max}/\tau$ adjacent time windows $((t-1)\tau,t\tau]$ with $t \in [1,...,n]$. Within each of them we extract the set of active nodes (of size $N_t$), the distribution of the hyperedge size $\Psi_t(m)$, the total number of interactions $n_t^{tot}$ and the total number of interactions in which each node is involved $n_t(i)$. Then we generate synthetic temporal hypergraphs $\mathcal{H}'$ with the same nodes of the empirical hypergraph, that within each temporal window $t$ have the same set of available nodes $N_t$, the same distribution $\Psi_t(m)$ of the hyperedge sizes of the empirical data and that, by an opportune tuning of the model parameters, reproduce quite well $n_t^{tot}$ and $n_t(i)$ $\forall i$. 
We consider three different models of temporal hypergraphs.
Then, we can perform temporal aggregations for both the empirical $\{\mathcal{H}_t \}_{t=1}^{t=n}$ and each synthetic $\{\mathcal{H}_t' \}_{t=1}^{t=n}$ hypergraphs. For instance, 
starting from data having a 5-minutes resolution, we generate synthetic hypergraphs with the same temporal resolution, and then we consider hypergraphs aggregated over 1-day time-windows for the analysis.

\subsubsection*{Activity-driven hypergraph (HAD)}
The higher-order activity-driven model (HAD) is the hypergraph generalization of the AD network \cite{Perra2012} and of the simplicial activity-driven model (SAD) \cite{petri2018simplicial}.
In this model, given a population of $N$ nodes, each node is assigned with an activity $a(i)$. In the discrete-time version of this model, in each time-step $\Delta t$ each node $i$ can activate with probability $a(i) \Delta t$. When a node activates, it generates a hyperedge of size $m$, drawn from the distribution $\Psi(m)$. The remaining $(m-1)$ nodes participating in the interaction are selected uniformly at random from the entire remaining population, i.e. each node is selected with probability $1/(N-1)$. At the following time-step all hyperedges are erased and the process continues iteratively. 
Here moreover, we take into account that the set of available nodes (of size $N_t$), the hyperedge size distribution $\Psi_t(m)$ and the activity of a node $a_t(i)$ can change over time.

The number of interactions in which a node is involved in the time window $t$ of extension $\tau$ is:
\begin{equation}
n_t(i)=a_t(i) \tau + \sum\limits_{j \neq i} a_t(j) \tau \frac{\langle m - 1 \rangle_t}{N_t-1},
\end{equation}
where the first term is due to the activation of the node $i$ itself and the second term to the activation of another node $j$. Moreover, $n_t^{tot}=\sum_{i} a_t(i) \tau$. Therefore, the HAD model replicates the $n_t(i)$ $\forall i$  and $n_t^{tot}$ of the empirical data set by fixing the activity of each node as:
\begin{equation}
a_t(i)=\frac{n_t(i)-\frac{\langle m-1 \rangle_t}{N_t-1} n_t^{tot}}{\tau \left(1- \frac{\langle m-1 \rangle_t}{N_t-1} \right)},
\label{eq:a_HAD}
\end{equation}
where $N_t$, $\Psi_t(m)$, $n_t(i)$ and $n_t^{tot}$ are fixed as in the empirical dataset. We set the time-step $\Delta t$ equal to the duration of the interactions in the empirical data set.

The model takes into account the hyperedge size distribution, the activity of each single node and their temporal evolution. The mechanism of hyperedges formation is uniform, random and without memory, therefore the generated temporal hypergraph structure is mean-field.

\subsubsection*{Activity-driven hypergraph with attractiveness (HADA)}
The higher-order activity-driven model with attractiveness (HADA) is a generalization of the AD network with attractiveness \cite{attr1,attr2,mancastroppa2022}, and it differs from the HAD model through the introduction of an attractiveness parameter which describes the propensity of nodes to attract active interactions. Given a population of $N$ nodes, each node is assigned with an activity $a(i)$ and an attractiveness $b(i)$: in each discrete time-step $\Delta t$ each node $i$ can activate with probability $a(i) \Delta t$. When a node $i$ activates, it generates a hyperedge of size $m$, drawn from the distribution $\Psi(m)$. The remaining $(m-1)$ nodes participating in the interaction are randomly selected from the population with probability proportional to their attractiveness, i.e. each node $j$ is selected with probability $ b(j)/\sum_{k \neq i} b(k)$. At the following time-step all the hyperedges are destroyed and the process is iterated. For simplicity, hereafter we will assume that $b(i)= a(i)$ $\forall i$, i.e. the most (less) active nodes are also the most (less) attractive ones, as observed in several real systems \cite{attr1,attr2}. The set of available nodes, the hyperedge size distribution and the activity of a node can change over time. 

The number of interactions in which a node is involved in the time window $t$ of extension $\tau$ is:
\begin{equation}
n_t(i)=a_t(i) \tau + \sum\limits_{j\neq i} a_t(j) \tau \frac{\langle m-1 \rangle_t a_t(i)}{\sum\limits_{k \neq j} a_t(k)},
\end{equation}
where the first term is due to the activation of the node itself and the second term to the activation of another node. The HADA model reproduces the $n_t(i)$ $\forall i$ observed in the empirical data, if the activity is:
\begin{equation}
a_t(i)= \frac{n_t(i)}{\tau \left( 1+ \langle m-1 \rangle_t \sum\limits_{j \neq i} \frac{a_t(j)}{n_t^{tot}/\tau-a_t(j)} \right)},
\label{eq:bHAD}
\end{equation}
where $N_t$, $\Psi_t(m)$, $n_t(i)$ and $n_t^{tot}$ are fixed as in the empirical dataset. When $N_t \gg 1$, we can approximate $a_t(i) \sim n_t(i)/\langle m \rangle_t \tau$ since $\sum_{j \neq i} a_t(j) \sim n_t^{tot}/\tau$: this holds for all the time windows of all the datasets considered. We set the time-step $\Delta t$ equal to the duration of the interactions in the empirical data set.

The model takes into account the hyperedge size distribution and the activity of each node, together with their temporal evolution; the hyperedges formation mechanism is still random and without memory, but favors interactions with high activity nodes. The generated temporal hypergraph has a progressive core-periphery structure: high-activity nodes compose the core, being densely connected to each other and to the rest of the population; nodes with progressively lower activity become gradually more peripheral, being increasingly less connected to each other and only connected to the nodes in the core. 

\subsubsection*{Activity-driven hypergraph with memory (HADAM)} 
The higher-order activity-driven model with memory differs from the HADA model for the introduction of a memory mechanism, analogous to that introduced in the AD network with memory \cite{Karsai2014,Ubaldi2016}. Given a population of $N$ nodes, each node is assigned an activity $a(i)$ and an attractiveness $b(i)$: in each discrete time-step $\Delta t$ each node $i$ can activate with probability $a(i) \Delta t$. Here we consider activities and attractiveness depending on time. At time $t$ moreover, we define the aggregated neighbourhood $\mathcal{N}_t(i)$ of $i$ as the set of nodes $i$ has interacted with in previous time steps.
When a node $i$ activates at time $t$, it generates a hyperedge of size $m$, drawn from the distribution $\Psi_t(m)$: 
\begin{itemize}
\item with probability $p_t(i)=1/(1+l_t(i))$, the $m-1$ nodes $i$ will interact with are selected among nodes that $i$ has not yet encountered, i.e. who do not belong to its neighbourhood $\mathcal{N}_t(i)$ at time $t$, where $l_t(i)=|\mathcal{N}_t(i)|$. In this case each node $j \notin \mathcal{N}_t(i)$ is selected with probability $b(j)/ \sum_{k \notin \mathcal{N}_t(i)} b(k)$;
\item with probability $(1-p_t(i))$, they are selected among nodes that $i$ has already met, i.e. who belongs to its neighbourhood $\mathcal{N}_t(i)$ at time $t$. In this case each node $j \in \mathcal{N}_t(i)$ is contacted with probability $\omega_{ij}^t b(j) / \sum_{k \in \mathcal{N}_t(i)} \omega_{ik}^t b(k)$, where $\omega_{ij}^t$ is the number of times that $i$ and $j$ have participated together in a hyperedge up to time $t$.
\end{itemize}
At the following time-step all the hyperedges are erased, the process continues iteratively and correlations are generated over time by the memory. For simplicity, hereafter we use $b_t(i)= a_t(i)$ $\forall i,t$ \cite{attr1,attr2}. 

In the HADAM model, we cannot determine the activity of the nodes in order to reproduce $n_t(i)$ as observed in the empirical data, since $n_t(i)$ depends on the full detailed history of contacts of $i$ up to time $t$. We fix the activities as in the HADA model, with Eq. \eqref{eq:bHAD}, and we have checked that this ansatz reproduces  well $n_t^{tot}$ and the average total degree in the aggregate snapshots. We set the time-step $\Delta t$ equal to the duration of the interactions in the empirical data set.

The model takes into account the hyperedge size distribution and the activity of each node, together with their temporal evolution. Initially, the hypergraph evolves as the HADA model since $p(i) \sim 1$ for all nodes. Then $p(i)$ decreases and memory effects become relevant: at first an active node generates hyperedges with both new and old contacts, and then preferentially with only nodes already met, selecting those contacted several times in the past. This memory-attractiveness mechanism favors dense interactions between groups of nodes with high activity and between groups of nodes that contact each other several times, thus generating a rich topological structure.\\

\section*{Data availability statement}
The data that support the findings of this study are publicly available. The APS data set can be requested at \url{https://journals.aps.org/datasets}; the SocioPattern data sets are available at \url{http://www.sociopatterns.org/}; the Contacts among Utah’s School-age Population data set at \url{https://royalsocietypublishing.org/doi/suppl/10.1098/rsif.2015.0279}; the email communications data set at \url{https://www.cs.cornell.edu/~arb/data/}; the Copenhagen Network Study data set at \url{https://doi.org/10.6084/m9.figshare.7267433}.

\section*{Acknowledgements}
M.M. and A.B. acknowledge support from the Agence Nationale de la Recherche (ANR) project DATAREDUX (ANR-19-CE46-0008).


\end{document}


\title{Supplementary Material for "The structural evolution of temporal hypergraphs through the lens of hyper-cores"}

\author{Marco Mancastroppa}
\affiliation{Aix-Marseille Univ, Universit\'e de Toulon, CNRS, CPT, Turing Center for Living Systems, 13009 Marseille, France}
\author{Iacopo Iacopini}
\affiliation{Network Science Institute, Northeastern University London, London, E1W 1LP, United Kingdom}
\author{Giovanni Petri} 
\affiliation{Network Science Institute, Northeastern University London, London, E1W 1LP, United Kingdom}
\affiliation{CENTAI, Corso Inghilterra 3, 10138 Turin, Italy}
\author{Alain Barrat}
\affiliation{Aix-Marseille Univ, Universit\'e de Toulon, CNRS, CPT, Turing Center for Living Systems, 13009 Marseille, France}

\maketitle



In this Supplementary Material we apply the characterization approach proposed in the main text to several further data sets, and we also report several further results. 

In Section \ref{sez:I} we present the main properties of the data sets; in Section \ref{sez:II} we show the results of the hyper-core decomposition on the empirical hypergraphs; in Section \ref{sez:III} we present the multi-scales topological analysis of the stability and evolution of empirical hypergraphs; in Section \ref{sez:IV} we show the proposed time-aggregated hypercoreness measures, the corresponding rankings and distributions; in Section \ref{sez:V} we analyze the structural contribution of the different classes of nodes/hyperedges over time; in Section \ref{sez:VI} we apply the proposed approach to the HAD, HADA, HADAM time-varying hypergraph models generated for some of the empirical data sets; in Section \ref{sez:VII} we apply the proposed approach to the APS scientific collaboration data set restricted to some specific journals; in Section \ref{sez:VIII} we consider different temporal scales, by applying the proposed approach to some empirical hypergraphs on a different temporal resolution than in the main text.

\section{Properties of the data sets}
\label{sez:I}

We consider several data sets that describe different systems of interactions: these systems present different statistical, structural and temporal properties. In Supplementary Table \ref{tab:dataset_time} for each data set we report the total time span covered, the interactions resolution due to the data collection and preprocessing procedure (see Methods), and the aggregation window considered. We also indicate the presence of node/hyperedges labels and the type of information they provide (e.g. school class, role in hospital). 

In Supplementary Table \ref{tab:nodes_labels}, for each data set, we report the list of available node labels: for each we indicate the number of nodes belonging to the corresponding category, and we provide some information on the interactions between categories. 

In Supplementary Table \ref{tab:APS_journals} we consider the APS data set and present the complete list of available hyperedges labels, i.e. journals published from 1893 to 2021: for each journal we indicate the period of publication, the total number of authors during this period, the number of papers published and the topics covered. 

Finally, in Supplementary Figs. \ref{fig:figure1}-\ref{fig:figure2} we show for each data set the temporal evolution of the number of active nodes $N_t$ and of the number of active hyperedges $E_t$.
We also show the distribution of the hyperedges size $\Psi_t(m)$ and the distribution of the total hyperdegree $P(D_2^{\mathcal{H}_t})$, i.e. of the total number of interactions of arbitrary size in which each node is involved $D_2^{\mathcal{H}_t}$, and their temporal evolution.

\begin{table}[]
\begin{tabular}{|p{2.5cm}|p{1.7cm}|p{2cm}|p{2cm}|p{2.7cm}|p{5.6cm}|}
\hline
\textbf{Data set} & \textbf{Total time span} & \textbf{Hyperedges resolution} & \textbf{Aggregation window} & \textbf{Labels} & \textbf{Notes}\\ \hline
\textbf{LyonSchool} & 2 days & 5 min & 1 hour & Node: class & The data were collected over two consecutive days: here we focus on the first day of data collection. See Supplementary Table \ref{tab:nodes_labels} for more details on the node labels.\\ \hline
\textbf{Utah\_elem} & 2 days & 5 min & 1 hour & Node: class & The data were collected over two consecutive days: here we focus on the first day of data collection. See Supplementary Table \ref{tab:nodes_labels} for more details on the node labels.\\ \hline
\textbf{SFHH} & 2 days & 5 min & 4 hours & - & The data were collected over two consecutive days: here we focus on the first day of data collection. \\ \hline
\textbf{LH10} & 96 hours & 5 min & 24 hours & Node: social role & See Supplementary Table \ref{tab:nodes_labels} for more details on the node labels.\\ \hline
\textbf{Thiers13} & 1 week & 5 min & 1 day & Node: class & The data were collected only during working days; the collection started on a Monday. See Supplementary Table \ref{tab:nodes_labels} for more details on the node labels.\\ \hline
\textbf{InVS15} & 2 weeks & 5 min & 1 day & Node: working department & The data were collected only during working days; the collection started on a Monday. See Supplementary Table \ref{tab:nodes_labels} for more details on the node labels.\\ \hline
\textbf{CopNS} & 4 weeks & 5 min & 1 day & - & The data were collected during both working days and weekends; the collection started on a Sunday.\\ \hline
\textbf{Email-EU} & 17 months & 1 sec & 1 month & - & The data were collected between Sep. 2003 and Feb. 2005; only few days were covered in Sep. 2003, thus we consider the first time window over the period Sep.-Oct. 2003. \\ \hline
\textbf{APS} & 80 years & 1 day & 5 years & Hyperedge: journal & The data cover the period 1893-2021, but here we focus on 1942-2021. See Supplementary Table \ref{tab:APS_journals} and its caption for more details on the hyperedge labels and on the preprocessing procedure.\\ \hline
\end{tabular}
\caption{\textbf{Data sets information.} The table provides for each data set considered: the total time span covered; the time resolution of group interactions; the duration of the aggregation windows considered; the presence of node or hyperedge labels.}
\label{tab:dataset_time}
\end{table}

\begin{table}[]
\begin{tabular}[t]{|p{2.1cm}|p{1.5cm}|p{1.5cm}|p{3cm}|}
\hline
\textbf{Data set} & \textbf{Labels} & \textbf{Number of nodes} & \textbf{Notes}\\ \thickhline
\multirow{9}*{\shortstack[l]{\textbf{Thiers13} \\ (class)}} & PSIst & 34 &  \multirow{9}*{\shortstack[l]{Each class has a \\ specialization; \\ different classes can \\ also interact e.g. \\ during lunch \cite{fournet2014,Mastrandrea2015}.}} \\ \cline{2-3}
	& PCst & 41 &  \\ \cline{2-3}
	& PC & 44 &  \\ \cline{2-3}	
	& MPst2 & 38 &  \\ \cline{2-3}
	& MPst1 & 29 &  \\ \cline{2-3}
	& MP & 35 & \\ \cline{2-3}	
	& 2BIO3 & 40 &  \\ \cline{2-3}
	& 2BIO2 & 34 &  \\ \cline{2-3}
	& 2BIO1 & 37 &  \\ \thickhline
\multirow{21}*{\shortstack[l]{\textbf{Utah\_elem} \\ (class)}} & 0 & 15 & \multirow{21}*{\shortstack[l]{Students stay with \\ their assigned class \\ during the day and \\ classes mix during \\ lunch, recess and an \\ assembly during the \\ morning of the first \\ day \cite{Toth2015}.}}  \\ \cline{2-3}
	& 1 & 8 &  \\ \cline{2-3}
	& 2 & 9 &  \\ \cline{2-3}	
	& 3 & 15 &  \\ \cline{2-3}
	& 4 & 17 &  \\ \cline{2-3}
	& 5 & 17 &  \\ \cline{2-3}	
	& 6 & 13 &  \\ \cline{2-3}
	& 7 & 17 &  \\ \cline{2-3}
	& 8 & 14 &  \\ \cline{2-3}	
	& 9 & 23 &  \\ \cline{2-3}
	& 10 & 19 & \\ \cline{2-3}
	& 11 & 18 & \\ \cline{2-3}	
	& 12 & 6 & \\ \cline{2-3}	
	& 13 & 19 & \\ \cline{2-3}
	& 14 & 17 & \\ \cline{2-3}
	& 15 & 17 &  \\ \cline{2-3}	
	& 16 & 23 & \\ \cline{2-3}
	& 17 & 19 & \\ \cline{2-3}
	& 18 & 18 & \\ \cline{2-3}	
	& 19 & 20 & \\ \cline{2-3}
	& 20 & 15 & \\ \hline
\end{tabular}
\hspace{0.1cm}
\begin{tabular}[t]{|p{2.1cm}|p{1.5cm}|p{1.5cm}|p{3.1cm}|}
\hline
\textbf{Data set} & \textbf{Labels} & \textbf{Number of nodes} & \textbf{Notes}\\ \thickhline
\multirow{5}*{\shortstack[l]{\textbf{LH10} \\ (role)}} & Patient & 29 & \multirow{5}*{\shortstack[l]{Each day different \\ teams of nurses and \\ doctors were active \\ with different daily \\ schedule \cite{cattuto2013}.}} \\ \cline{2-3}
	& Paramed & 28 &  \\ \cline{2-3}
	& Med & 11 &  \\ \cline{2-3}	
	& Admin & 8 &  \\
	&  &  &  \\ \thickhline
\multirow{12}*{\shortstack[l]{\textbf{LyonSchool} \\ (class)}} & teachers &  10 & \multirow{12}{*}{\shortstack[l]{Each class has an \\ assigned room and \\ an assigned teacher; \\ two/three classes \\ have breaks at the \\ same time in a \\ shared place; lunches \\ are served in a \\ shared canteen in \\ two consecutive \\ turns \cite{barrat2011}.}} \\ \cline{2-3}
	& cpb & 25 &\\ \cline{2-3}
	& cpa & 23 &\\ \cline{2-3}	
	& cm2b & 24 &\\ \cline{2-3}
	& cm2a & 22 & \\ \cline{2-3}
	& cm1b & 23 & \\ \cline{2-3}	
	& cm1a & 21 & \\ \cline{2-3}
	& ce2b & 22 &\\ \cline{2-3}
	& ce2a & 23 &\\ \cline{2-3}	
	& ce1b & 26 &\\ \cline{2-3}
	& ce1a & 23 & \\
	&  &  &  \\ \thickhline
\multirow{13}*{\shortstack[l]{\textbf{InVS15} \\ (working dep.)}} & SSI & 9 & \multirow{13}*{\shortstack[l]{Individuals mainly \\ work in their \\ department, but can \\ mix during lunch and \\ meetings; a big event \\ occurred the morning \\ of the first day \cite{genois2015,Genois2018}.}} \\ \cline{2-3}
	& SRH & 9 & \\ \cline{2-3}
	& SFLE & 17 & \\ \cline{2-3}	
	& SDOC & 4 & \\ \cline{2-3}
	& SCOM & 7 &  \\ \cline{2-3}
	& DST & 24 & \\ \cline{2-3}	
	& DSE & 34 & \\ \cline{2-3}
	& DMI & 60 & \\ \cline{2-3}
	& DMCT & 33 & \\ \cline{2-3}	
	& DISQ & 18 & \\ \cline{2-3}
	& DG & 2  & \\ \cline{2-3}
	& DCAR & 15 &    \\
	&  &  &  \\ \hline
\end{tabular}
\caption{\textbf{Node labels information.} The table provides for each data set the node labels available, the number of nodes belonging to each class over the full observation period and some information on the contacts patterns between the different classes.}
\label{tab:nodes_labels}
\end{table}

\begin{table}[]
\begin{tabular}{|p{2.5cm}|p{2cm}|p{1.7cm}|p{1.7cm}|p{9cm}|}
\hline
\textbf{Journal} & \textbf{Years} & \textbf{Number of authors} & \textbf{Number of papers} & \textbf{Topics}\\ \hline
PRI & 1893-1912 & 212 & 191 & All of physics\\ \hline
PR & 1913-1969 & 23041 & 26434 & All of physics \\ \hline
RMP & 1929-actual & 4542 & 1684 & Full range of applied, fundamental, and interdisciplinary physics research topics\\ \hline
PRL & 1958-actual & 161632 & 114303 & Full range of applied, fundamental, and interdisciplinary physics research topics\\ \hline
PRA & 1970-actual & 73221 & 73744 & Atomic, molecular, and optical physics, foundations of quantum mechanics, and quantum information\\ \hline
PRB & 1970-actual & 168909 & 184231 &  Full range of condensed matter, materials physics, and related sub-fields\\ \hline
PRC & 1970-actual & 36221 & 36371 & Experimental and theoretical results in all aspects of nuclear physics\\ \hline
PRD & 1970-actual & 51591 & 75431 & Experimental and theoretical results in particle physics, field theory, gravitation, cosmology, and astrophysics\\ \hline
PRE & 1993-actual & 69200 & 56046 & Wide range of traditional and interdisciplinary physics topics (statistical physics, non-linear dynamics, networks)\\ \hline
PRSTAB & 1998-2015 & 5590 & 2030 & All topics in accelerator science, applications, and technology\\ \hline
PRAB & 2016-actual & 4722 & 1238 & All topics in accelerator science, applications, and technology\\ \hline
PRSTPER & 2005-2015 & 535 & 322 & All topics in experimental and theoretical physics education research, at all educational levels, from elementary through graduate education\\ \hline
PRPER & 2016-actual & 1046 & 560 & All topics in experimental and theoretical physics education research, at all educational levels, from elementary through graduate education\\ \hline
PRX & 2011-actual & 9136 & 2034 & Full spectrum of subject areas in physics\\ \hline
PRAPPLIED & 2014-actual & 18327 & 4277 & Experimental and theoretical applications of physics (interactions with other sciences, engineering and industry)\\ \hline
PRFLUIDS & 2016-actual & 6355 & 2805 & All aspects of fluid dynamics research\\ \hline
PRMATERIALS & 2017-actual & 14132 & 3205 & Wide range of topics on materials research\\ \hline
PRRESEARCH & 2019-actual & 12135 & 3453 & Full spectrum of research topics of interest to the physics community\\ \hline
PRXQUANTUM & 2020-actual & 1276 & 263 & Multidisciplinary quantum information science and technology\\ \hline
\end{tabular}
\caption{\textbf{APS journals information.} The table provides for each journal of the American Physical Society (APS): the period of active publication; the total number of authors; the total number of papers published; the topics covered. See Supplementary Fig. \ref{fig:figure37}, for details on the temporal evolution of the basic properties of some specific journals, and for an estimate of the overlap between the sets of authors of the various journals over time. Note that we only consider papers with at least two co-authors and we also consider the APS data set after the preprocessing procedure described in the Methods. The entire APS data set before any preprocessing covers the period 1893-2021 and consists of 700035 entries corresponding to papers and 496310 distinct author names. The preprocessing procedure addresses the following issues (see Methods): 10071 paper entries have missing information and are therefore removed; 248926 author names are contracted (see Methods), for 72587 contracted names we have identified a unique compatible extended name, we have reassigned all the corresponding papers to the author with that extended name and removed the contracted name, for 160929 contracted names there were no compatible extended names (they do not constitute a problem and are considered as additional unique authors), for 15410 contracted names multiple compatible extended names were available, preventing to uniquely identify the corresponding author (we consider these contracted names as additional unique authors, see Methods).}
\label{tab:APS_journals}
\end{table}

\begin{figure*}
    \centering
    \includegraphics[width=0.95\textwidth]{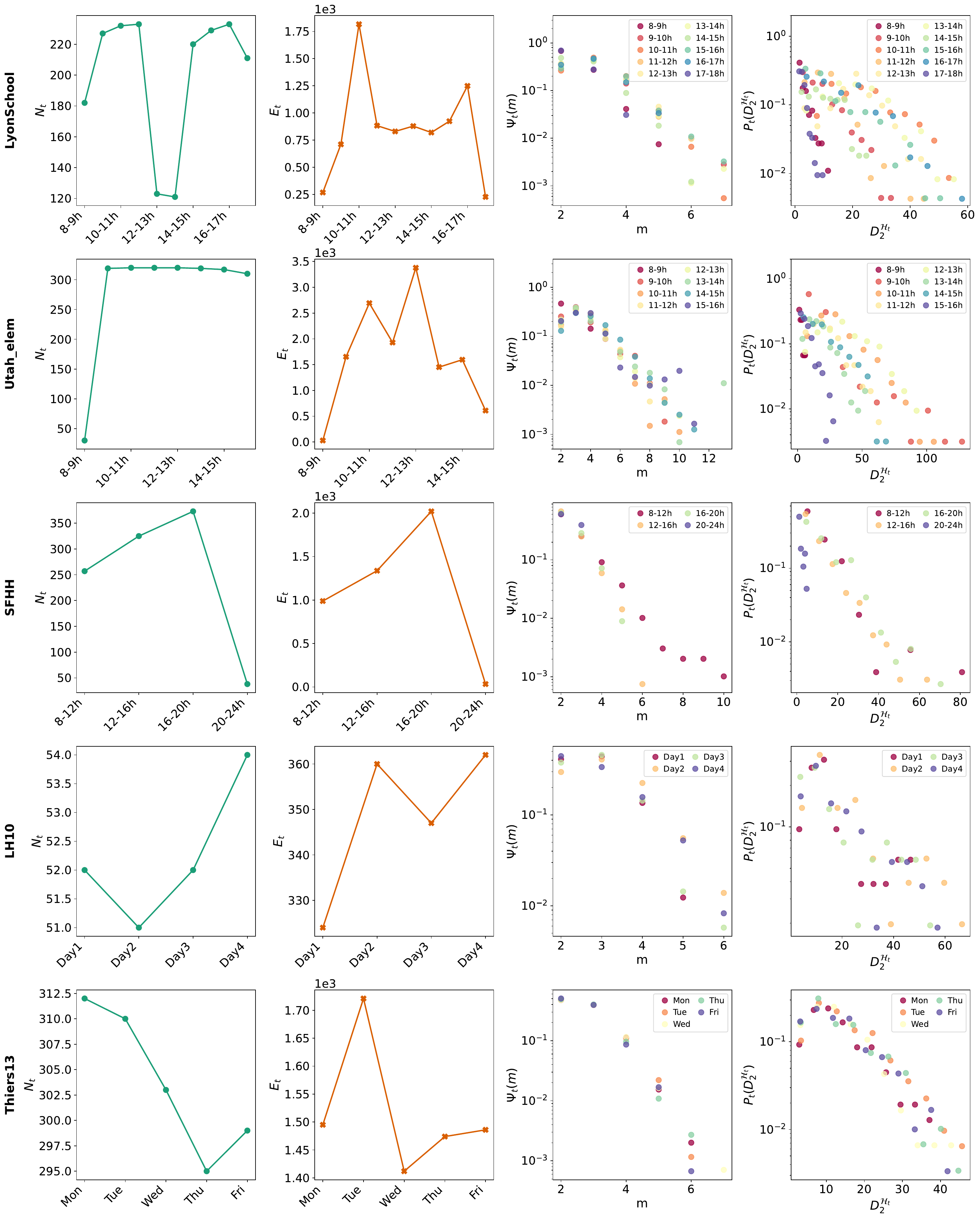}
    \caption{\textbf{Evolution of the properties of the data sets - I.} Each row corresponds to a different data set and for each of them we show: the temporal evolution of the number of active nodes $N_t$ and of the number of active hyperedges $E_t$, calculated in each static snapshot $t$; the distribution of hyperedges size $\Psi_t(m)$ and the distribution of the total degree $P_t(D_2^{\mathcal{H}_t})$ in some time windows, where $D_2^{\mathcal{H}_t}(i)$ is the total number of interactions of arbitrary size in which the node $i$ is involved in the static snapshot $\mathcal{H}_t$.
    }
    \label{fig:figure1}
\end{figure*}

\begin{figure*}
    \centering
    \includegraphics[width=0.95\textwidth]{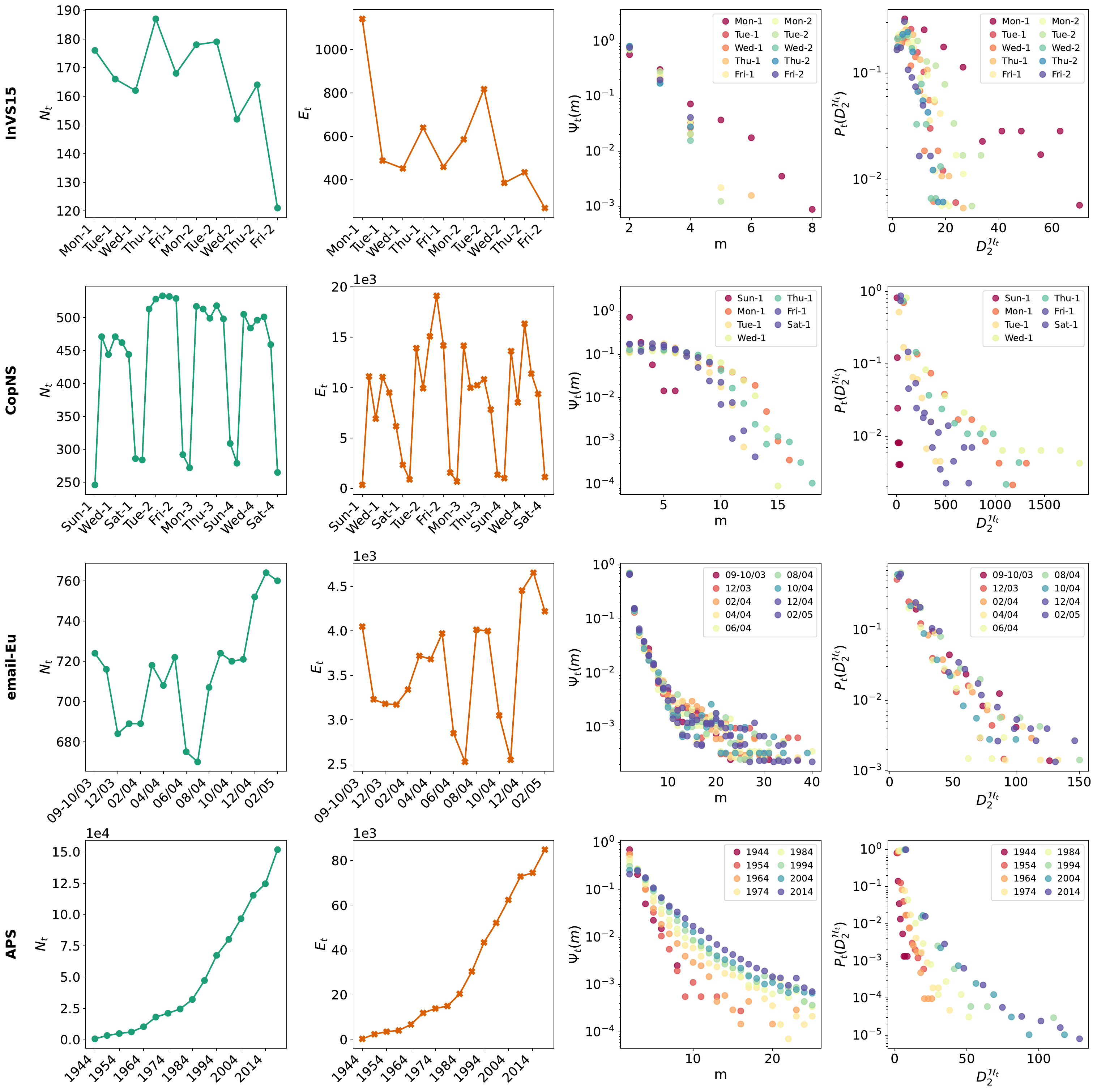}
    \caption{\textbf{Evolution of the properties of the data sets - II.} Same as in Supplementary Fig. \ref{fig:figure1}, for the remaining data sets.
    }
    \label{fig:figure2}
\end{figure*}

\clearpage
\newpage

\section{Hyper-core decomposition of empirical temporal hypergraphs}
\label{sez:II}

\begin{figure*}
    \centering
    \includegraphics[width=\textwidth]{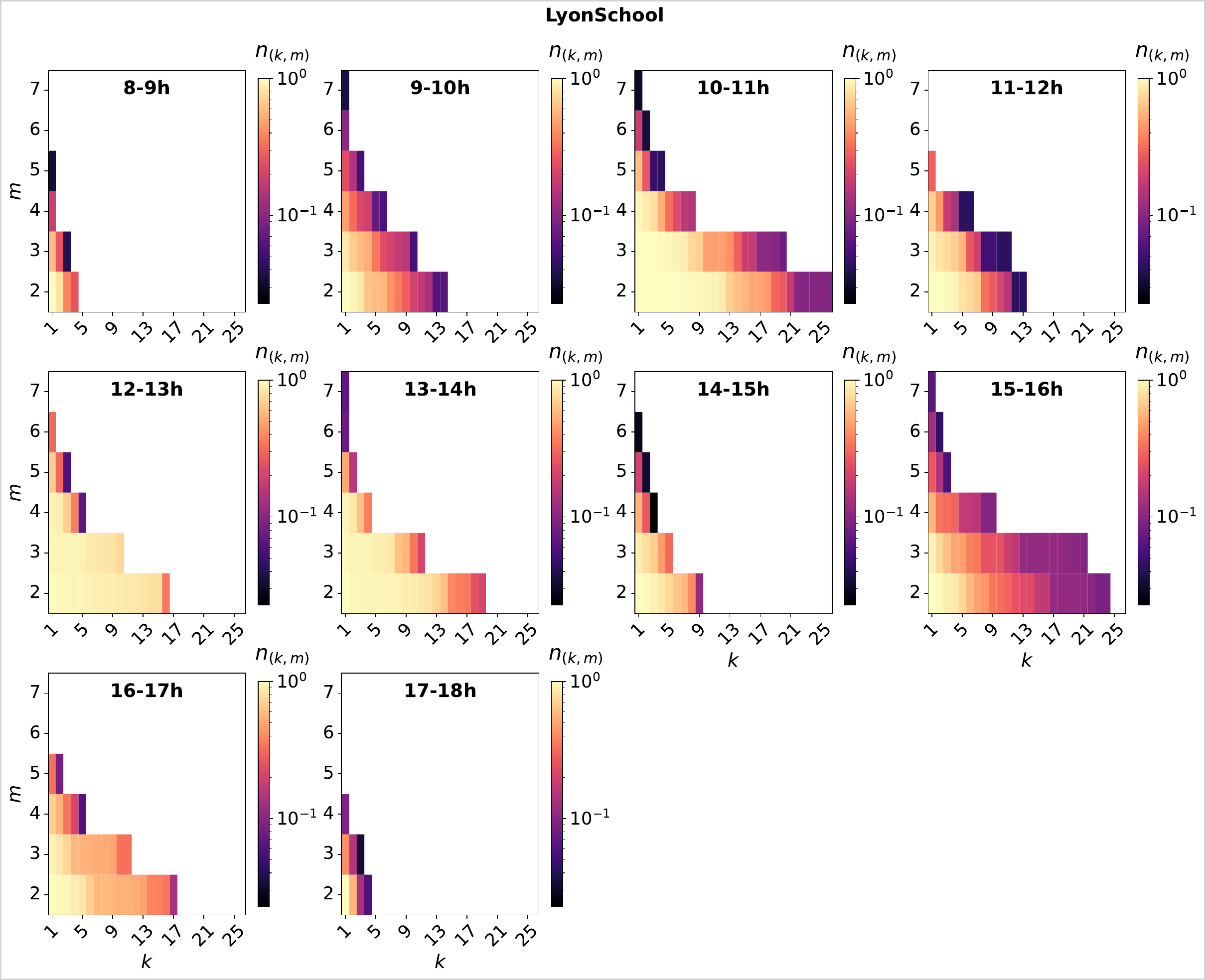} \\[0.5cm]
    \includegraphics[width=\textwidth]{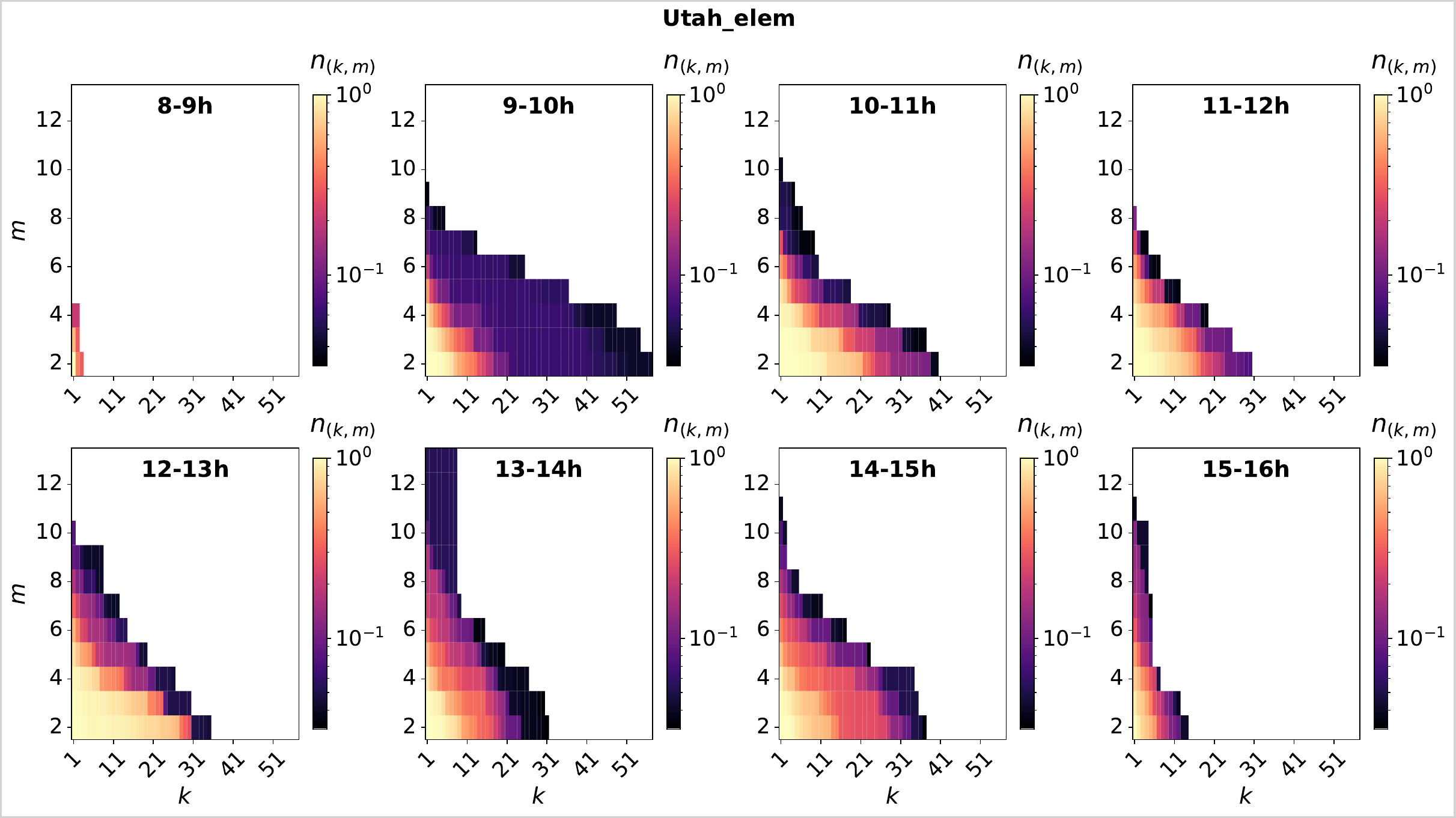}
    \caption{\textbf{Hyper-core structure evolution - I.} We show the fraction of nodes $n_{(k,m)}$ in the $(k,m)$-core as a function of $k$ and $m$ for each time window. Here we consider the LyonSchool and Utah\_elem data sets.
    }
    \label{fig:figure3}
\end{figure*}

\begin{figure*}
    \centering
    \includegraphics[width=\textwidth]{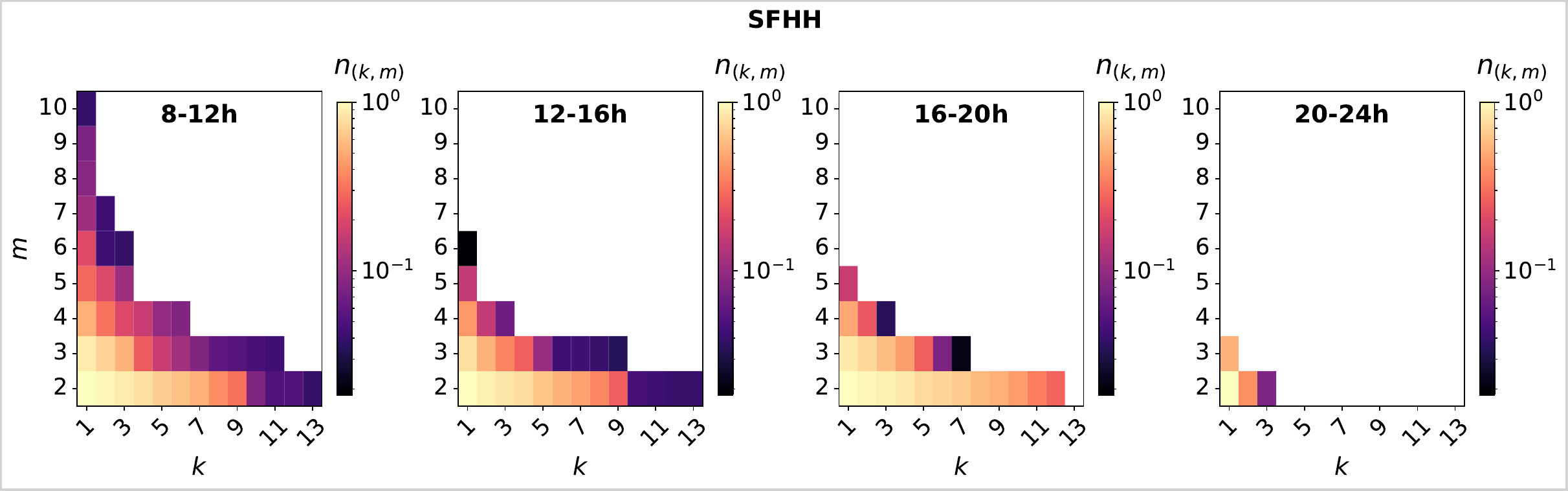} \\[0.5cm]
    \includegraphics[width=\textwidth]{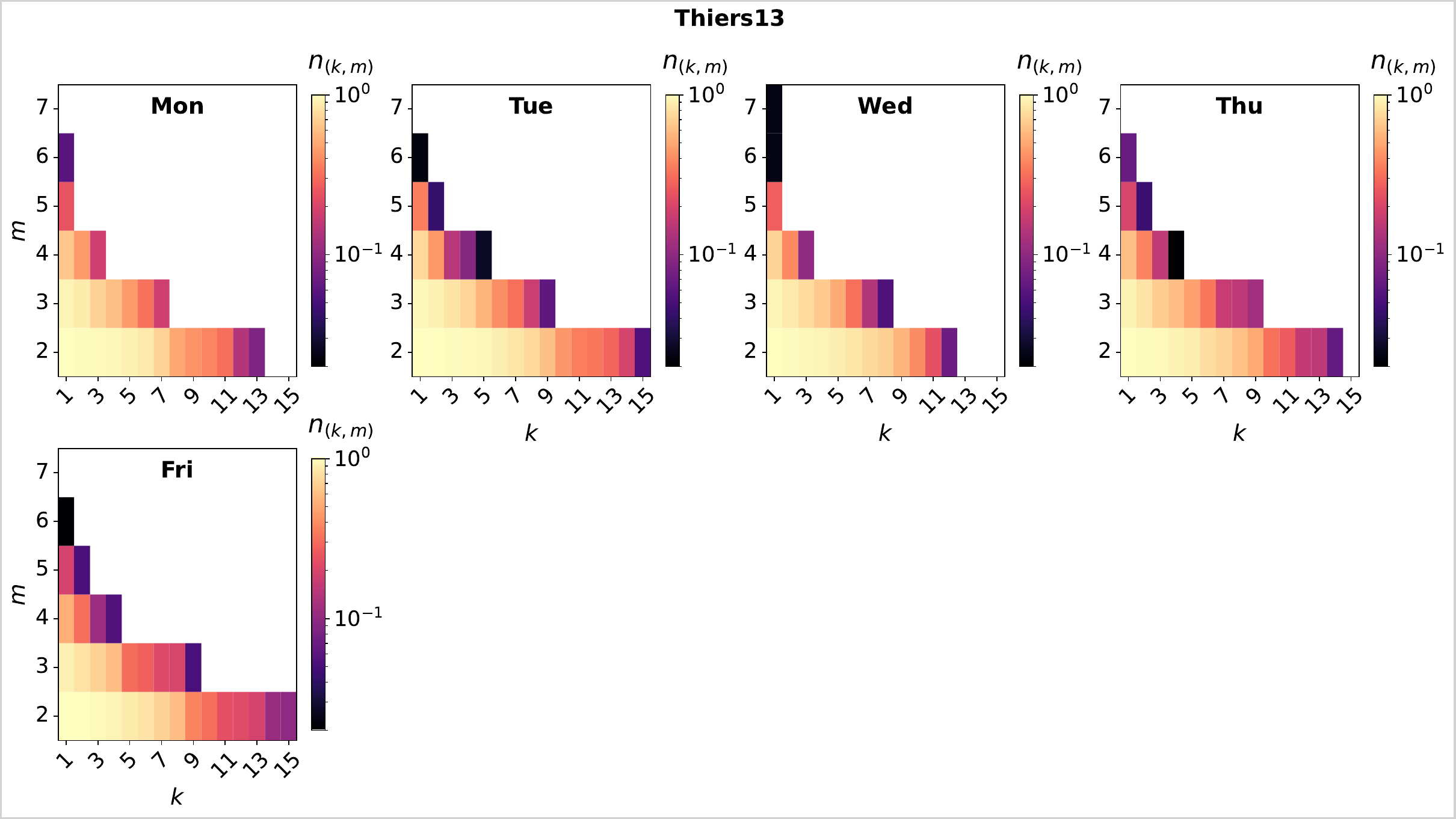}    
    \caption{\textbf{Hyper-core structure evolution - II.} As in Supplementary Fig. \ref{fig:figure3}, but here we consider the SFHH and Thiers13 data sets.
    }
    \label{fig:figure4}
\end{figure*}

\begin{figure*}
    \centering
    \includegraphics[width=\textwidth]{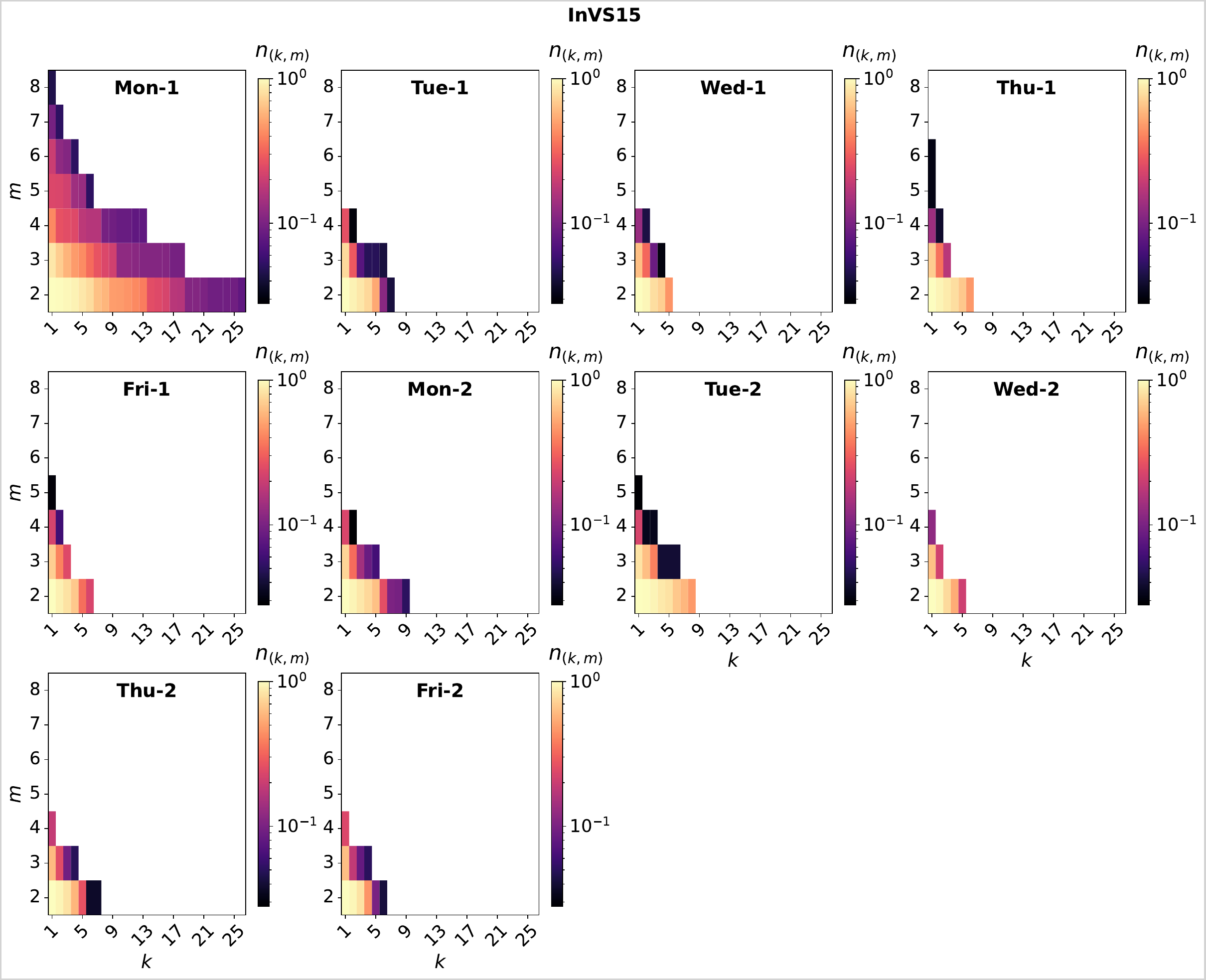}
    \caption{\textbf{Hyper-core structure evolution - III.} As in Supplementary Fig. \ref{fig:figure3}, but here we consider the InVS15 data set.
    }
    \label{fig:figure5}
\end{figure*}

\begin{figure*}
    \centering
    \includegraphics[width=0.83\textwidth]{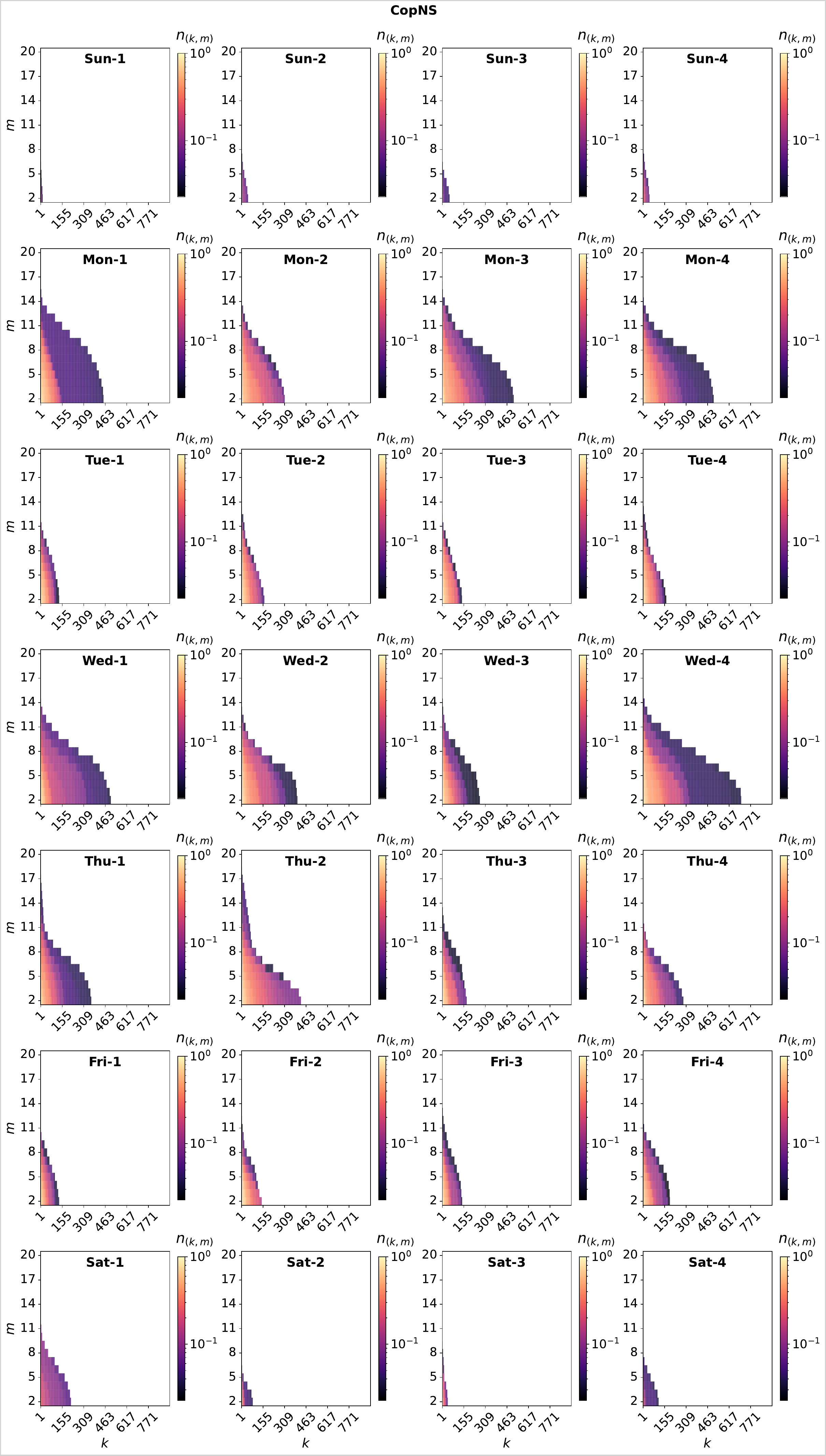}
    \caption{\textbf{Hyper-core structure evolution - IV.} As in Supplementary Fig. \ref{fig:figure3}, but here we consider the CopNS data set.
    }
    \label{fig:figure6}
\end{figure*}

\begin{figure*}
    \centering
    \includegraphics[width=\textwidth]{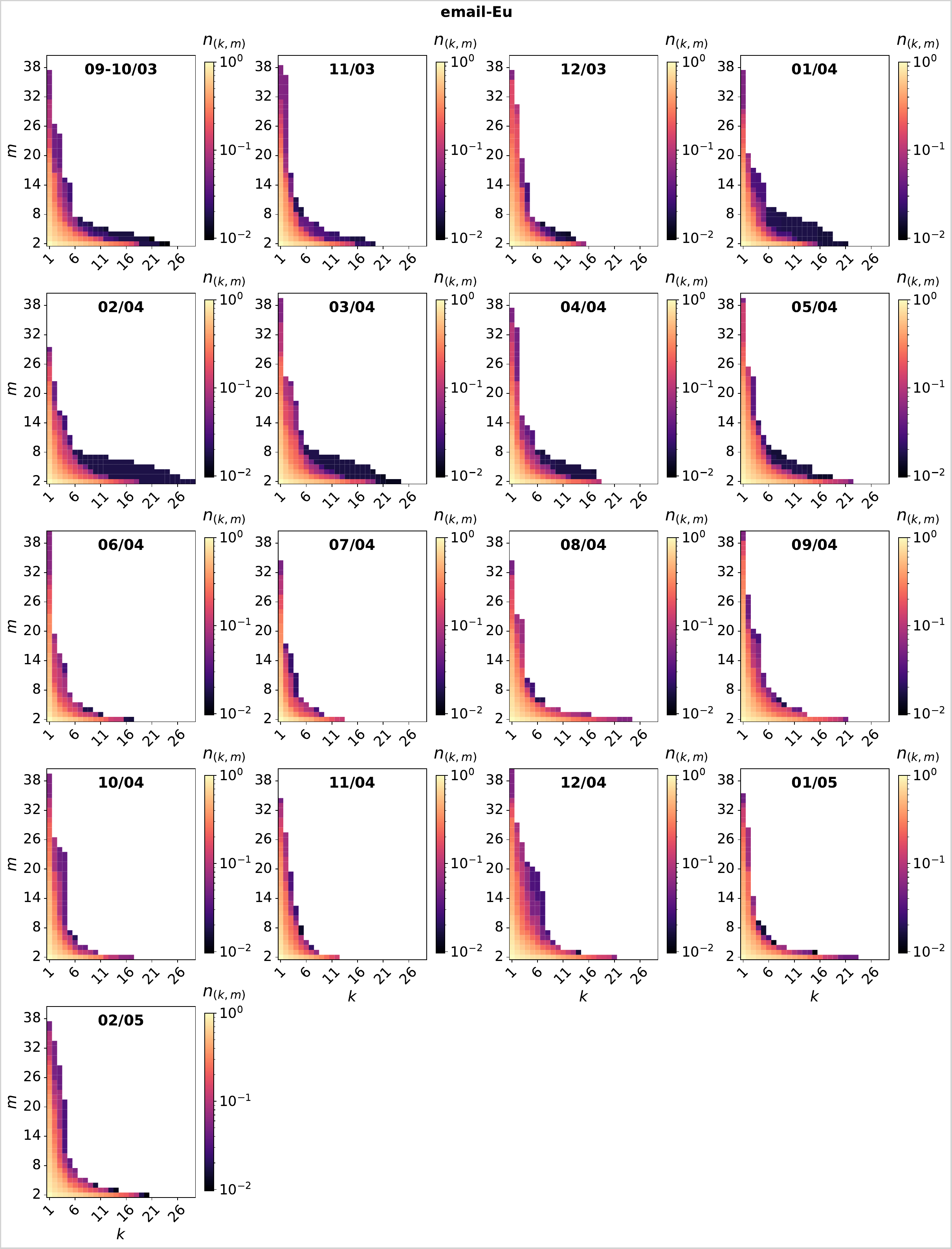}
    \caption{\textbf{Hyper-core structure evolution - V.} As in Supplementary Fig. \ref{fig:figure3}, but here we consider the email-EU data set.
    }
    \label{fig:figure7}
\end{figure*}

In this Section we present the results of the $(k,m)$-core decomposition on the empirical temporal hypergraphs not reported in the main text: given an empirical hypergraph represented with a sequence of static snapshots, we show the $(k, m)$-cores relative population size $n_{(k,m)}$ as a function of $k$ and $m$ for each time window (Supplementary Figs. \ref{fig:figure3}-\ref{fig:figure7}).

\clearpage
\newpage

\clearpage
\newpage

\section{Multi-scales hyper-core structure evolution}
\label{sez:III}

\begin{figure*}
    \centering
    \includegraphics[width=\textwidth]{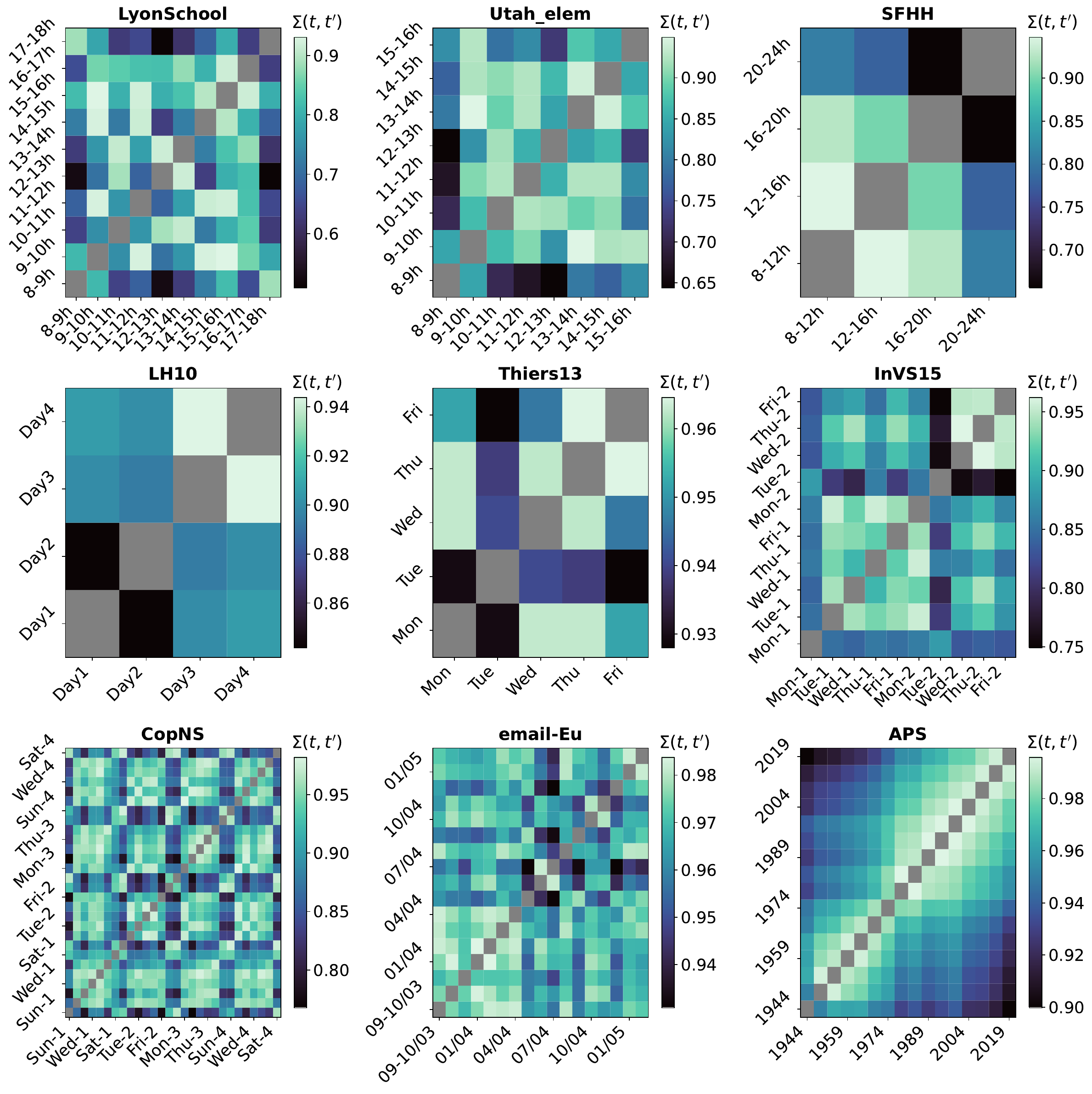}
    \caption{\textbf{Stability of hyper-core filling profiles - $\Sigma$.} For each data set, we show through a colormap the root-mean-square deviation similarity $\Sigma(t,t')$ between the filling profiles $n_{(k,m)}(t)$ and $n_{(k,m)}(t')$ in the time windows $t$ and $t'$ -- the grey diagonal corresponds to $\Sigma=1$.
    }
    \label{fig:figure8}
\end{figure*}

\begin{figure*}
    \centering
    \includegraphics[width=\textwidth]{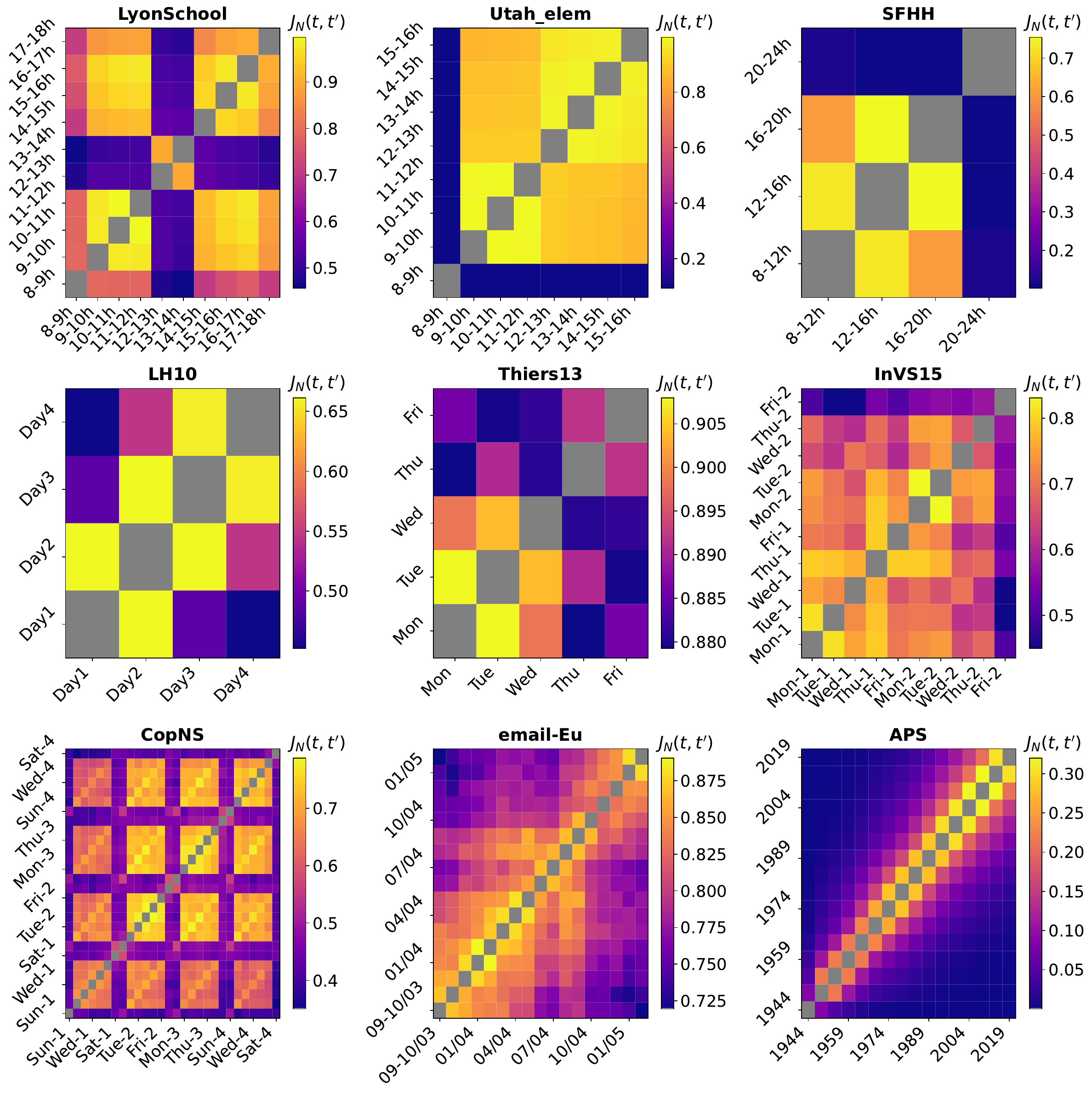}
    \caption{\textbf{Stability of the population - $J_N$.} For each data set, we show through a colormap the Jaccard similarity $J_N(t,t')$ between the sets of nodes composing the whole population at time $t$ and $t'$ -- the grey diagonal corresponds to $J_N=1$.
    }
    \label{fig:figure9}
\end{figure*}

\begin{figure*}
    \centering
    \includegraphics[width=\textwidth]{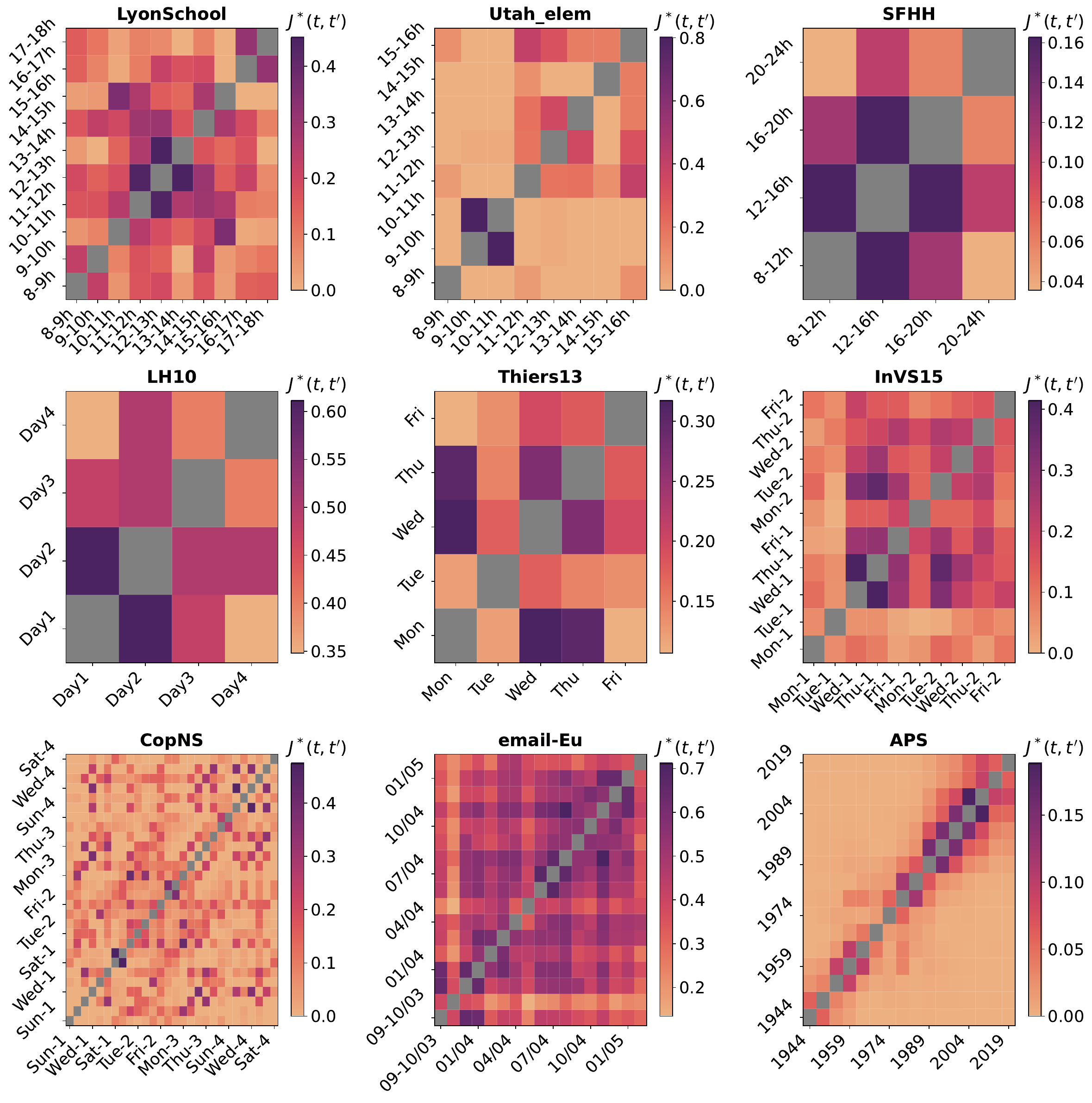}
    \caption{\textbf{Stability of central hyper-core composition - $J^*$.} For each data set, we show through a colormap the Jaccard similarity $J^*(t,t')$ between the sets of nodes belonging to the most central hyper-cores, i.e. to the $(k_{max}^m,m)$-cores $\forall m$, at time $t$ and $t'$ -- the grey diagonal corresponds to $J^*=1$.
    }
    \label{fig:figure10}
\end{figure*}

\begin{figure*}
    \centering
    \includegraphics[width=\textwidth]{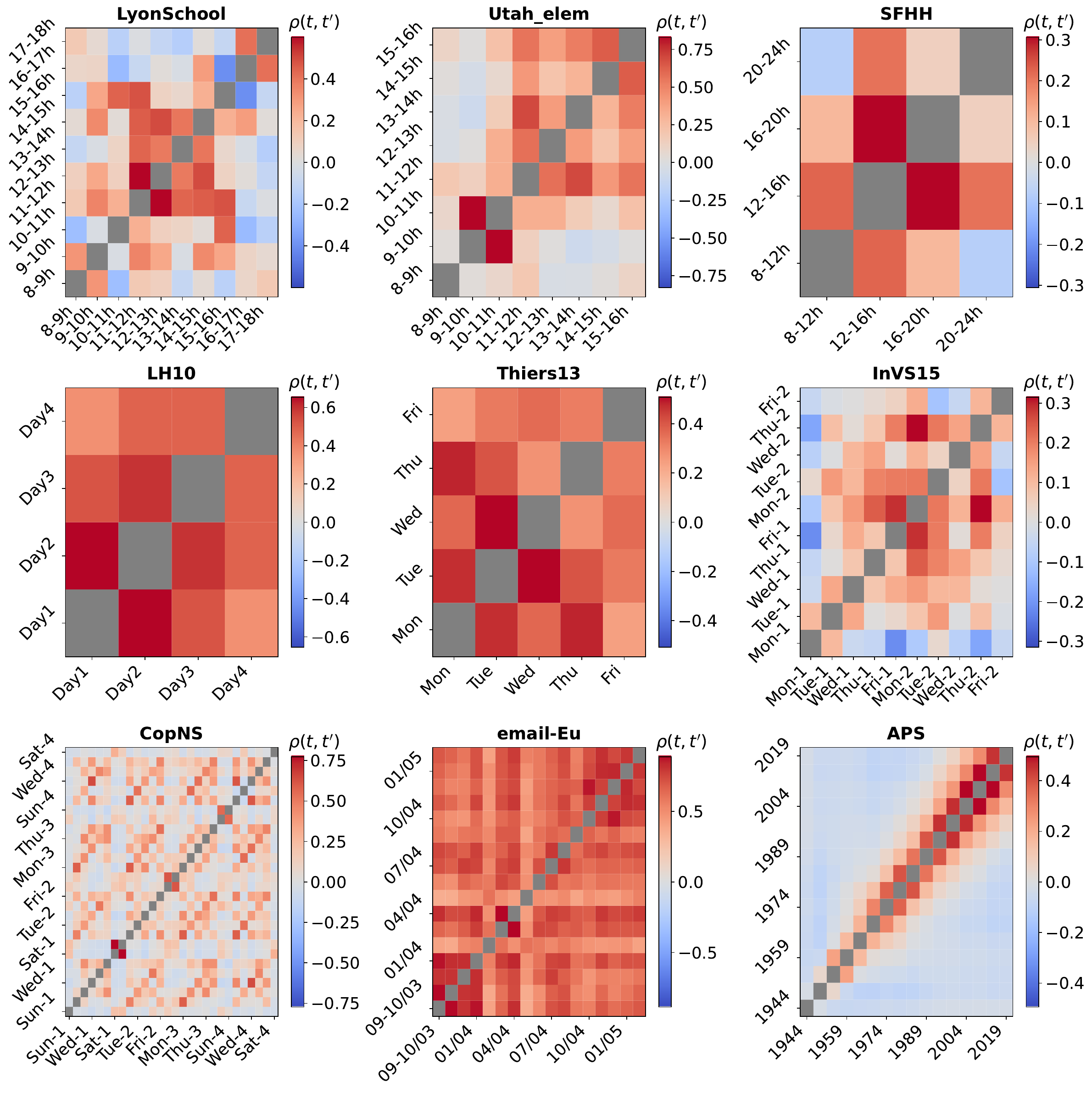}
    \caption{\textbf{Stability of instantaneous hypercoreness rankings - $\rho$.} For each data set, we show through a colormap the Pearson correlation coefficient $\rho(t,t')=\rho(R(i,t),R(i,t'))$ between the nodes hypercoreness at time $t$ and $t'$, considering all the nodes that are active in at least one of the snapshots -- the grey diagonal corresponds to $\rho=1$. 
    }
    \label{fig:figure11}
\end{figure*}

\begin{figure*}
    \centering
    \includegraphics[width=\textwidth]{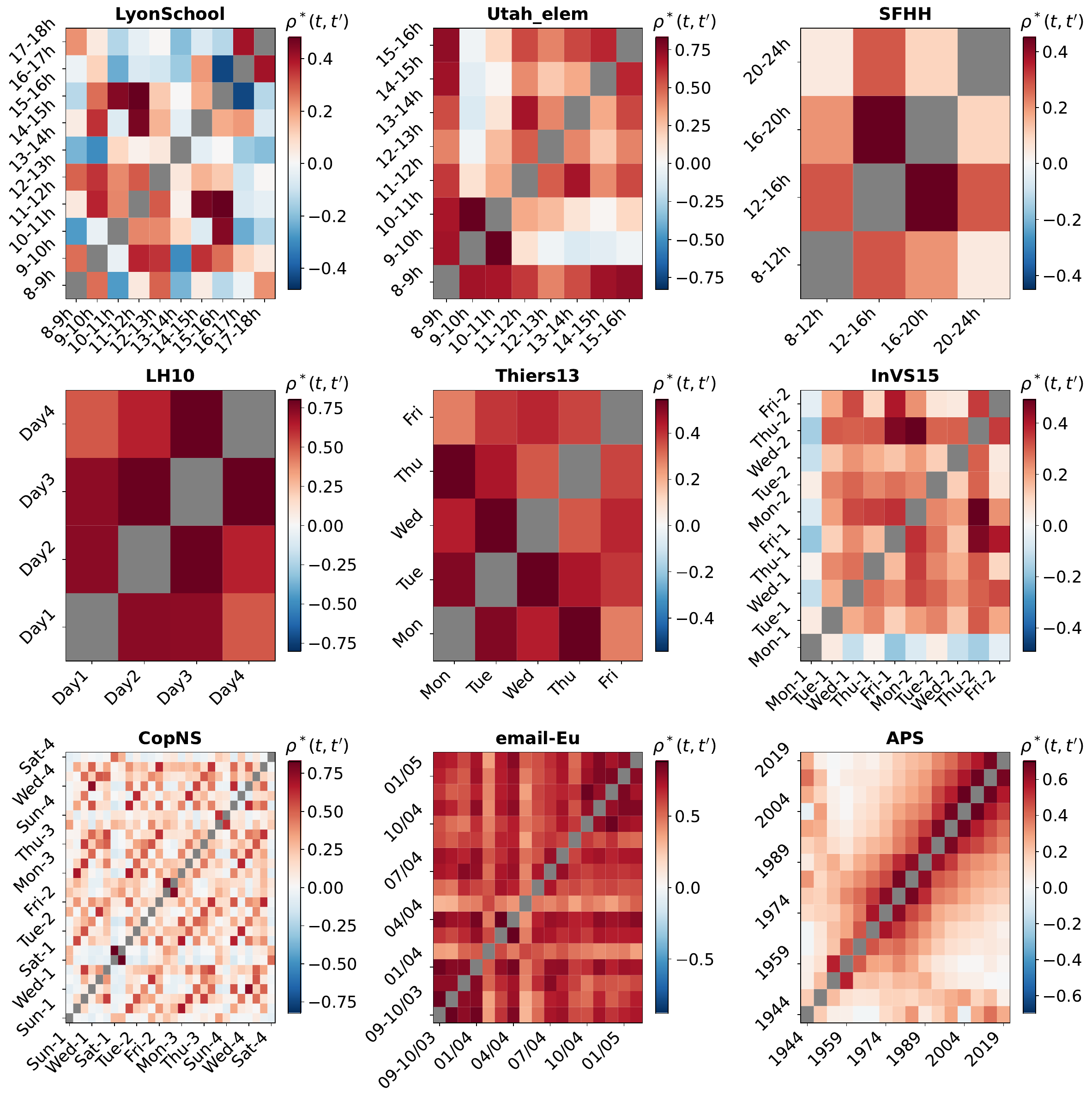}
    \caption{\textbf{Stability of instantaneous hypercoreness rankings - $\rho^*$.} For each data set, we show through a colormap the Pearson correlation coefficient $\rho^*(t,t')=\rho^*(R(i,t),R(i,t'))$ between the nodes hypercoreness at time $t$ and $t'$, considering only the nodes that are active in both the snapshots -- the grey diagonal corresponds to $\rho^*=1$.
    }
    \label{fig:figure12}
\end{figure*}

\begin{figure*}
    \centering
    \includegraphics[width=\textwidth]{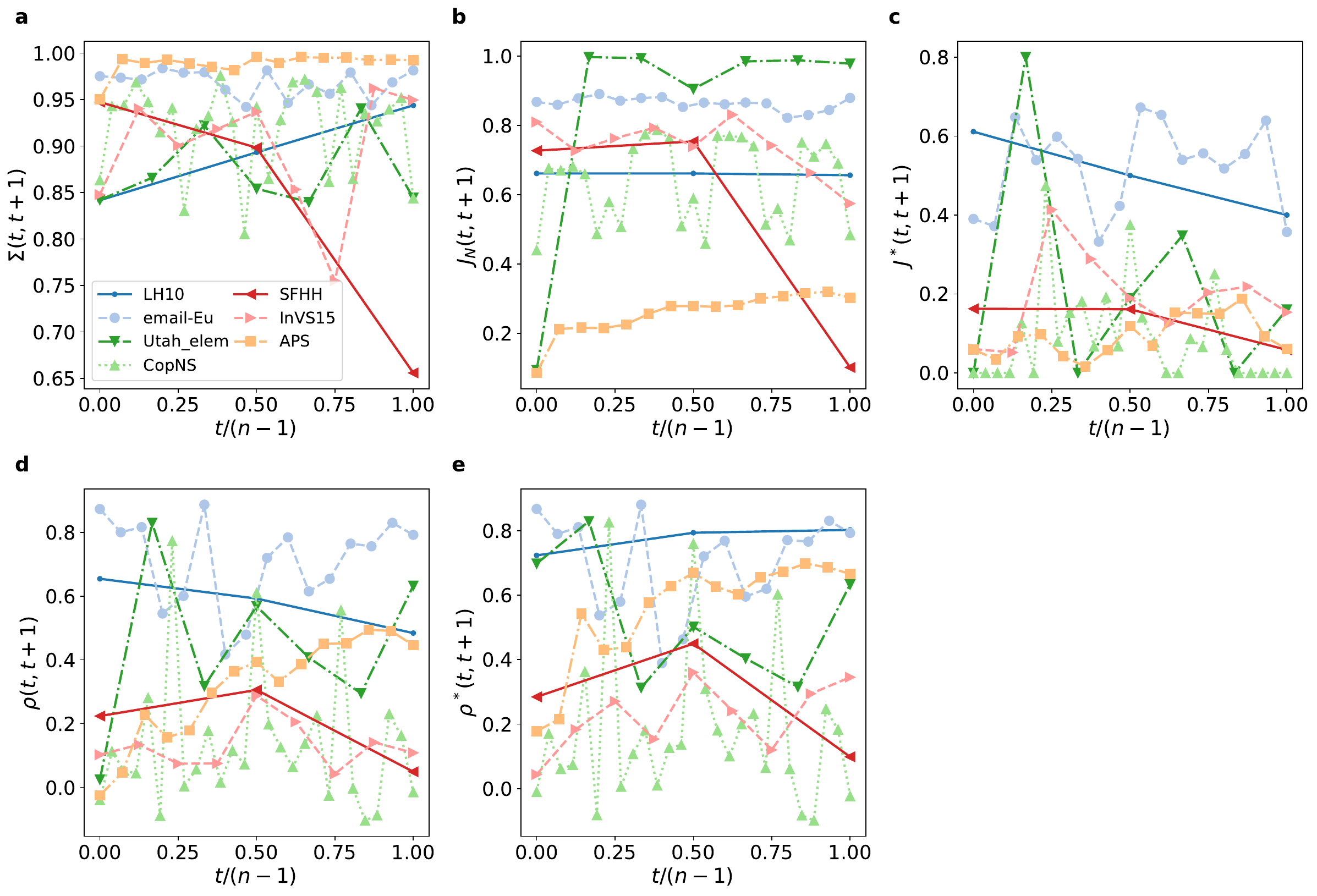}
    \caption{\textbf{Comparison of the stability of different data sets.} In panel \textbf{a} we show the similarity $\Sigma(t,t+1)$ of the hyper-core filling profiles $n_{(k,m)}$ between two adjacent snapshots, $t$ and $t+1$, as a function of time $t$. Panel \textbf{b} reports the similarity $J_N(t,t+1)$ between the sets of nodes composing the entire population in two consecutive time windows, as a function of time $t$. In panel \textbf{c} we show the similarity $J^*(t,t+1)$ between the set of nodes composing the most central hyper-cores, i.e. $(k_{max}^m,m)$-cores $\forall m$, in two adjacent snapshots, as a function of time $t$. Panels \textbf{d} and \textbf{e} report the correlation between the hypercoreness of the nodes in two adjacent time windows as a function of time $t$, respectively considering all the nodes that are active in at least one of the two snapshots, $\rho(t,t+1)$, and considering only the nodes that are active in both snapshots, $\rho^*(t,t+1)$. We consider some data sets representative of specific classes of systems and with different behaviors: LH10 and email-EU, which present high stability; SFHH and InVS15, which present low stability; Utah\_elem and CopNS, which present phases of high stability alternating with phases of low stability; APS, which presents high structural similarity, low stability in the central cores and in the hypercoreness ranking, but with temporal consolidation of the rankings. We rescale the time $t$ with $(n-1)=(t_{max}/\tau-1)$, in order to compare different data sets covering different time intervals.
    }
    \label{fig:figure13}
\end{figure*}

\begin{figure*}
    \centering
    \includegraphics[width=0.93\textwidth]{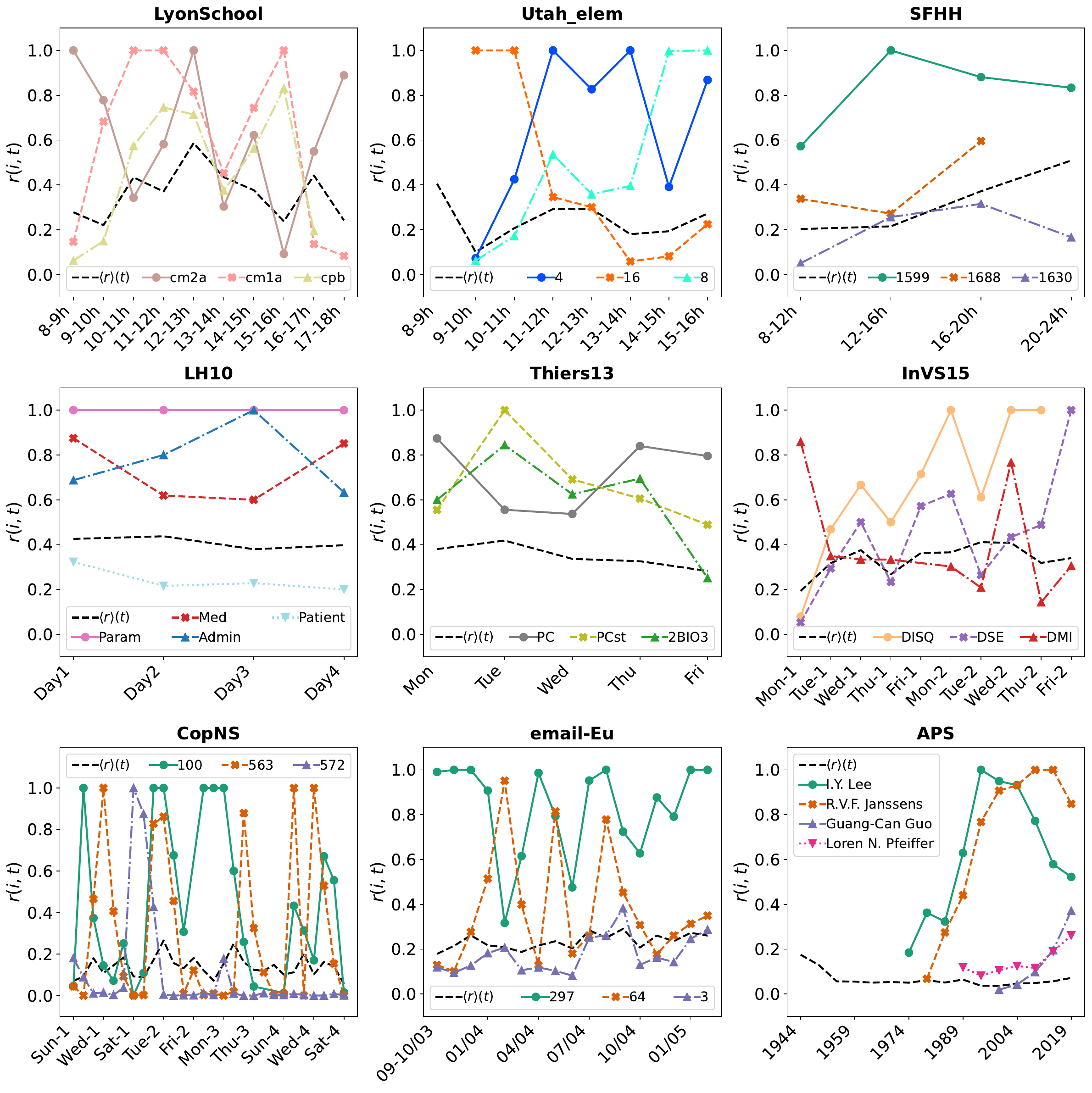}
    \caption{\textbf{Hypercoreness evolution.} For each data set, we show the temporal evolution of the hypercoreness $r(i,t)$ for several nodes, chosen among different nodes classes, if available (see Supplementary Tables \ref{tab:dataset_time}-\ref{tab:APS_journals}), and selected to illustrate different hypercoreness trajectory evolutions over time. 
    We also plot the mean $\langle r \rangle (t)$ value, averaging only on active nodes. In the legend of each panel we indicate the node \texttt{id} or the corresponding hyperedge label. We consider: in LyonSchool the nodes \texttt{id=1452} (cm2a), \texttt{id=1500} (cm1a) and \texttt{id=1765} (cpb); in Utah\_elem the nodes \texttt{id=143} (4), \texttt{id=287} (16) and \texttt{id=337} (8); in SFHH the nodes \texttt{id=1599}, \texttt{id=1688} and \texttt{id=1630}; in LH10 the nodes \texttt{id=1210} (Param), \texttt{id=1144} (Med), \texttt{id=1098} (Admin)  and \texttt{id=1383} (Patient); in Thiers13 the nodes \texttt{id=492} (PC), \texttt{id=254} (PCst) and \texttt{id=122} (2BIO3); in InVS15 the nodes \texttt{id=267} (DISQ), \texttt{id=510} (DSE) and \texttt{id=74} (DMI); in CopNS the nodes \texttt{id=100}, \texttt{id=563} and \texttt{id=572}; in email-EU the nodes \texttt{id=297}, \texttt{id=64} and \texttt{id=3}; in APS the authors I.Y. Lee, R.V.F. Janssens, Guang-Can Guo and Loren N. Pfeiffer.  
    } 
    \label{fig:figure14}
\end{figure*}

In this Section we present the analysis on the multi-scales hyper-core structure evolution for all the empirical temporal hypergraphs and we also compare them. For each empirical dataset, we consider two snapshots $t$, $t'$ and we show: the stability of the hyper-core filling profiles $n_{(k,m)}$, $\Sigma(t,t')$ (Supplementary Fig. \ref{fig:figure8}); the stability of whole population composition, $J_N(t,t')$ (Supplementary Fig. \ref{fig:figure9}); the stability of the central hyper-cores composition, $J^*(t,t')$ (Supplementary Fig. \ref{fig:figure10}); the stability of the instantaneous hypercoreness rankings, considering nodes active at least in $t$ or $t'$, $\rho(t,t')$ (Supplementary Fig. \ref{fig:figure11}) and considering nodes active in both $t$ and $t'$, $\rho^*(t,t')$ (Supplementary Fig. \ref{fig:figure12}). We also compare the stability of different data sets for the various measures considered (i.e. at the various topological scales), by considering adjacent snapshots, $t$ and $t+1$, and rescaling the time $t$ with $(n-1)=(t_{max}/\tau-1)$ in order to make the data sets comparable (Supplementary Fig. \ref{fig:figure13}). Finally, for each data set we show the temporal evolution of the hypercoreness $r(i,t)=R(i,t)/\max_i\{R(i,t)\}$ for several nodes, which are chosen among different nodes classes, if available (see Supplementary Tables \ref{tab:dataset_time}-\ref{tab:APS_journals}), and are selected to show different hypercoreness trajectory evolutions over time (Supplementary Fig. \ref{fig:figure14}). 

\clearpage
\newpage

\newpage

\clearpage
\newpage

\section{Time-aggregated hypercoreness}
\label{sez:IV}

\begin{figure*}
    \centering
    \includegraphics[width=\textwidth]{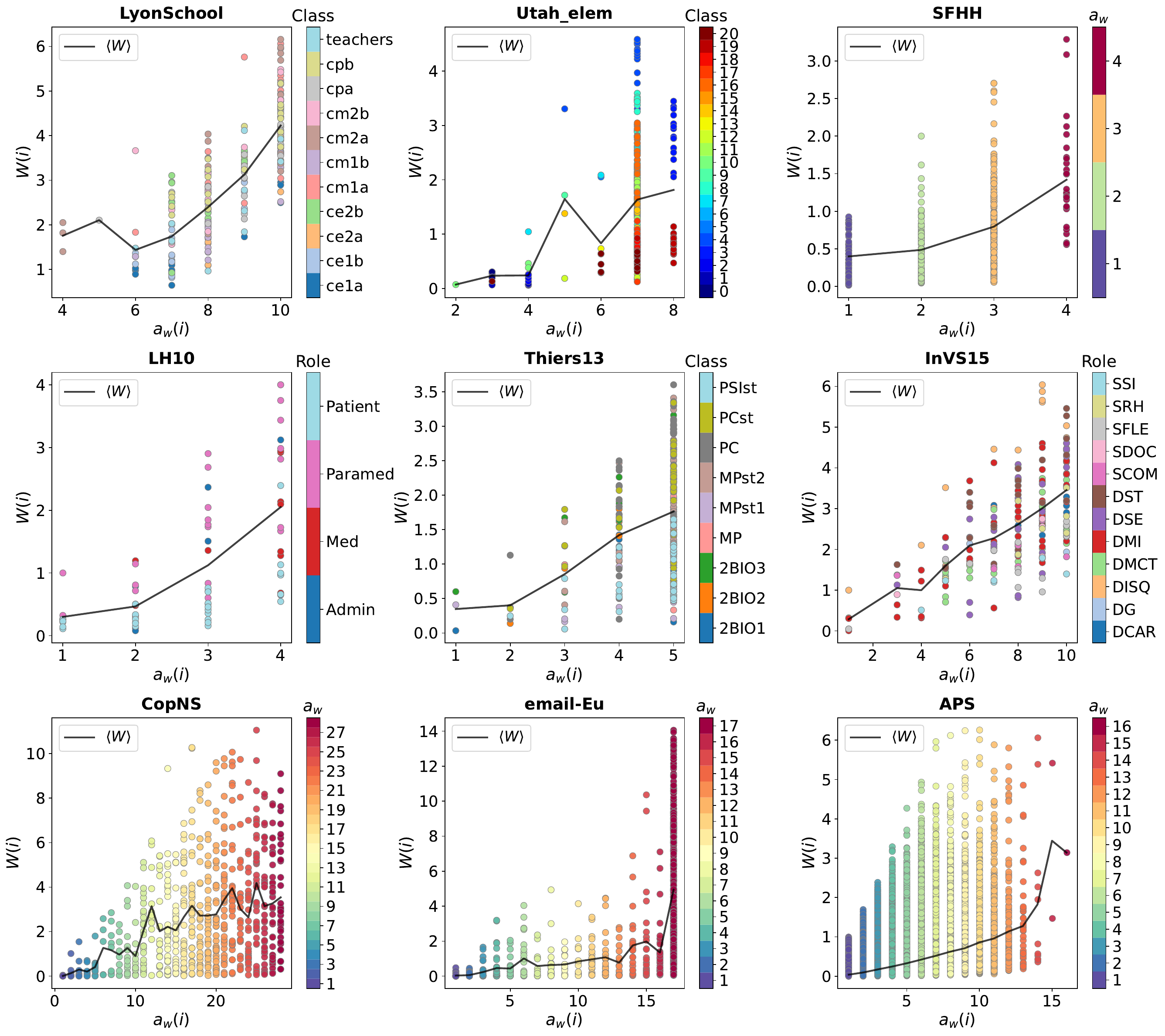}
    \caption{\textbf{Time-aggregated centrality measures - I.} For each data set, we show a scatter plot of the aggregated hypercoreness $W(i)$ as a function of the snapshot activity $a_w(i)$ for all nodes $i$, and the average aggregated hypercoreness $\langle W \rangle$ as a function of $a_w$. The points are colored according to the node activity snapshot $a_w(i)$ (SFHH, CopNS, email-EU, APS) or to the node label (LyonSchool, Utah\_elem, LH10, Thiers13, InVS15).
    }
    \label{fig:figure15}
\end{figure*}

\begin{figure*}
	\centering
    \includegraphics[width=\textwidth]{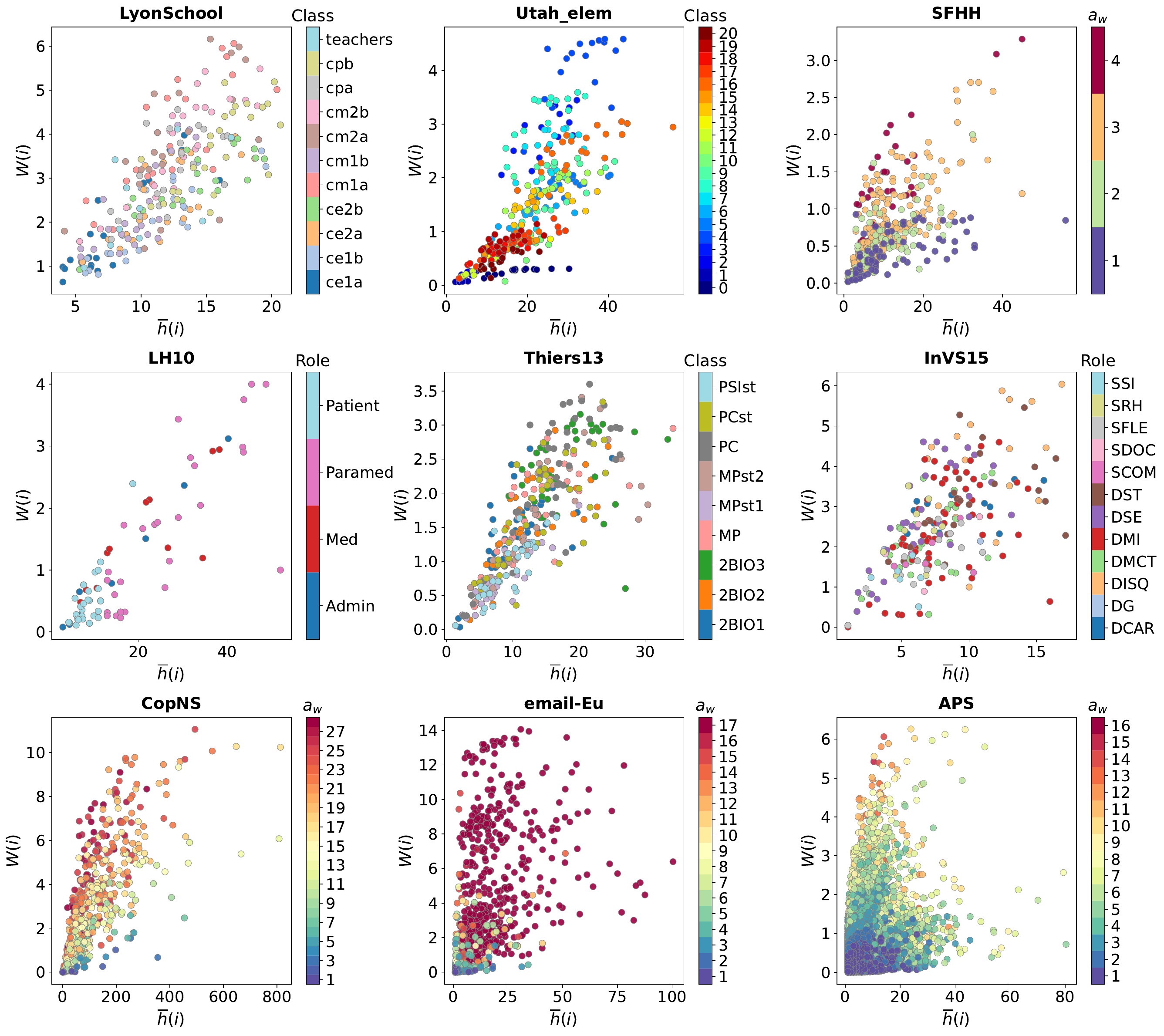}
    \caption{\textbf{Time-aggregated centrality measures - II.} For each data set, we show a scatter plot of the aggregated hypercoreness $W(i)$ as a function of the average number of interactions per active window $\overline{h}(i)$ for all nodes $i$. The points are colored according to the node activity snapshot $a_w(i)$ (SFHH, CopNS, email-EU, APS) or to the node class (LyonSchool, Utah\_elem, LH10, Thiers13, InVS15).
    } 
    \label{fig:figure16}
\end{figure*}

\begin{figure*}
	\centering
    \includegraphics[width=\textwidth]{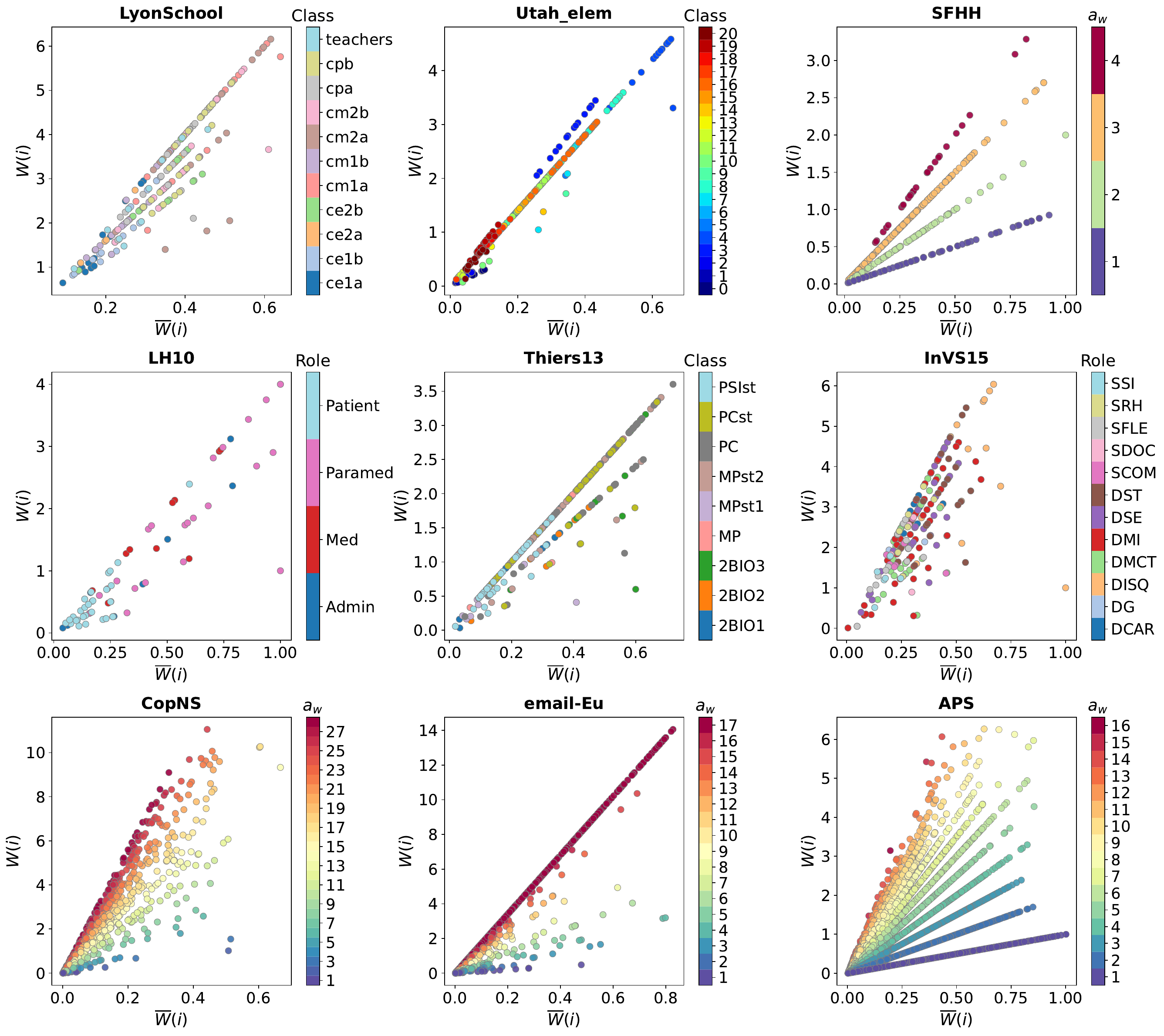}
    \caption{\textbf{Time-aggregated centrality measures - III.} For each data set, we show a scatter plot of the aggregated hypercoreness $W(i)$ as a function of the activity-averaged hypercoreness $\overline{W}(i)$ for all nodes $i$. The points are colored according to the node activity snapshot $a_w(i)$ (SFHH, CopNS, email-EU, APS) or to the node class (LyonSchool, Utah\_elem, LH10, Thiers13, InVS15).
    } 
    \label{fig:figure17}
\end{figure*}

\begin{figure*}
    \centering
    \includegraphics[width=\textwidth]{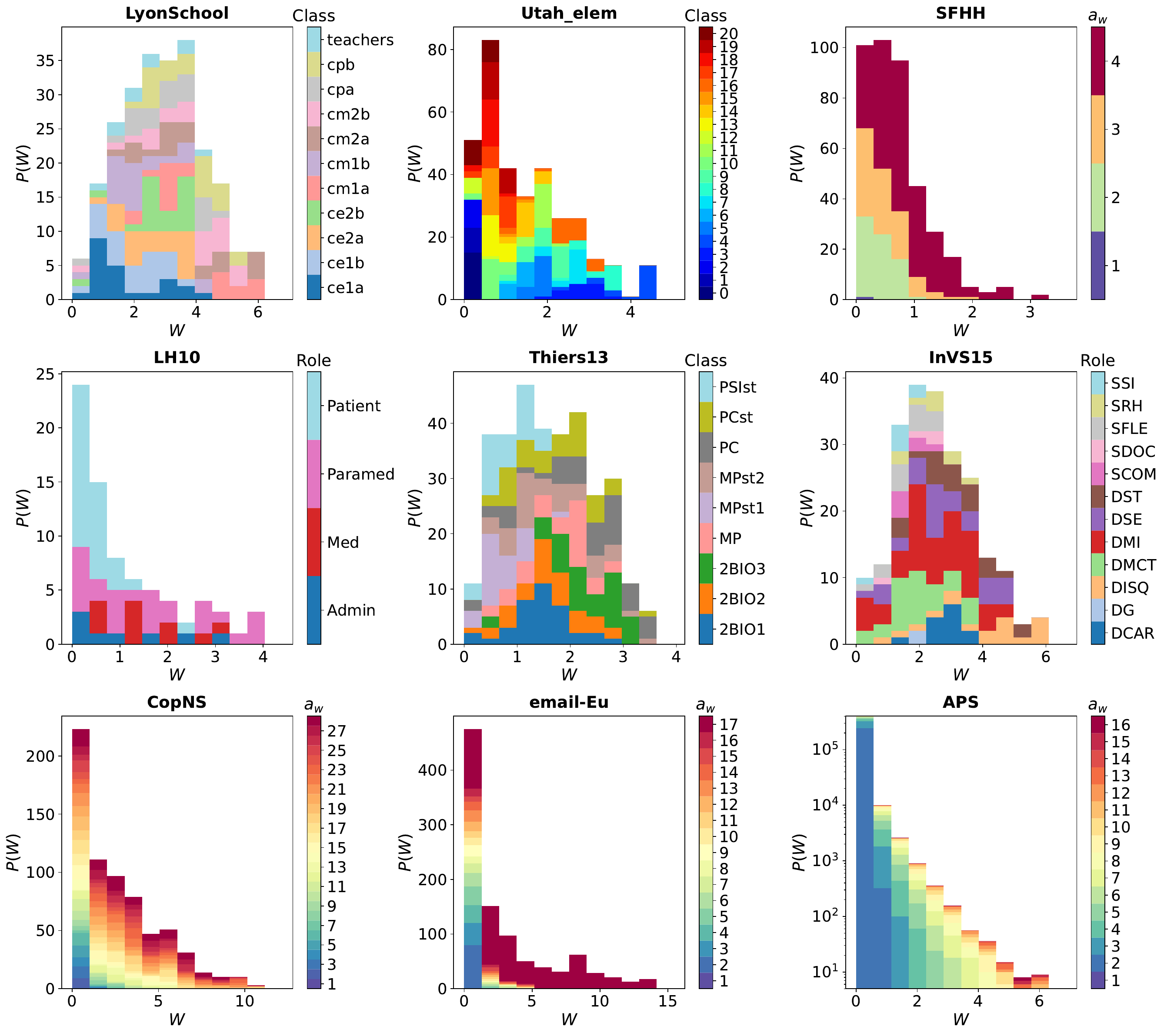}
    \caption{\textbf{Distribution of the aggregated hypercoreness.} For each data set, we show through histograms the number of nodes $P(W)$ with aggregated hypercoreness $W$: within each bar we distinguish the relative frequency of nodes belonging to each class $a_w$ (SFHH, CopNS, email-EU, APS) or to each node label (LyonSchool, Utah\_elem, LH10, Thiers13, InVS15), through stacked bars.
    }
    \label{fig:figure18}
\end{figure*}

\begin{figure*}
    \centering
    \includegraphics[width=\textwidth]{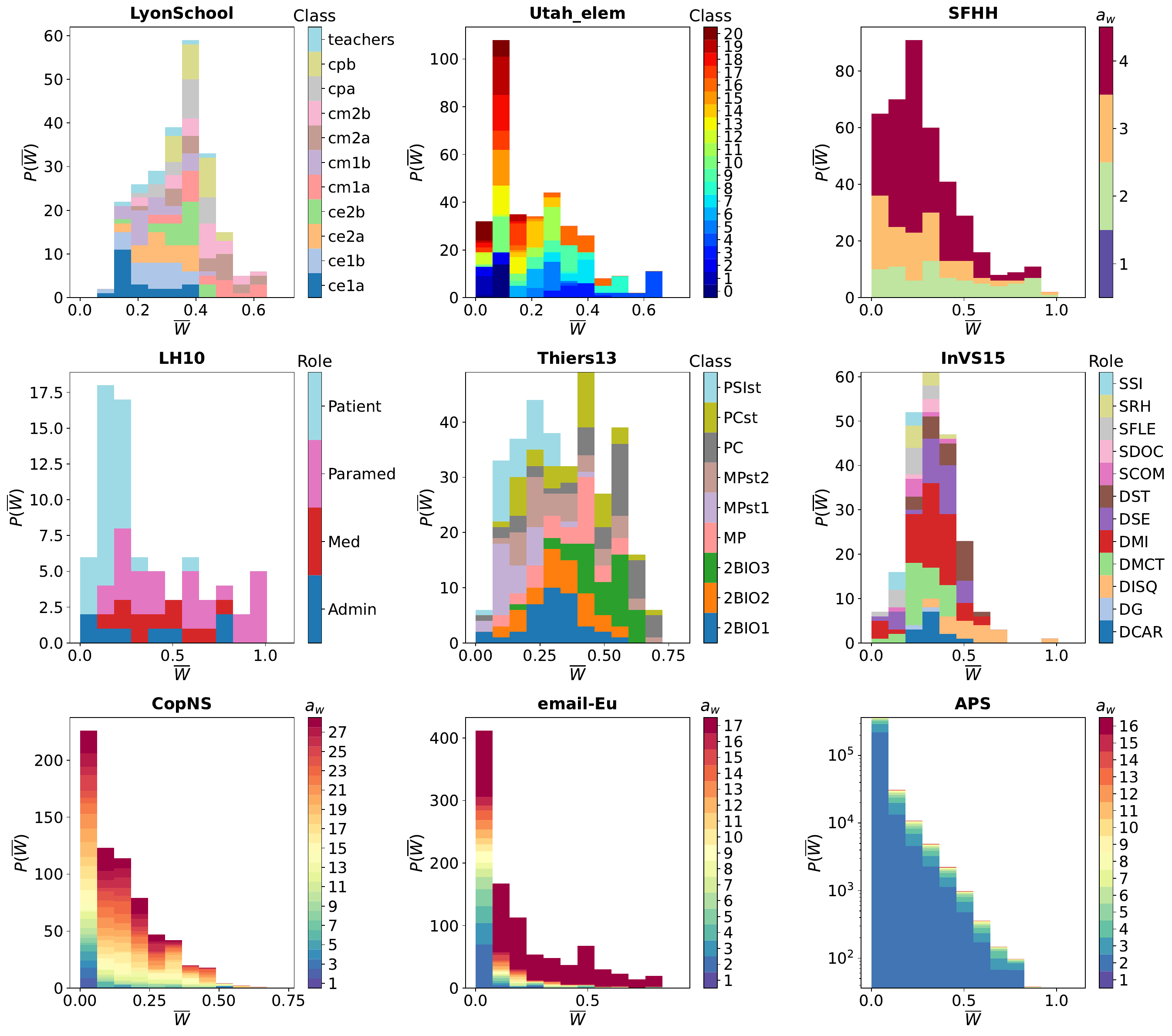}
    \caption{\textbf{Distribution of the activity-averaged hypercoreness.} For each data set, we show through histograms the number of nodes $P(\overline{W})$ with activity-averaged hypercoreness $\overline{W}$: within each bar we distinguish the relative frequency of nodes belonging to each class $a_w$ (SFHH, CopNS, email-EU, APS) or to each node label (LyonSchool, Utah\_elem, LH10, Thiers13, InVS15), through stacked bars.
    }
    \label{fig:figure19}
\end{figure*}

In this Section we consider the time-aggregated hypercoreness measures proposed, their correlations and distributions. For each empirical dataset, we show through a scatter-plot the aggregated hypercoreness $W(i)$ as a function of: the snapshot activity $a_w(i)$ (Supplementary Fig. \ref{fig:figure15}); the average number of interactions per active window $\overline{h}(i)$ (Supplementary Fig. \ref{fig:figure16}); the activity-averaged hypercoreness $\overline{W}(i)$ (Supplementary Fig. \ref{fig:figure17}). We also show the distribution $P(W)$ of the aggregated hypercoreness (Supplementary Fig. \ref{fig:figure18}) and the distribution $P(\overline{W})$ of the activity-averaged hypercoreness (Supplementary Fig. \ref{fig:figure19}).

\clearpage
\newpage

\newpage

\clearpage
\newpage

\section{Prevalent class in hyper-cores}
\label{sez:V}

In this Section we focus on the structural contribution of the different classes of nodes/hyperedges (i.e., labels) in the hyper-core structure and its temporal evolution. We consider each data set not shown in the main text with node/hyperedge labels (see Supplementary Tables \ref{tab:dataset_time}-\ref{tab:APS_journals}), for each time window we show the dominant role of the nodes that compose each $(k,m)$-core (see Supplementary Figs. \ref{fig:figure20}-\ref{fig:figure21}), considering a dominant role if its frequency exceeds 0.5. Furthermore, we consider a randomized version of each empirical hypergraph, generating 50 randomized realizations for each time window (see Methods), and we consider the average frequency of each node/hyperedge label in each $(k,m)$-core: we only report the APS data set (see Supplementary Figs. \ref{fig:figure22}), where in some time windows a non-trivial pattern is present; all the other data sets present a trivial pattern in which it is not possible to identify a dominant role of the nodes in any $(k,m)$-core and in any time window, therefore we do not report these figures which are simply characterized by all white panels. Finally, for each data set we show the temporal evolution of the relative frequency of the various node labels in the top positions of the instantaneous hypercoreness ranking or the evolution of the relative frequency of the hyperedge labels in the most central hyper-cores (see Supplementary Figs. \ref{fig:figure23}): we consider both the original data set and the randomized version of the hypergraph, obtained by generating 50 randomized realizations in each window temporal (see Methods) and averaging the labels frequencies observed in each realization. 

\clearpage
\newpage

\begin{figure*}
    \centering
    \includegraphics[width=0.95\textwidth]{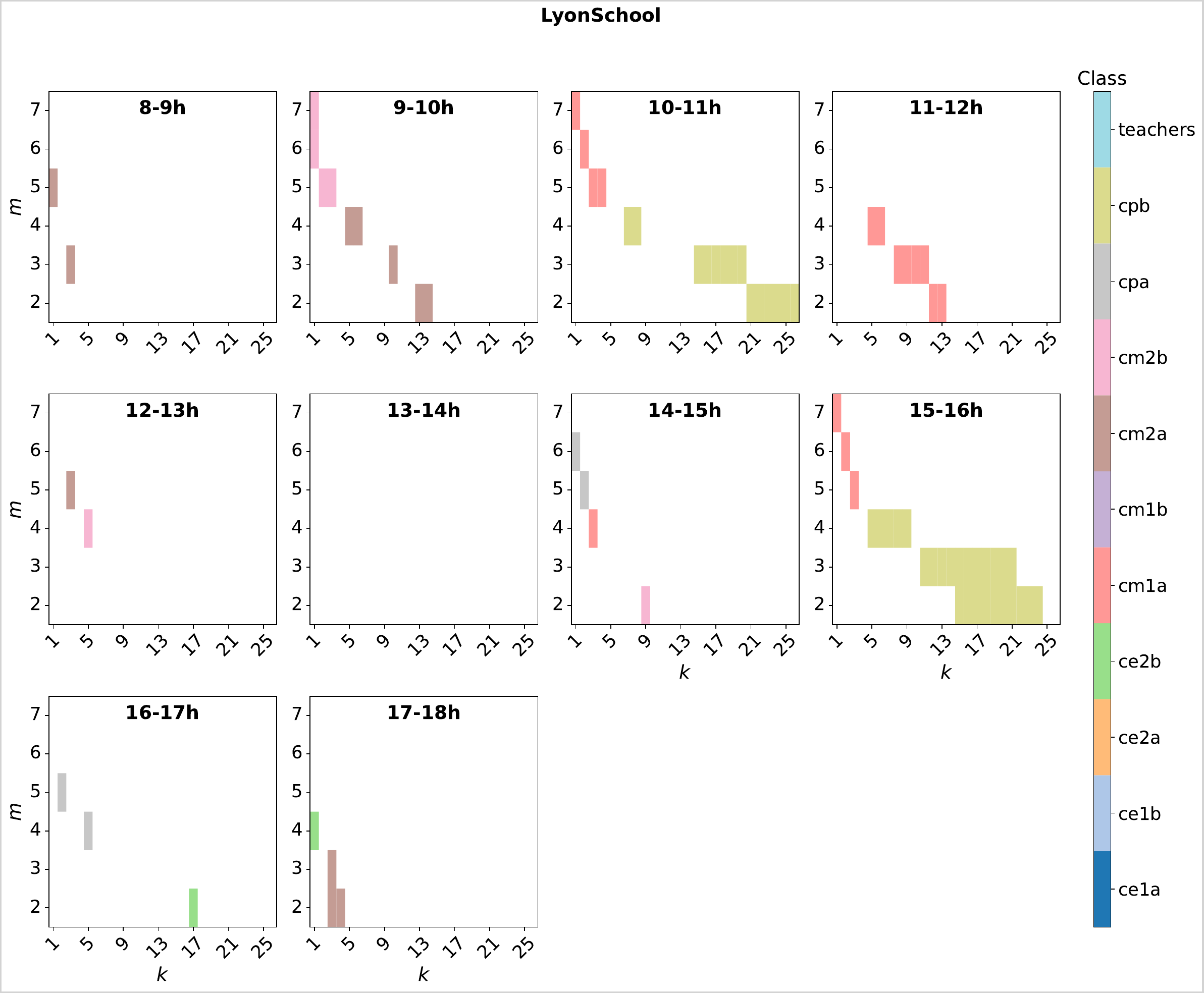} \\[0.2cm]
    \includegraphics[width=0.95\textwidth]{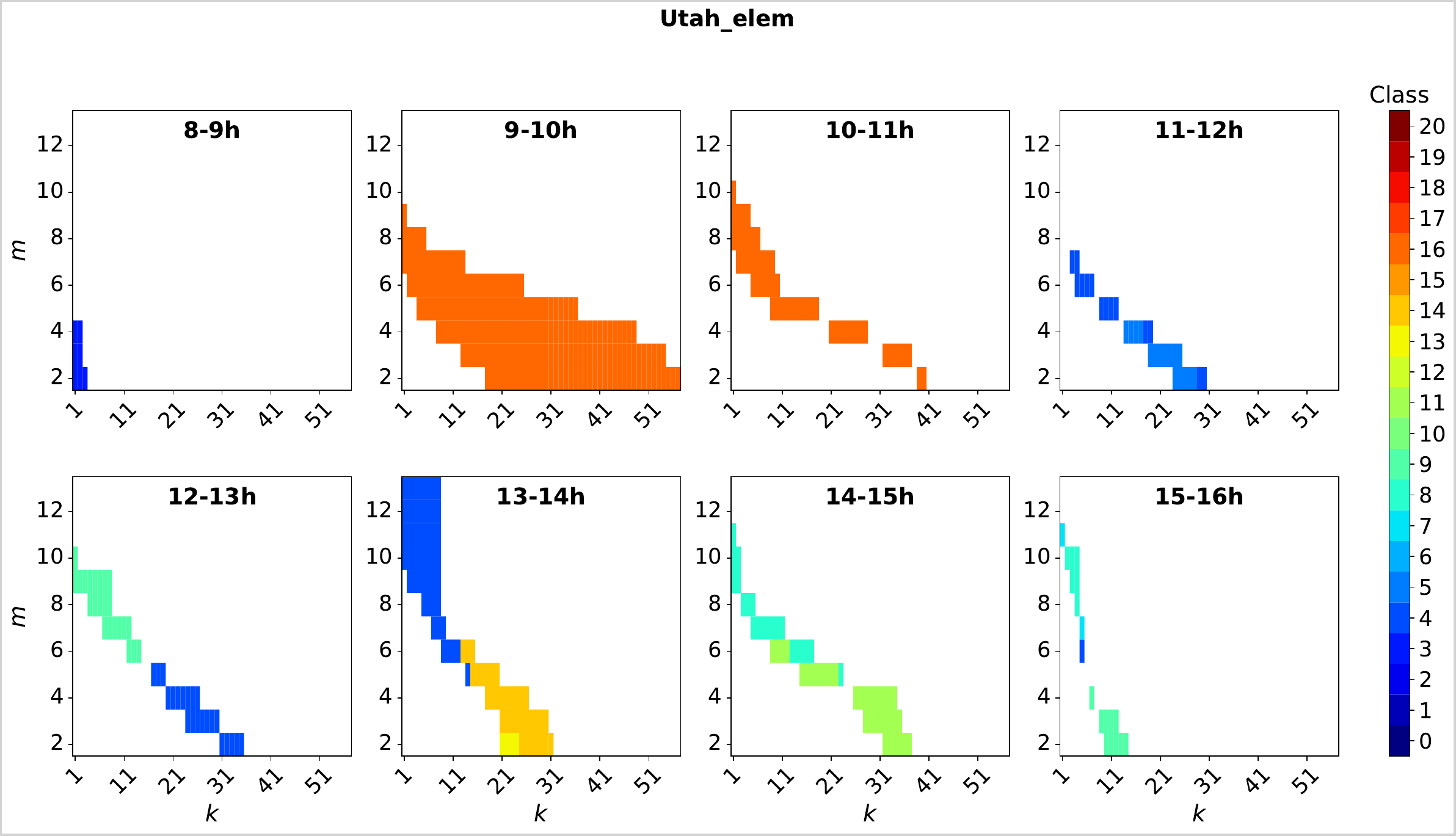}
    \caption{\textbf{Prevalent node role in hyper-cores - I.} For each time window we show the prevalent node role in each $(k,m)$-hyper-core, as the most frequent label in the core: we use a color code for identifying roles and we consider a role dominant only if its frequency is larger than 0.5. In white are indicated hyper-cores which are empty or where does not exist a dominant role. Here we consider the LyonSchool and Utah\_elem data sets.
    }
    \label{fig:figure20}
\end{figure*}

\begin{figure*}
    \centering
    \includegraphics[width=0.95\textwidth]{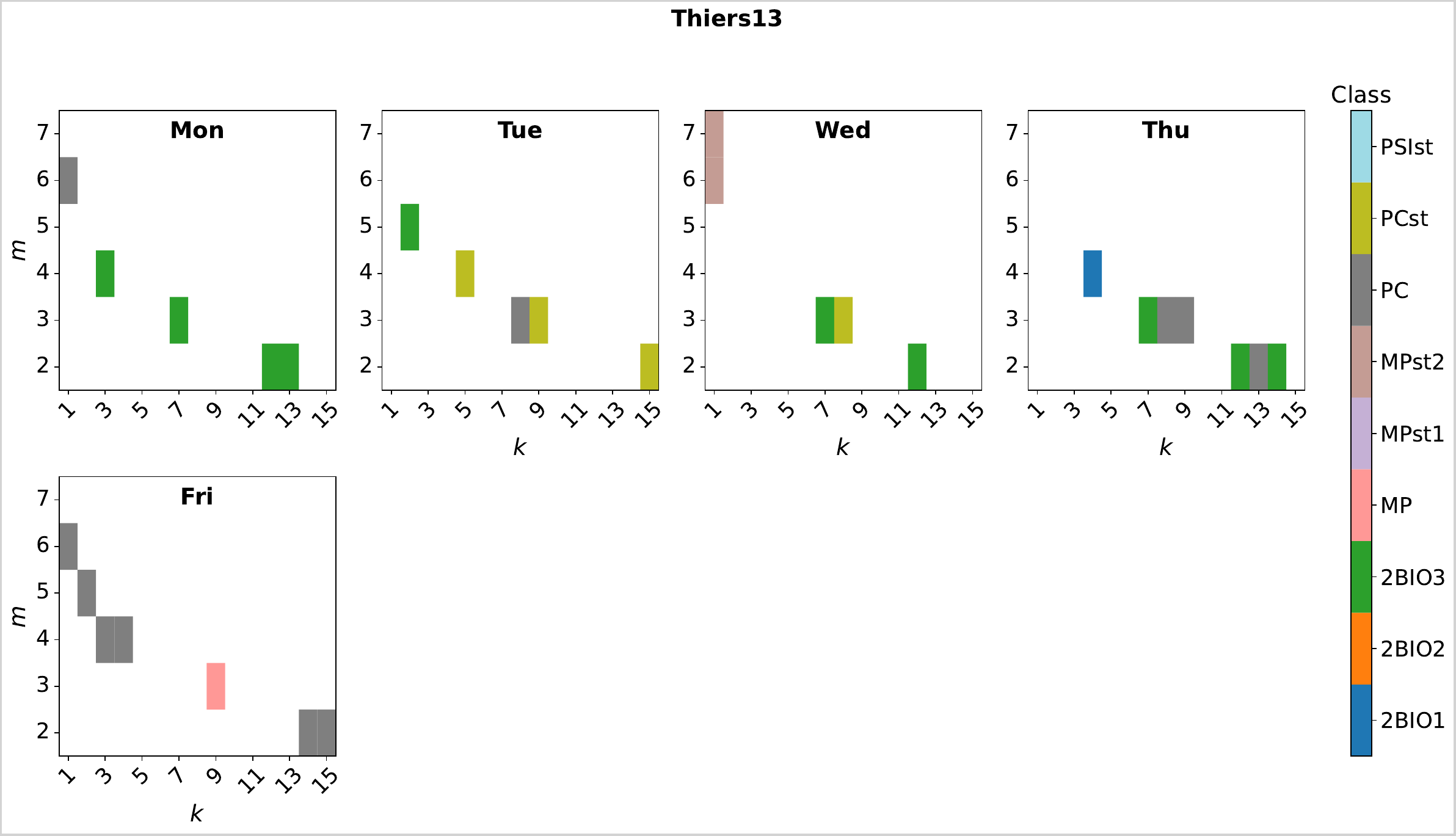} \\[0.5cm]
    \includegraphics[width=0.95\textwidth]{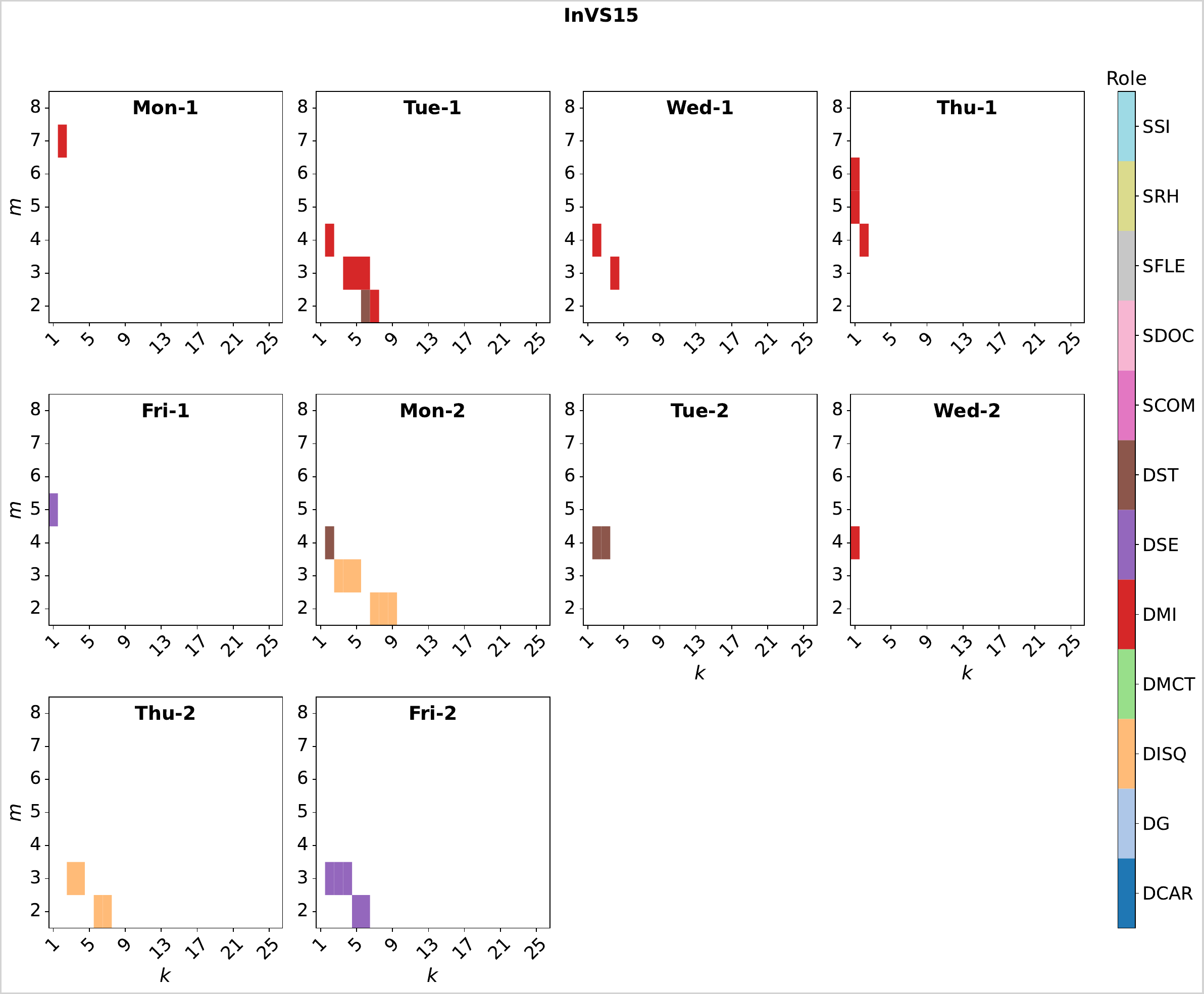}
    \caption{\textbf{Prevalent node role in hyper-cores - II.} As in Supplementary Fig. \ref{fig:figure20}, but here we consider the Thiers13 and InVS15 data sets.
    }
    \label{fig:figure21}
\end{figure*}

\begin{figure*}
    \centering
    \includegraphics[width=\textwidth]{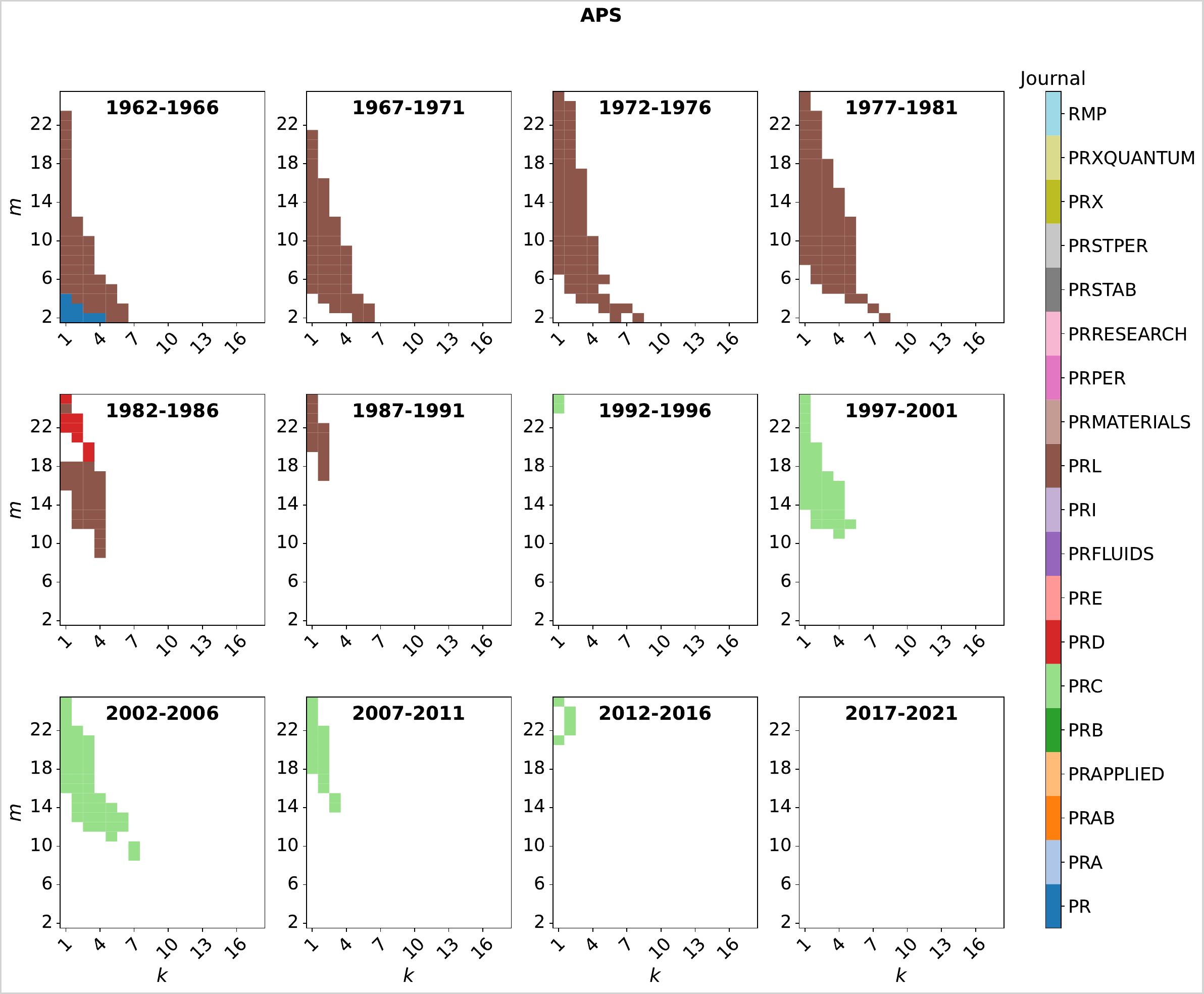}
    \caption{\textbf{Prevalent journal in hyper-cores of the reshuffled APS hypergraph.} As in Supplementary Fig. \ref{fig:figure20}, but here we consider a randomized version of the APS hypergraph, by generating 50 randomized realization in each time window (see Methods) and considering the average frequency of each label in each $(k,m)$-core. In white are indicated hyper-cores which are empty or where does not exist a dominant role.
    }
    \label{fig:figure22}
\end{figure*}

\begin{figure*}
    \centering
    \includegraphics[width=\textwidth]{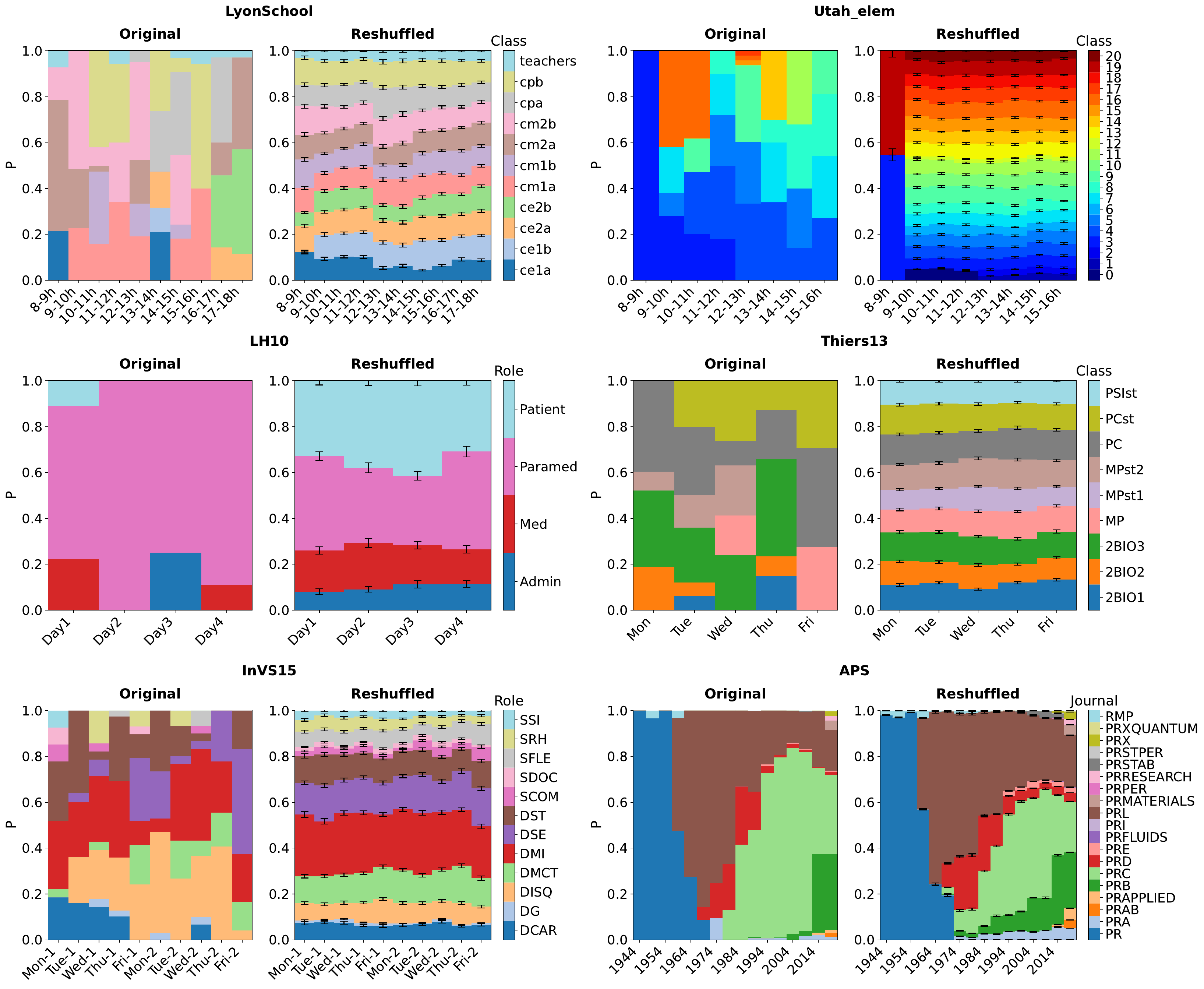}
    \caption{\textbf{Prevalent labels in relevant high-order structures.} For the LyonSchool, Utah\_elem, LH10, Thiers13 and InVS15 data sets we show the temporal evolution of the relative frequency $P$ of the various node labels within the top 15\% positions of the instantaneous hypercoreness ranking, through a stacked bar chart. For the APS data set we show the temporal evolution of the relative frequency $P$ of the various hyperedge labels within the most central hyper-cores, i.e. $(k_{max}^m,m)$-cores $\forall m$, through a stacked bar chart. We consider both the original data set and a randomized version of the hypergraph, by generating 50 randomized realizations in each time window (see Methods) and considering the average frequency of each label, also reporting error bars which indicate the standard error of the mean.
    }
    \label{fig:figure23}
\end{figure*}

\clearpage
\newpage

\section{Time-varying hypergraph models validation}
\label{sez:VI}

\begin{figure*}[h]
    \centering
    \includegraphics[width=\textwidth]{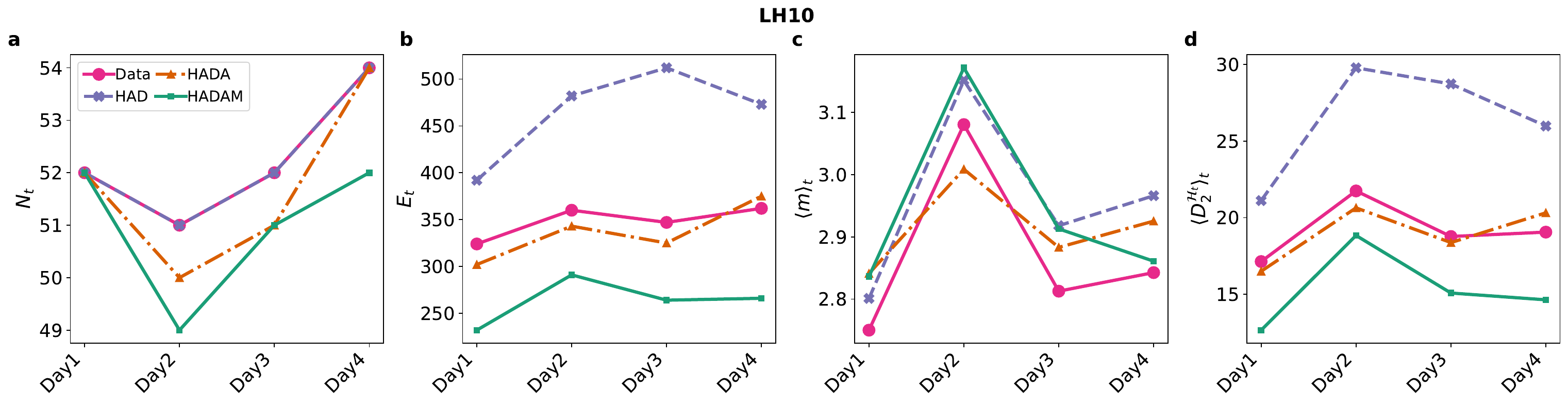}
    \caption{\textbf{Temporal hypergraphs models properties - I.} For the LH10 data set, we show the temporal evolution of the number of active nodes $N_t$, of the number of active hyperedges $E_t$, of the average hyperedge size $\langle m \rangle_t$ and of the average total hyperdegree $\langle D_2^{\mathcal{H}_t} \rangle_t$, calculated in each static snapshot $t$. We consider both the original data set and the corresponding HAD, HADA and HADAM hypergraphs models.
    }
    \label{fig:figure24}
\end{figure*}

\begin{figure*}[h]
    \centering
    \includegraphics[width=\textwidth]{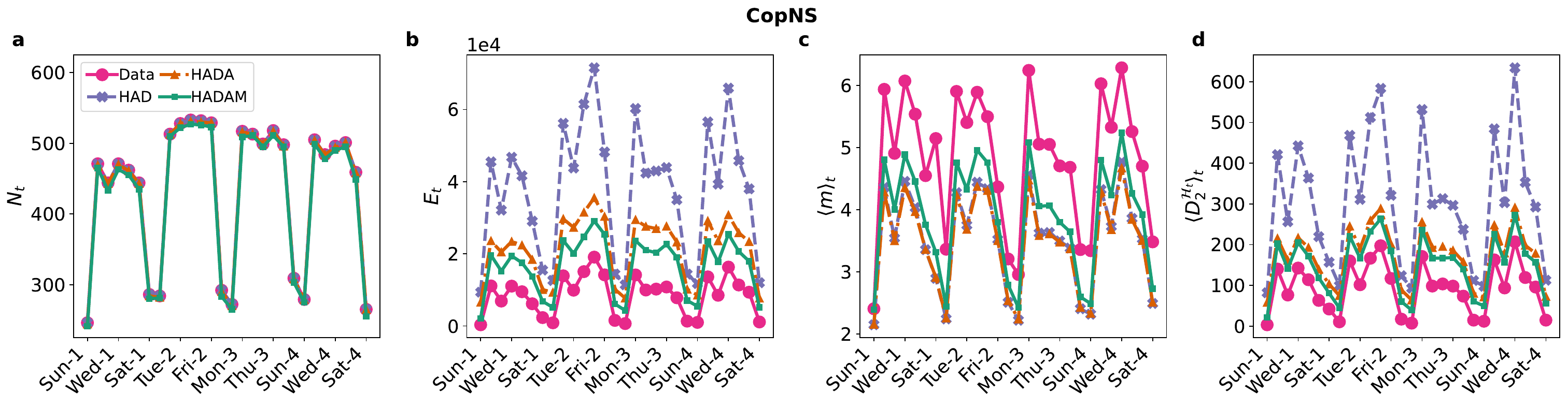}
    \caption{\textbf{Temporal hypergraphs models properties - II.} Same as Supplementary Fig. \ref{fig:figure24}, but here we consider the CopNS data set.
    }
    \label{fig:figure25}
\end{figure*}

\begin{figure*}
    \centering
    \includegraphics[width=\textwidth]{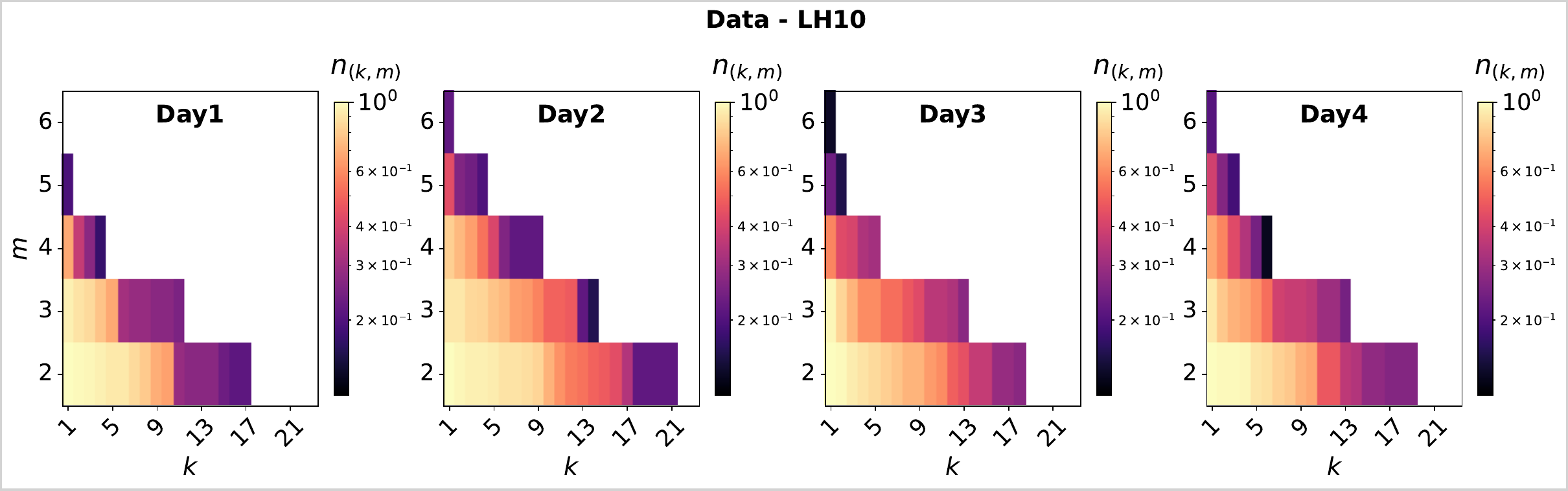} \\ [0.5cm]
    \includegraphics[width=\textwidth]{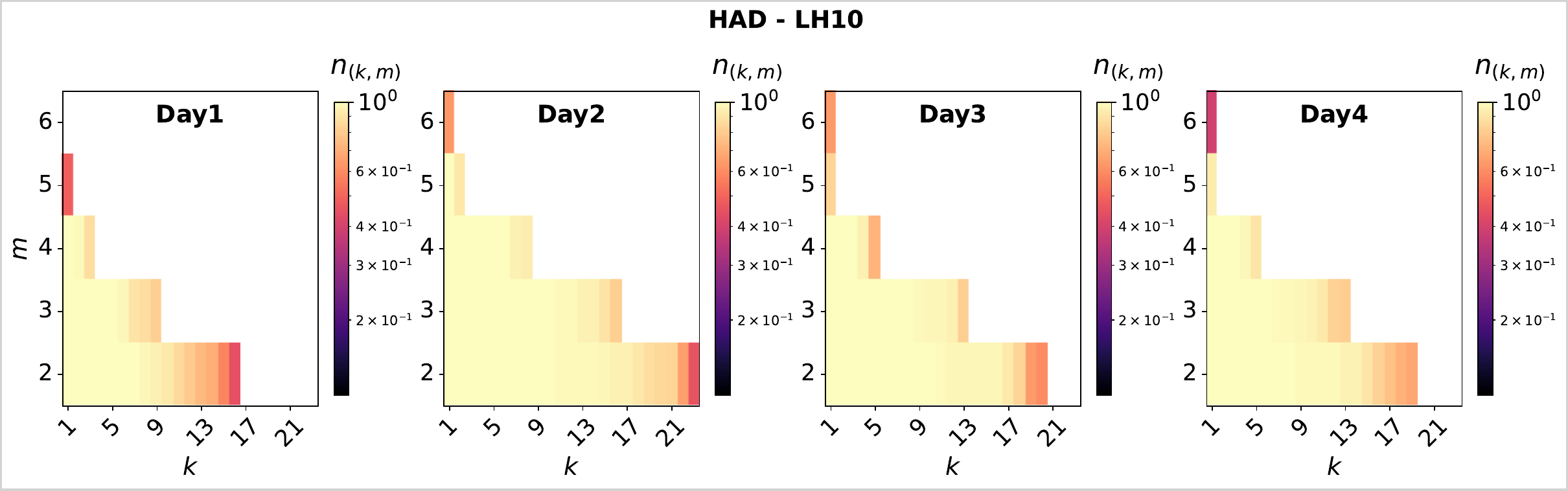} \\ [0.5cm]
    \includegraphics[width=\textwidth]{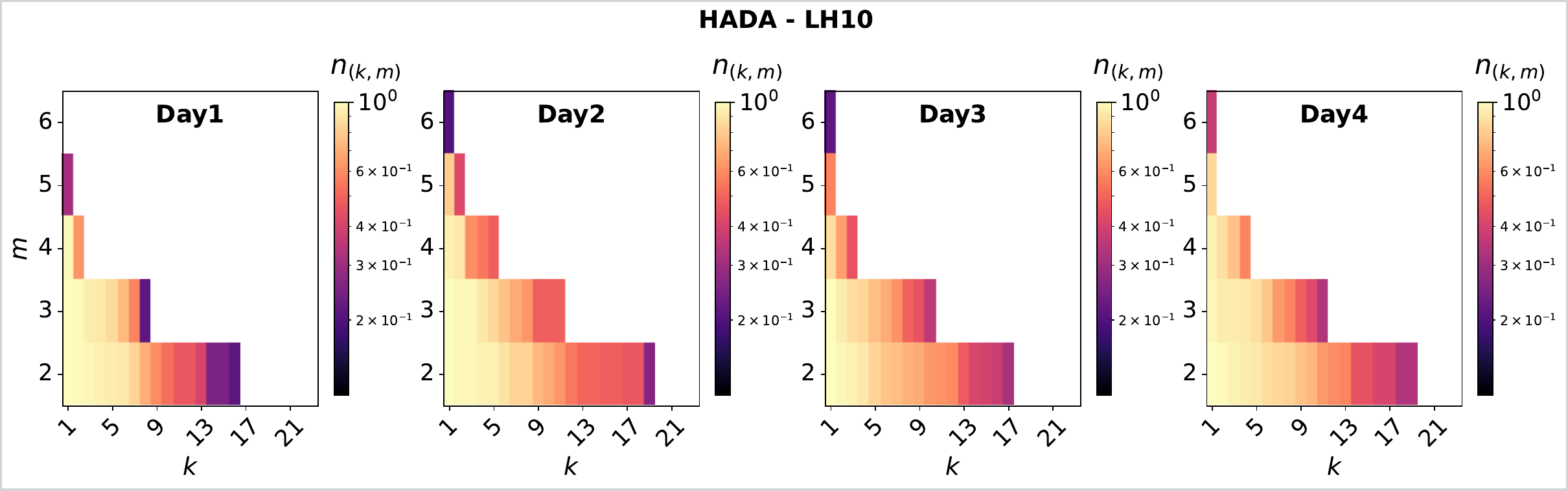} \\ [0.5cm]
    \includegraphics[width=\textwidth]{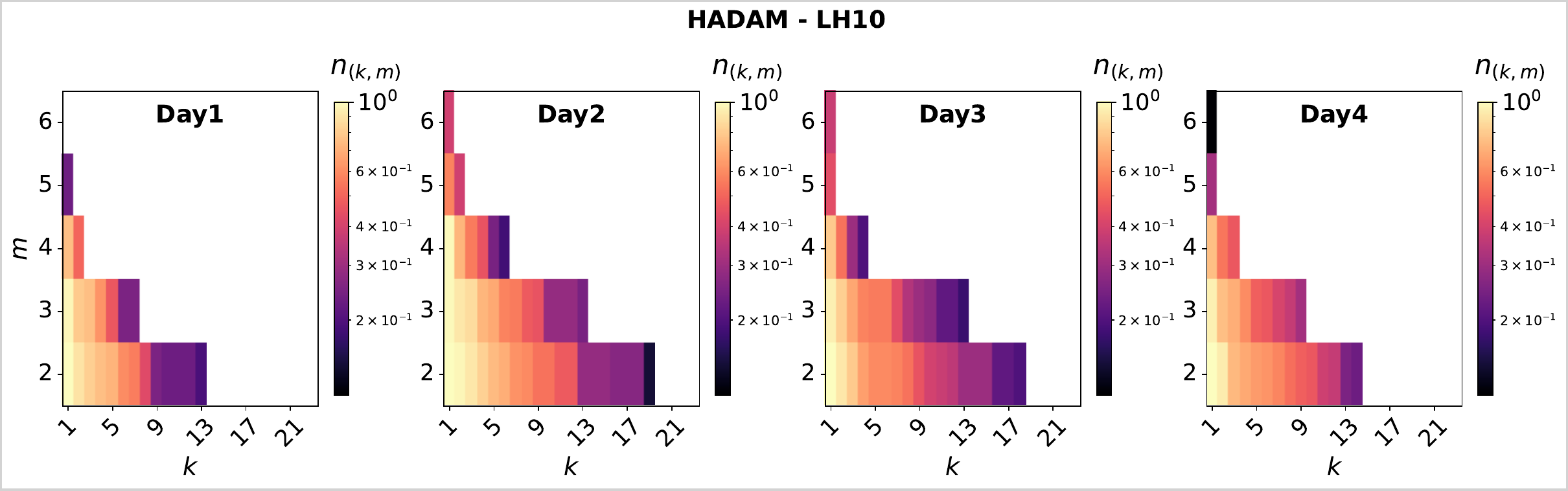} \\ [0.5cm]
    \caption{\textbf{Hyper-core structure evolution in synthetic hypergraphs - I.} We show the fraction of nodes $n_{(k,m)}$ in the $(k,m)$-core as a function of $k$ and $m$ for each time window. Each row correspond or to the original data set or to each of the considered hypergraph models (HAD, HADA and HADAM). Here we consider the LH10 data set.
    }
    \label{fig:figure26}
\end{figure*}

\begin{figure*}
    \centering
    \includegraphics[width=0.81\textwidth]{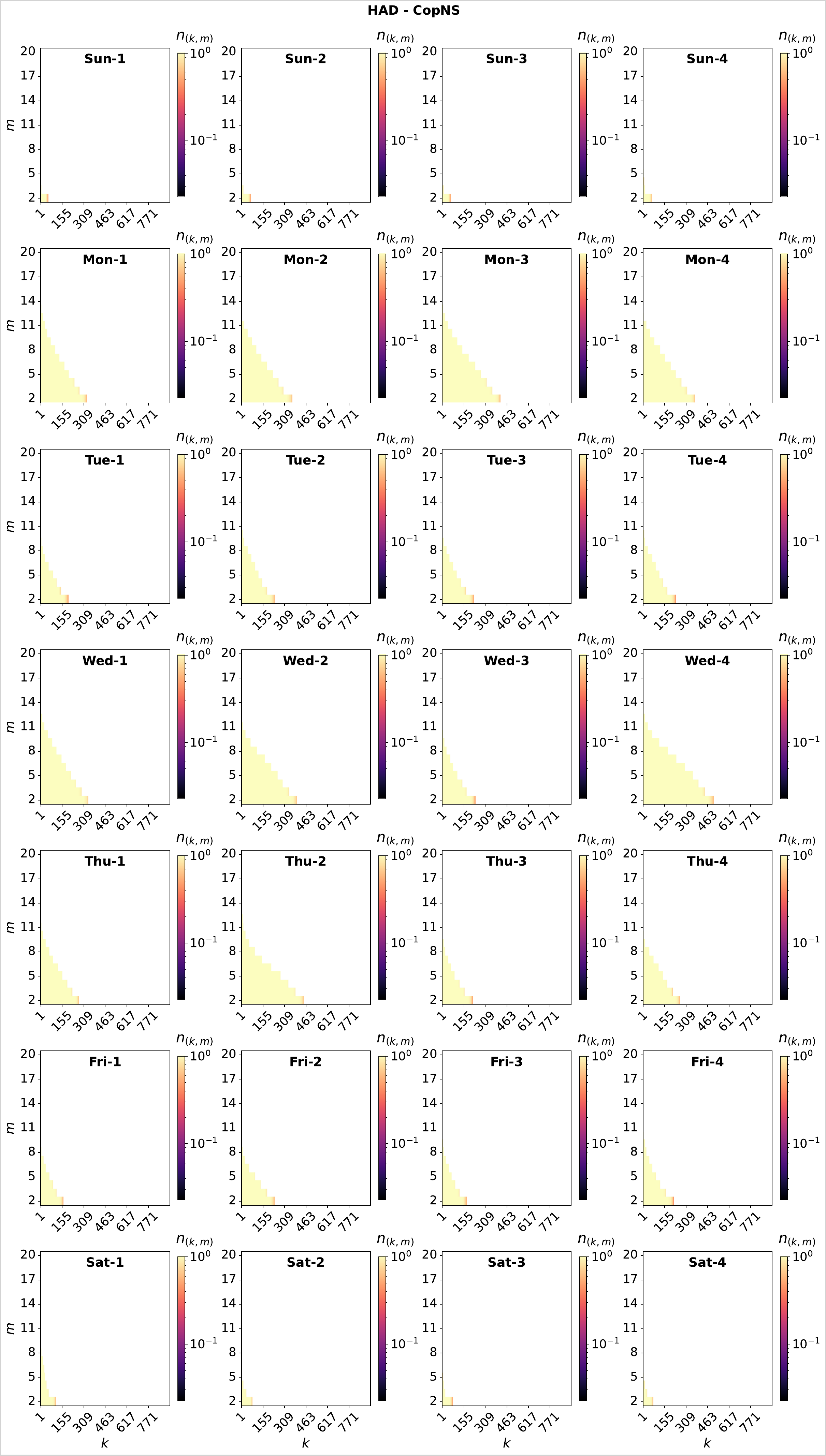}
    \caption{\textbf{Hyper-core structure evolution in synthetic hypergraphs - II.} Same as Supplementary Fig. \ref{fig:figure26}, but here we consider the CopNS data set and the HAD model.
    }
    \label{fig:figure27}
\end{figure*}

\begin{figure*}
    \centering
    \includegraphics[width=0.81\textwidth]{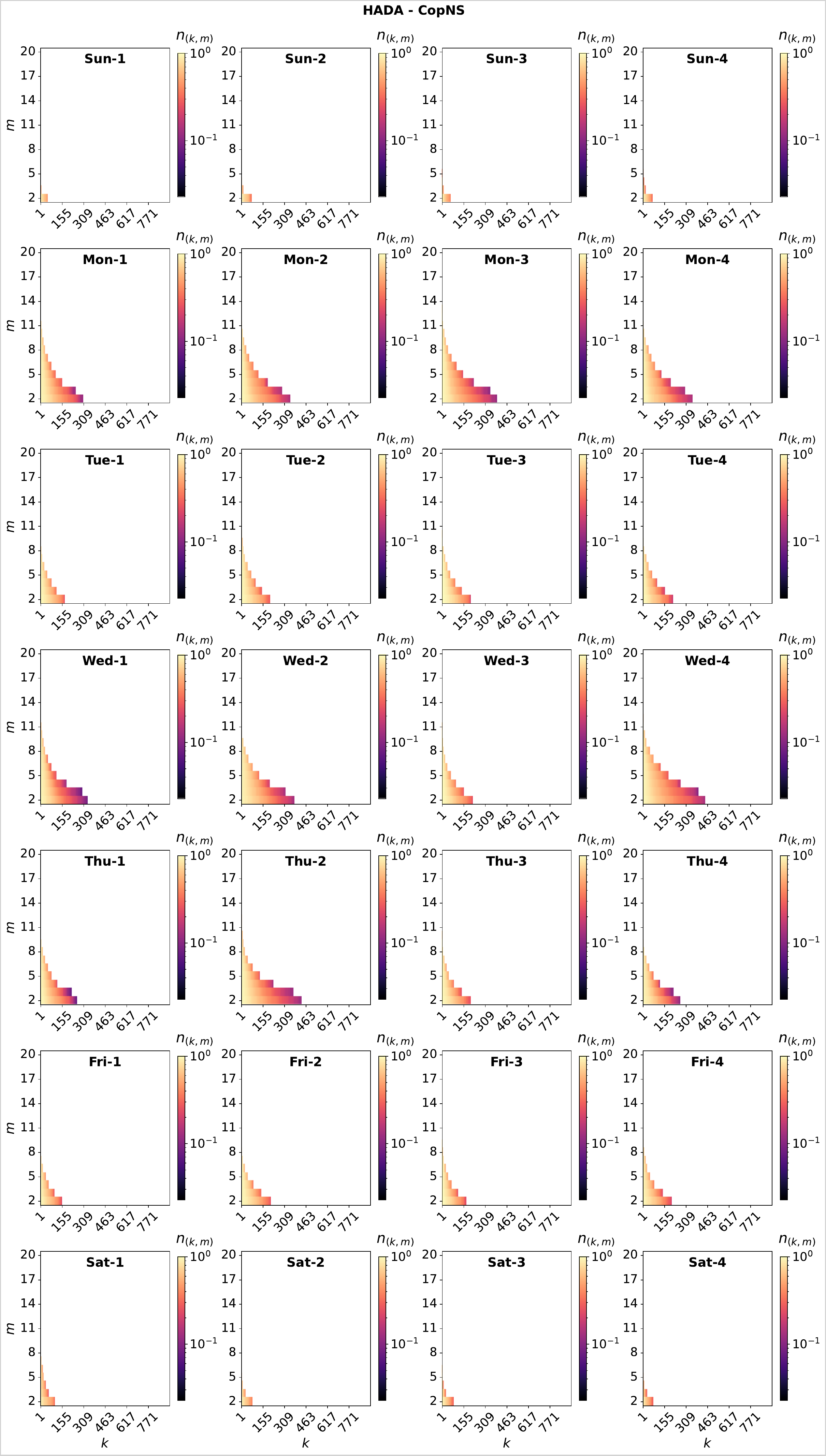}
    \caption{\textbf{Hyper-core structure evolution in synthetic hypergraphs - III.} Same as Supplementary Fig. \ref{fig:figure26}, but here we consider the CopNS data set and the HADA model.
    }
    \label{fig:figure28}
\end{figure*}

\begin{figure*}
    \centering
    \includegraphics[width=0.81\textwidth]{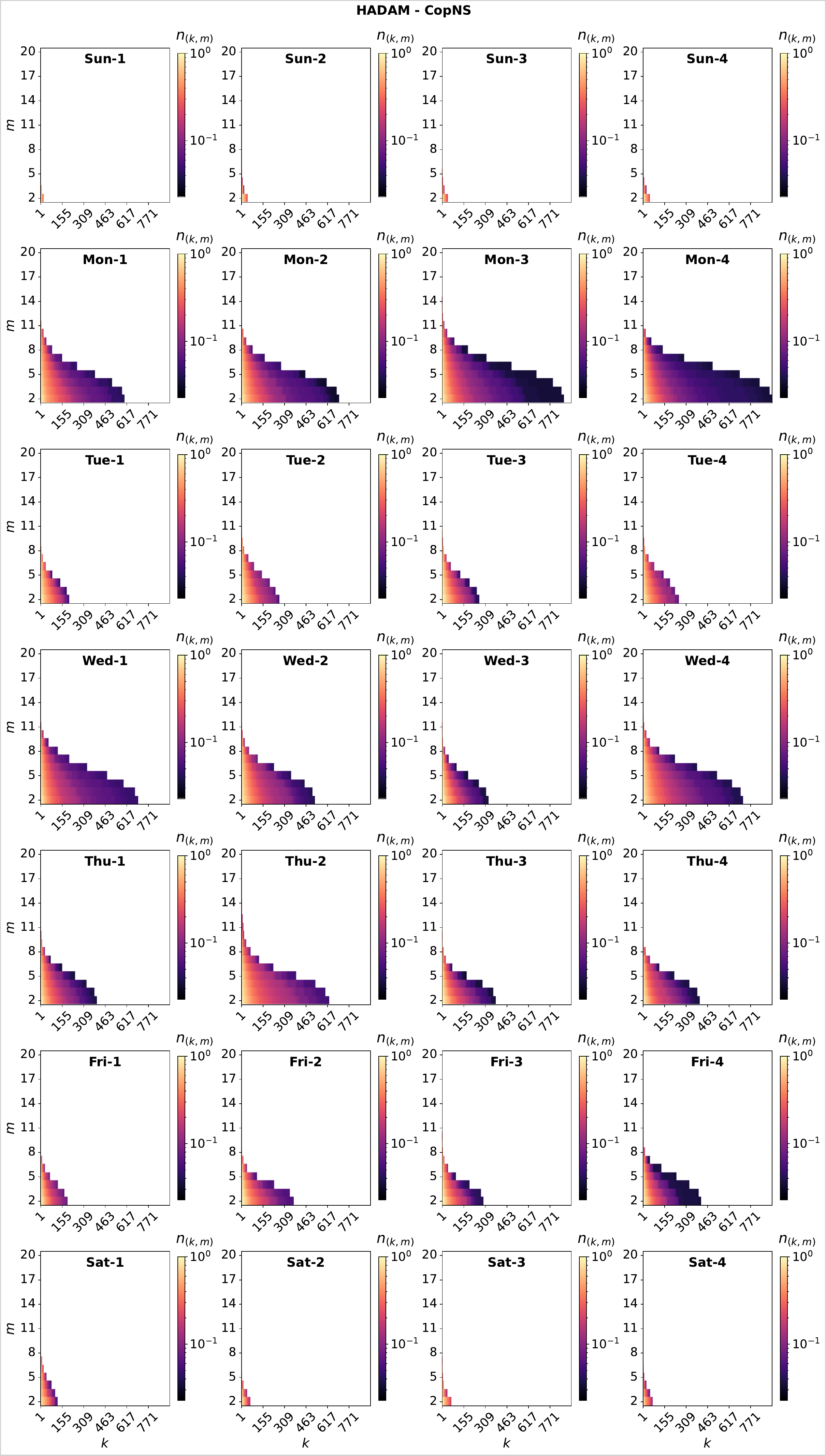}
    \caption{\textbf{Hyper-core structure evolution in synthetic hypergraphs - IV.} Same as Supplementary Fig. \ref{fig:figure26}, but here we consider the CopNS data set and the HADAM model.
    }
    \label{fig:figure29}
\end{figure*}

\begin{figure*}
    \centering
    \includegraphics[width=\textwidth]{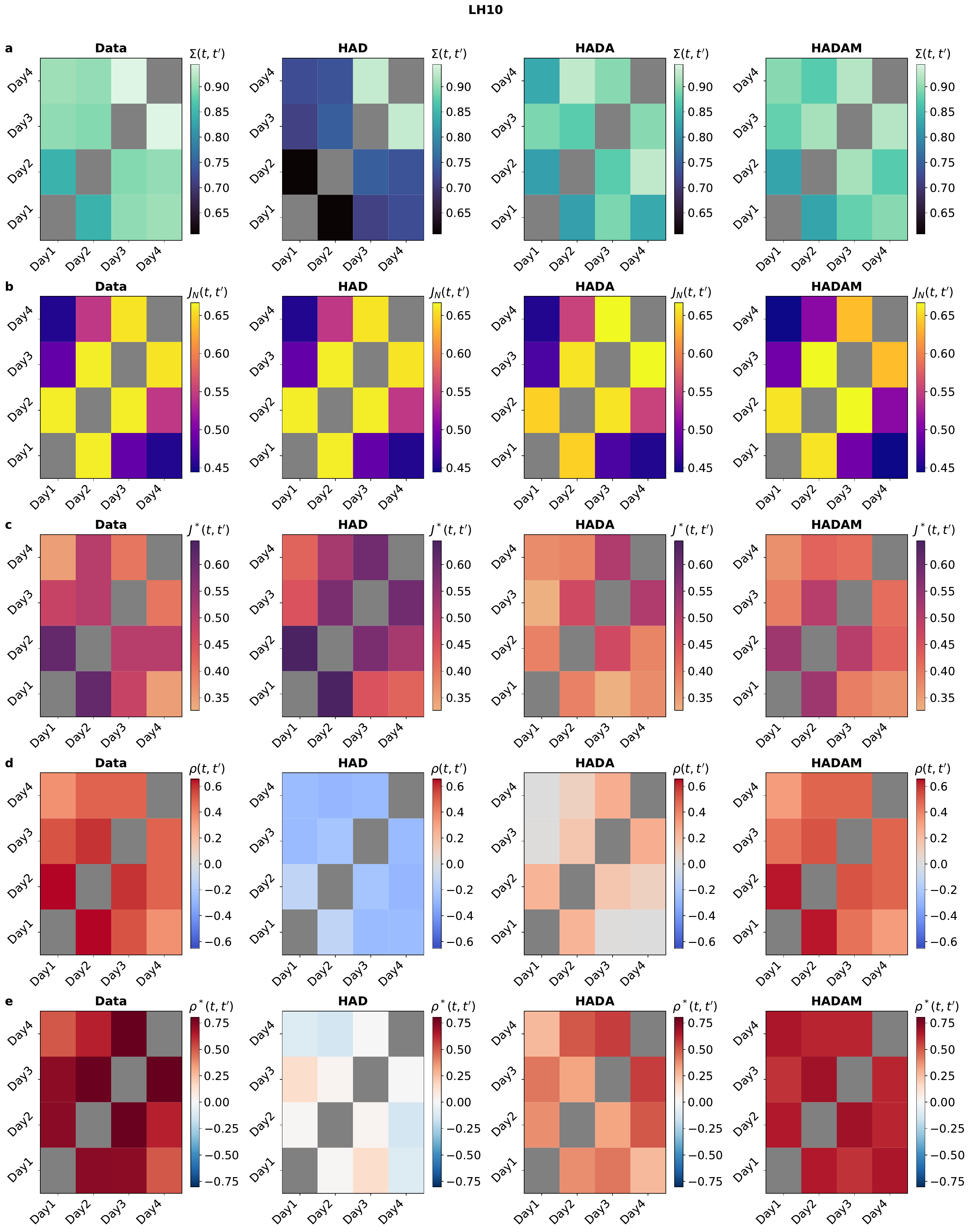}
    \caption{\textbf{Multi-scales stability of temporal hypergraphs models - I.} For the empirical hypergraph and for each of the considered models (HAD, HADA, HADAM), we consider two time windows $t$ and $t'$ and we show through a colormap: the root-mean-square deviation similarity $\Sigma(t,t')$ between the filling profiles $n_{(k,m)}(t)$ and $n_{(k,m)}(t')$ (panel \textbf{a}); the Jaccard similarity $J_N(t,t')$ between the sets of nodes composing the whole population at time $t$ and $t'$ (panel \textbf{b}); the Jaccard similarity $J^*(t,t')$ between the sets of nodes belonging to the most central hyper-cores, i.e. to the $(k_{max}^m,m)$-cores $\forall m$, at time $t$ and $t'$ (panel \textbf{c}); the Pearson correlation coefficient between the nodes hypercoreness rankings at time $t$ and $t'$, considering all the nodes that are active in at least one of the snapshots, $\rho(t,t')$ (panel \textbf{d}), and considering only the nodes that are active in both the snapshots, $\rho^*(t,t')$ (panel \textbf{e}). Here we consider the LH10 data set.
	}
    \label{fig:figure30}
\end{figure*}

\begin{figure*}
    \centering
    \includegraphics[width=\textwidth]{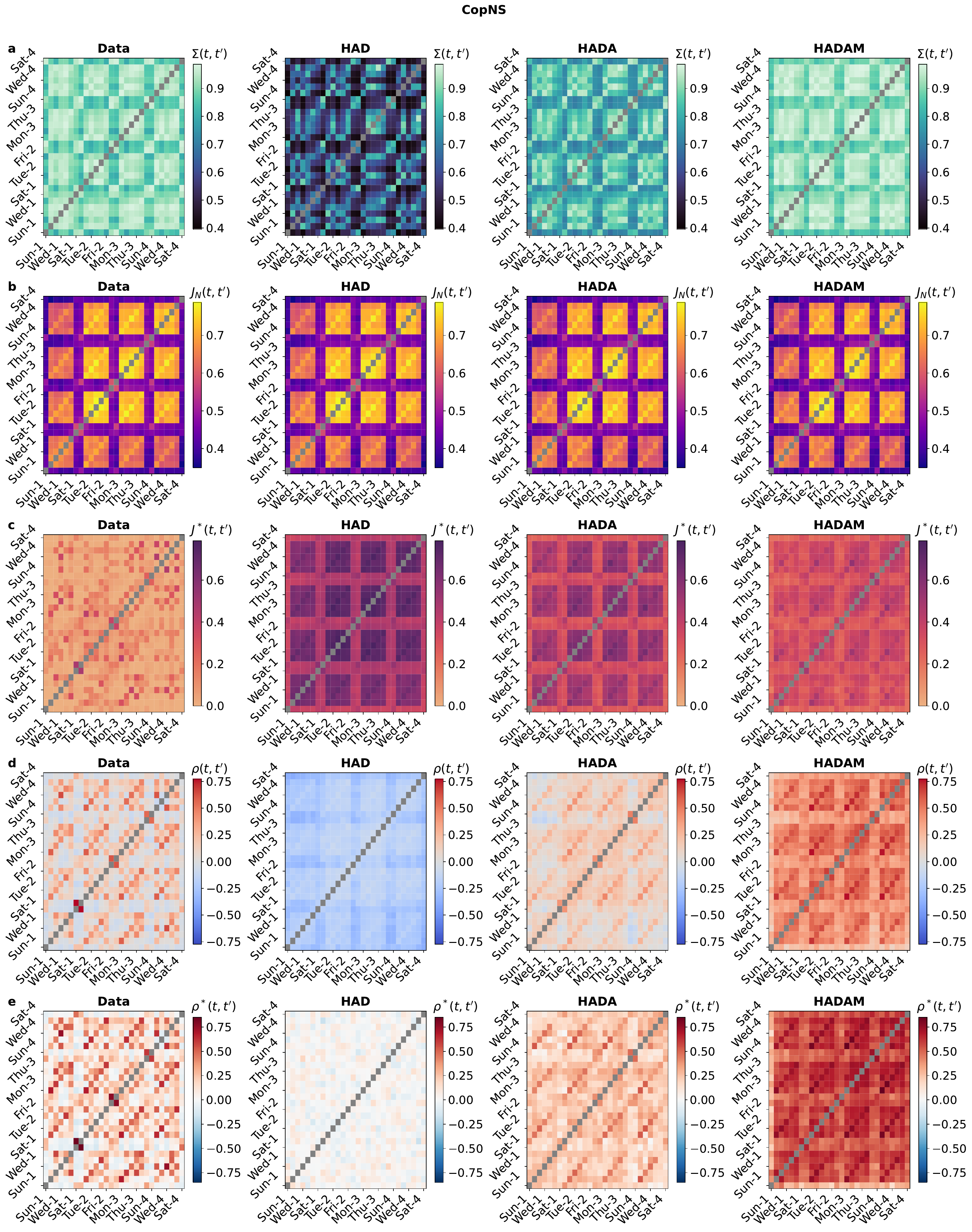}
    \caption{\textbf{Multi-scales stability of temporal hypergraphs models - II.} Same as Supplementary Fig. \ref{fig:figure30}, but here we consider the CopNS data set.
	}
    \label{fig:figure31}
\end{figure*}

\begin{figure*}
    \centering
    \includegraphics[width=\textwidth]{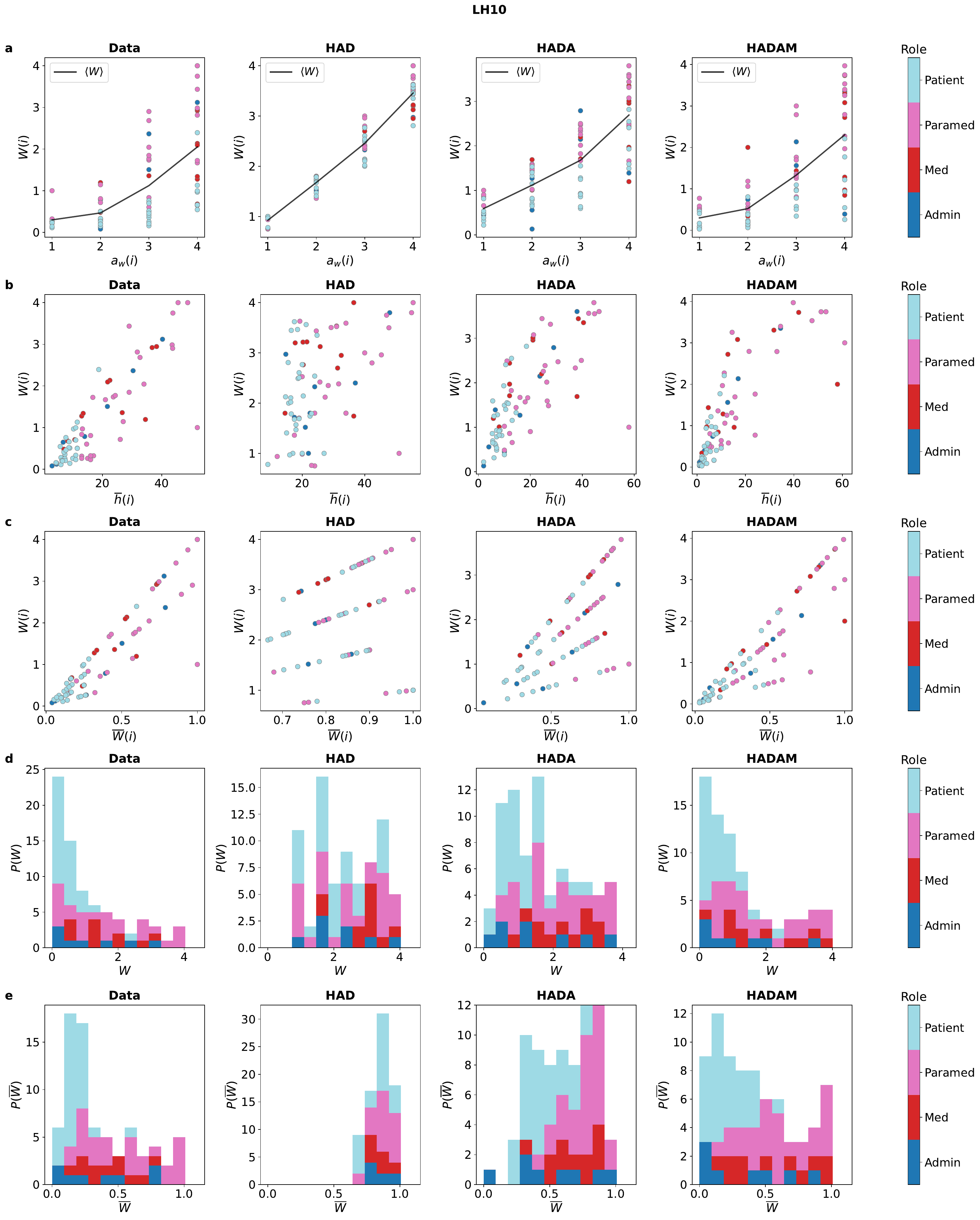}
    \caption{\textbf{Time-aggregated centrality measures in temporal hypergraphs models - I.} For the empirical hypergraph and for each of the considered models (HAD, HADA, HADAM), we show a scatter plot of the aggregated hypercoreness $W(i)$ for all nodes $i$ as a function of their: snapshot activity $a_w(i)$ (panel \textbf{a}); average number of interactions when active $\overline{h}(i)$ (panel \textbf{b}); activity-averaged hypercoreness $\overline{W}(i)$ (panel \textbf{c}). In panel \textbf{a} we also show the average aggregated hypercoreness $\langle W \rangle$ as a function of $a_w$. In panels \textbf{d},\textbf{e} we show through histograms respectively the number of nodes $P(W)$ with aggregated hypercoreness $W$ and the number of nodes $P(\overline{W})$ with activity-averaged hypercoreness $\overline{W}$. Here we consider the LH10 data set: in panels \textbf{a}-\textbf{c} the points are colored according to the node social role; in panels \textbf{d},\textbf{e} within each bar we distinguish the relative frequency of nodes belonging to each social role, through stacked bars.
    }
    \label{fig:figure32}
\end{figure*}

\begin{figure*}
    \centering
    \includegraphics[width=\textwidth]{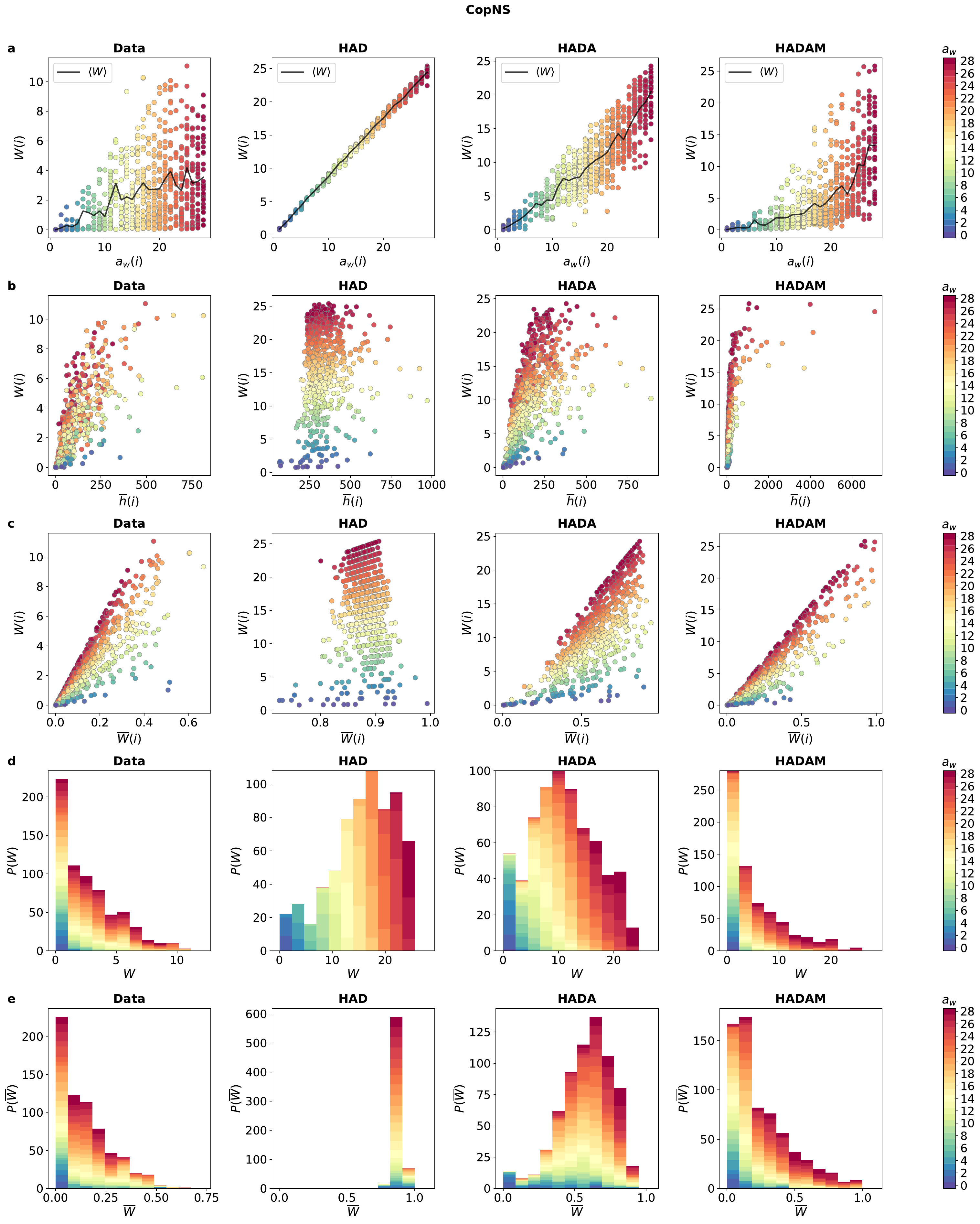}
    \caption{\textbf{Time-aggregated centrality measures in temporal hypergraphs models - II.} Same as Supplementary Fig. \ref{fig:figure32}, but here we consider the CopNS data set: in panels \textbf{a}-\textbf{c} the points are colored according to the node activity snapshot $a_w$; in panels \textbf{d},\textbf{e} within each bar we distinguish the relative frequency of nodes belonging to each $a_w$ class, through stacked bars.
    }
    \label{fig:figure33}
\end{figure*}

\begin{figure*}
    \centering
    \includegraphics[width=\textwidth]{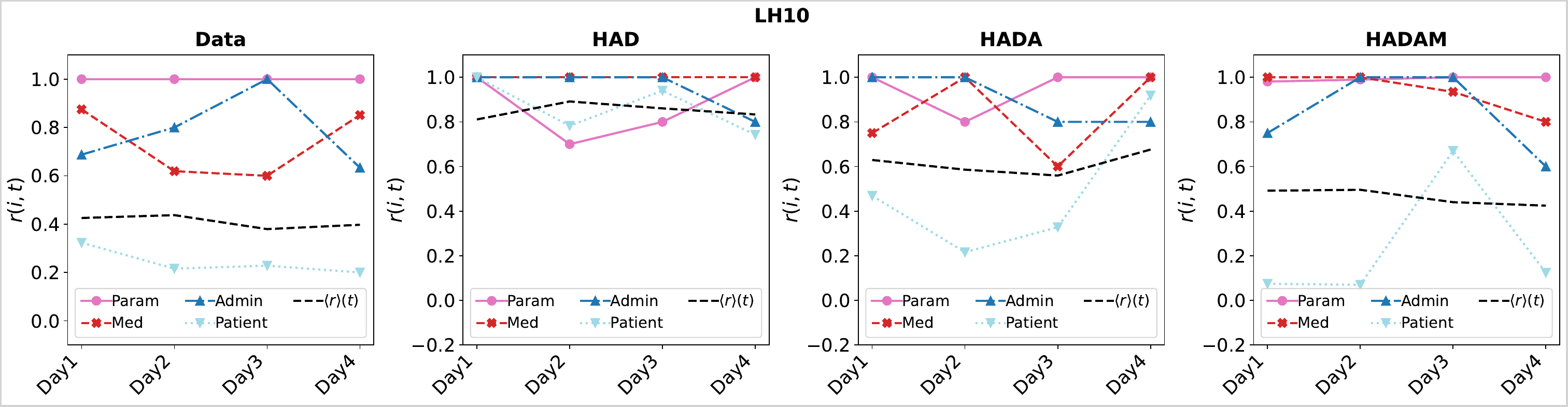} \\ [0.5cm]
    \includegraphics[width=\textwidth]{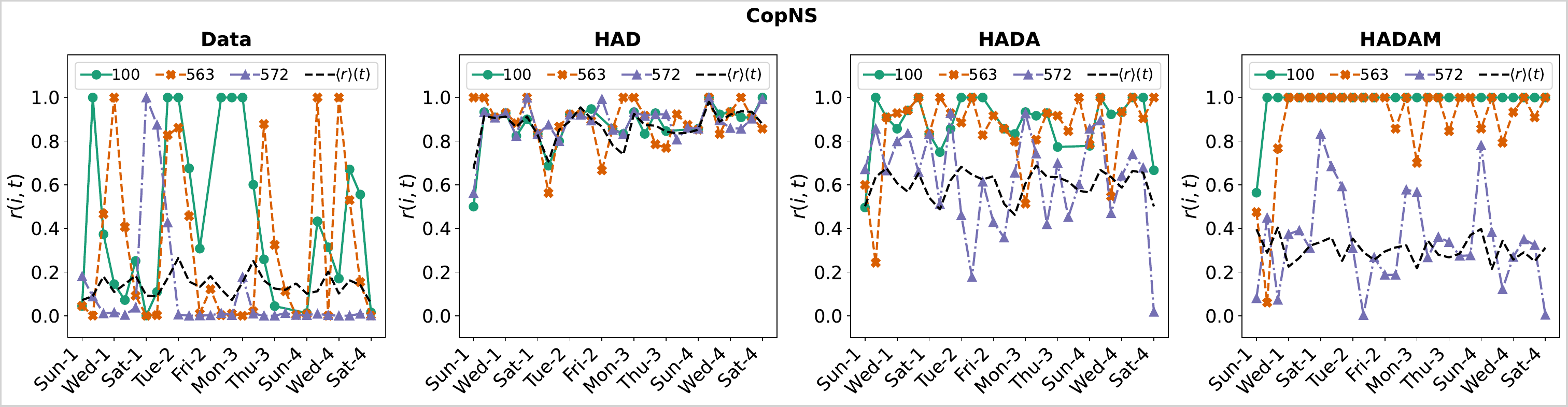}
    \caption{\textbf{Hypercoreness evolution in synthetic hypergraphs.} For each data set, we show the temporal evolution of the hypercoreness $r(i,t)$ for some nodes and we plot the mean $\langle r \rangle (t)$ value, by averaging only on active nodes. We consider the same nodes in the empirical hypergraph and in the corresponding HAD, HADA and HADAM models. In the legend of each panel we indicate the node \texttt{id} or the corresponding label. We consider: in LH10 the nodes \texttt{1210} (Param), \texttt{1144} (Med), \texttt{1098} (Admin), \texttt{1383} (Patient); in CopNS the nodes \texttt{100}, \texttt{563}, \texttt{572}.
    }
    \label{fig:figure34}
\end{figure*}

\begin{figure*}
    \centering
    \includegraphics[width=\textwidth]{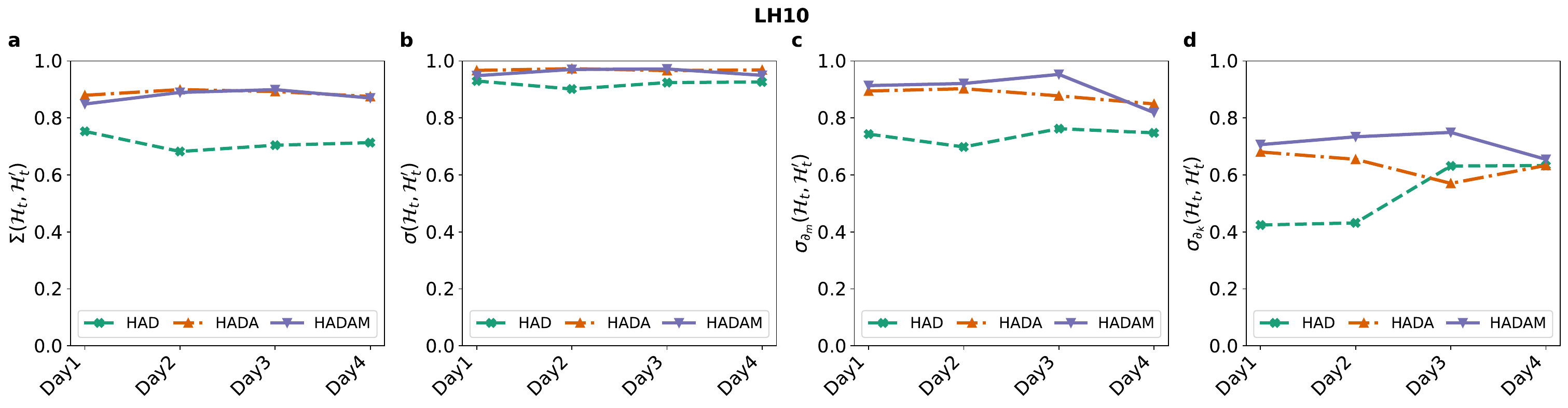}
    \caption{\textbf{Similarity between empirical hypergraph and models - I.} For each time window $t$ we consider the empirical hypergraph $\mathcal{H}_t$ and each of the corresponding synthetic hypergraphs $\mathcal{H}_t'$. We show the evolution of: the root-mean-square deviation similarity $\Sigma(\mathcal{H}_t,\mathcal{H}_t')$ (panel \textbf{a}), the cosine similarities $\sigma(\mathcal{H}_t,\mathcal{H}_t')$ (panel \textbf{b}), $\sigma_{\partial_m}(\mathcal{H}_t,\mathcal{H}_t')$ (panel \textbf{c}) and $\sigma_{\partial_k}(\mathcal{H}_t,\mathcal{H}_t')$ (panel \textbf{d}). Here we consider the LH10 data set.
    }
    \label{fig:figure35}
\end{figure*}

\begin{figure*}
    \centering
    \includegraphics[width=\textwidth]{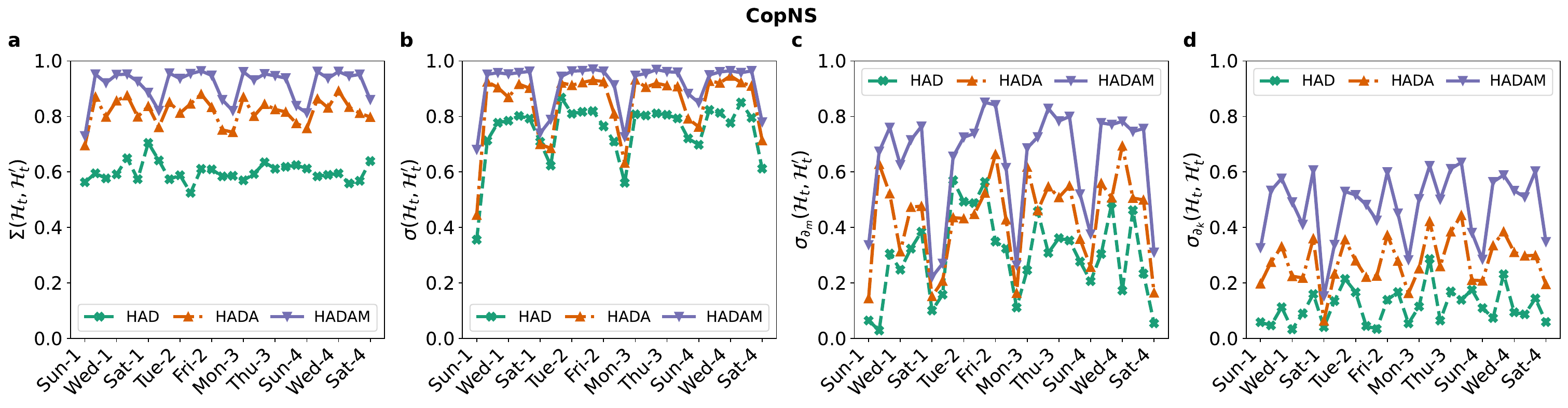}
    \caption{\textbf{Similarity between empirical hypergraph and models - II.} Same as Supplementary Fig. \ref{fig:figure35}, but here we consider the CopNS data set.
    }
    \label{fig:figure36}
\end{figure*}

In this Section we show the results of the proposed multi-scales approach applied to different temporal hyperegraphs models. We consider the LH10 and CopNS datasets, and the corresponding HAD, HADA and HADAM models (see Methods). We compare the original hypergaph and all the models considering: their main fundamental properties (Supplementary Figs. \ref{fig:figure24}, \ref{fig:figure25}); the hyper-cores filling profiles (Supplementary Figs. \ref{fig:figure26}-\ref{fig:figure29}); the multi-scales stability in the structural evolution (Supplementary Figs. \ref{fig:figure30}, \ref{fig:figure31}); the time-aggregated hypercoreness measures, their correlations and distributions (Supplementary Figs. \ref{fig:figure32}, \ref{fig:figure33}); the microscopic behavior of some specific nodes (Supplementary Figs. \ref{fig:figure34}). Finally, Supplementary Figs. \ref{fig:figure35}, \ref{fig:figure36} compare the filling profile $n_{(k,m)}(t)$ of the $(k,m)$-cores of the empirical hypergraph $\mathcal{H}_t$ with the filling profiles $n_{(k,m)}'(t)$ of the different models considered $\mathcal{H}_t'$ in each temporal snapshot $t$, using different similarity measures. We evaluate the root-mean-square deviation similarity $\Sigma(\mathcal{H}_t,\mathcal{H}_t')$, defined as in Eq. (2) of the main text considering $\mathcal{A}=\mathcal{H}_t$ and $\mathcal{B}=\mathcal{H}_t'$. We also consider the cosine similarity $\sigma(\mathcal{A},\mathcal{B})$ between the matrices $a_{(k,m)}$, $b_{(k,m)}$ associated to the general hypergraphs $\mathcal{A}$ and $\mathcal{B}$ \cite{fournet2014}:
\begin{equation}
{\small \sigma(\mathcal{A},\mathcal{B})=\frac{\sum\limits_{k=1}^{\overline{K}} \sum\limits_{m=2}^{\overline{M}} a_{(k,m)} b_{(k,m)}}{\sqrt{\sum\limits_{k=1}^{\overline{K}} \sum\limits_{m=2}^{\overline{M}} a_{(k,m)}^2} \sqrt{\sum\limits_{k=1}^{\overline{K}} \sum\limits_{m=2}^{\overline{M}} b_{(k,m)}^2}},}
\end{equation}
where $\sigma(\mathcal{A},\mathcal{B}) \in [0,1]$, $\overline{K}=\max\limits_m\{\max\{k_{max}^m(\mathcal{A}),k_{max}^m(\mathcal{B})\}\}$, $\overline{M}=\max\{M_{\mathcal{A}},M_{\mathcal{B}}\}$. We consider three versions of the cosine similarity: $\sigma(\mathcal{H}_t,\mathcal{H}_t')$, between $n_{(k,m)}(t)$, $n_{(k,m)}'(t)$, which estimates whether there is a proportionality in the fillings of the $(k,m)$-cores in the empirical hypergraph and in the models; $\sigma_{\partial_k}(\mathcal{H}_t,\mathcal{H}_t')$ between the partial derivative matrices $\partial_k n_{(k,m)}(t)$, $\partial_k n_{(k,m)}'(t)$, and $\sigma_{\partial_m}(\mathcal{H}_t,\mathcal{H}_t')$ between the partial derivative matrices $\partial_m n_{(k,m)}(t)$, $\partial_m n_{(k,m)}'(t)$, which estimate the similarity in the rate with which the filling of the hyper-cores changes when respectively $k$ and $m$ vary.

\clearpage
\newpage

\section{Scientific collaborations within specific APS journals}
\label{sez:VII}

\begin{figure*}[h]
    \centering
    \includegraphics[width=\textwidth]{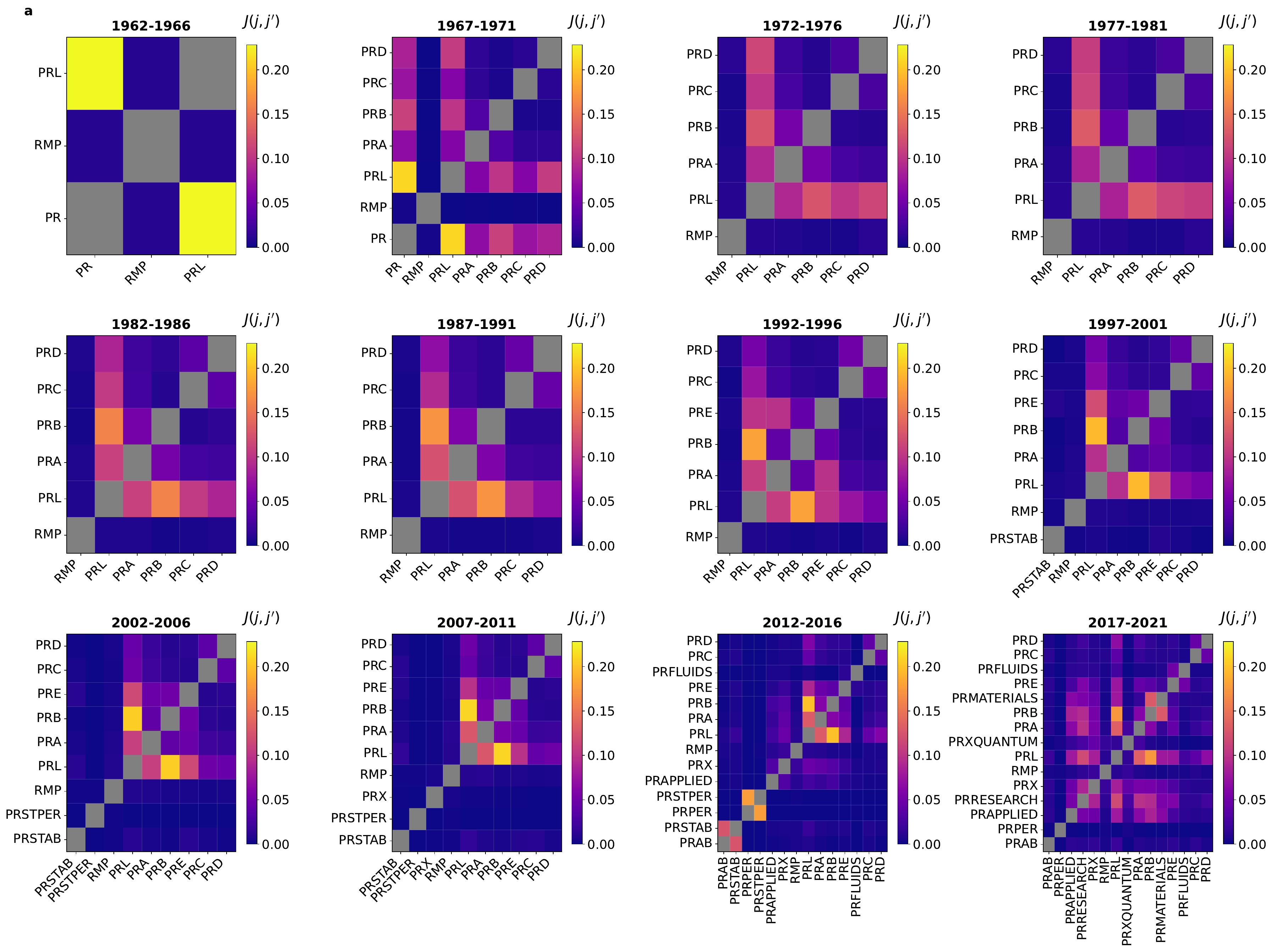} \\
    \includegraphics[width=\textwidth]{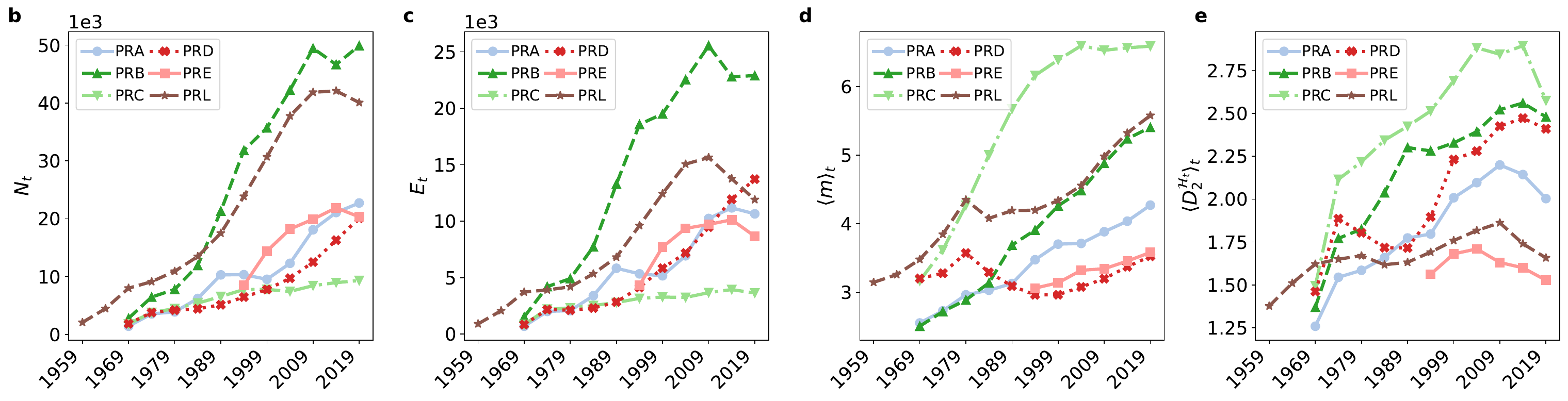}
    \caption{\textbf{APS journals properties.} In panel \textbf{a}, for each time window, we show the matrix $J(j,j')$, obtained by calculating the Jaccard similarity between the sets of authors who published in the journals $j$ and $j'$ during that period; in grey we indicate $J=1$. We build the temporal hypergraphs considering only the papers published in specific journals separately, for each of them we show the temporal evolution of: the number of active nodes $N_t$ (panel \textbf{b}), the number of active hyperedges $E_t$ (panel \textbf{c}), the average size of interactions $\langle m \rangle_t$ (panel \textbf{d}) and the average number of interactions of arbitrary size in which each node is involved $\langle D_2^{\mathcal{H}_t} \rangle_t$ (panel \textbf{e}).
    }
    \label{fig:figure37}
\end{figure*}

\begin{figure*}
    \centering
    \includegraphics[width=\textwidth]{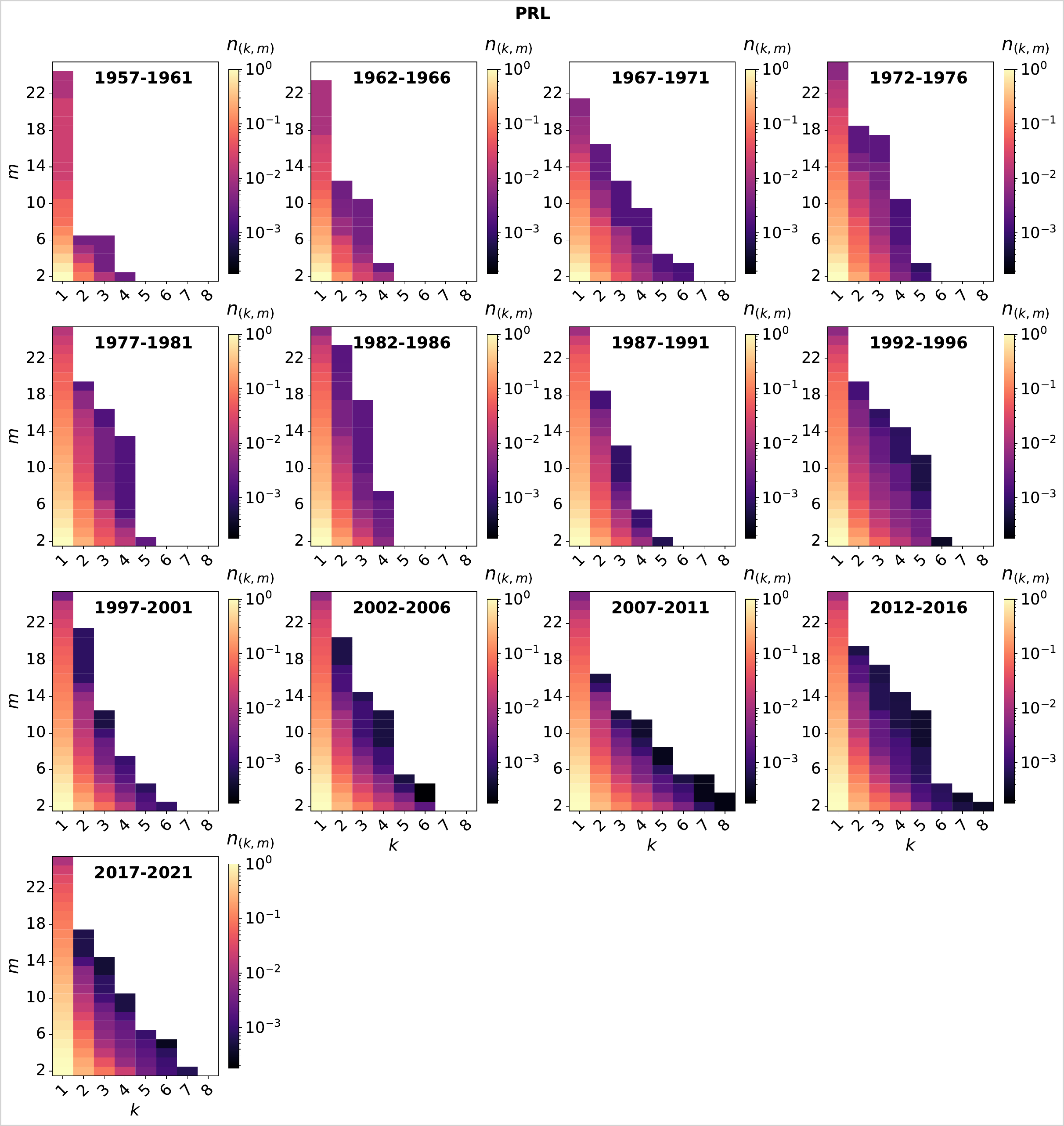}
    \caption{\textbf{Hyper-core structure evolution for specific APS journals - I.} We build the temporal hypergraph of scientific collaborations from the APS data set by considering only the papers published in a specific journal, then we consider the snapshot representation with 5-years resolution. We show the fraction of nodes $n_{(k,m)}$ in the $(k,m)$-core as a function of $k$ and $m$ for each time window. Here we consider the PRL journal.
    }
    \label{fig:figure38}
\end{figure*}

\begin{figure*}
    \centering
    \includegraphics[width=\textwidth]{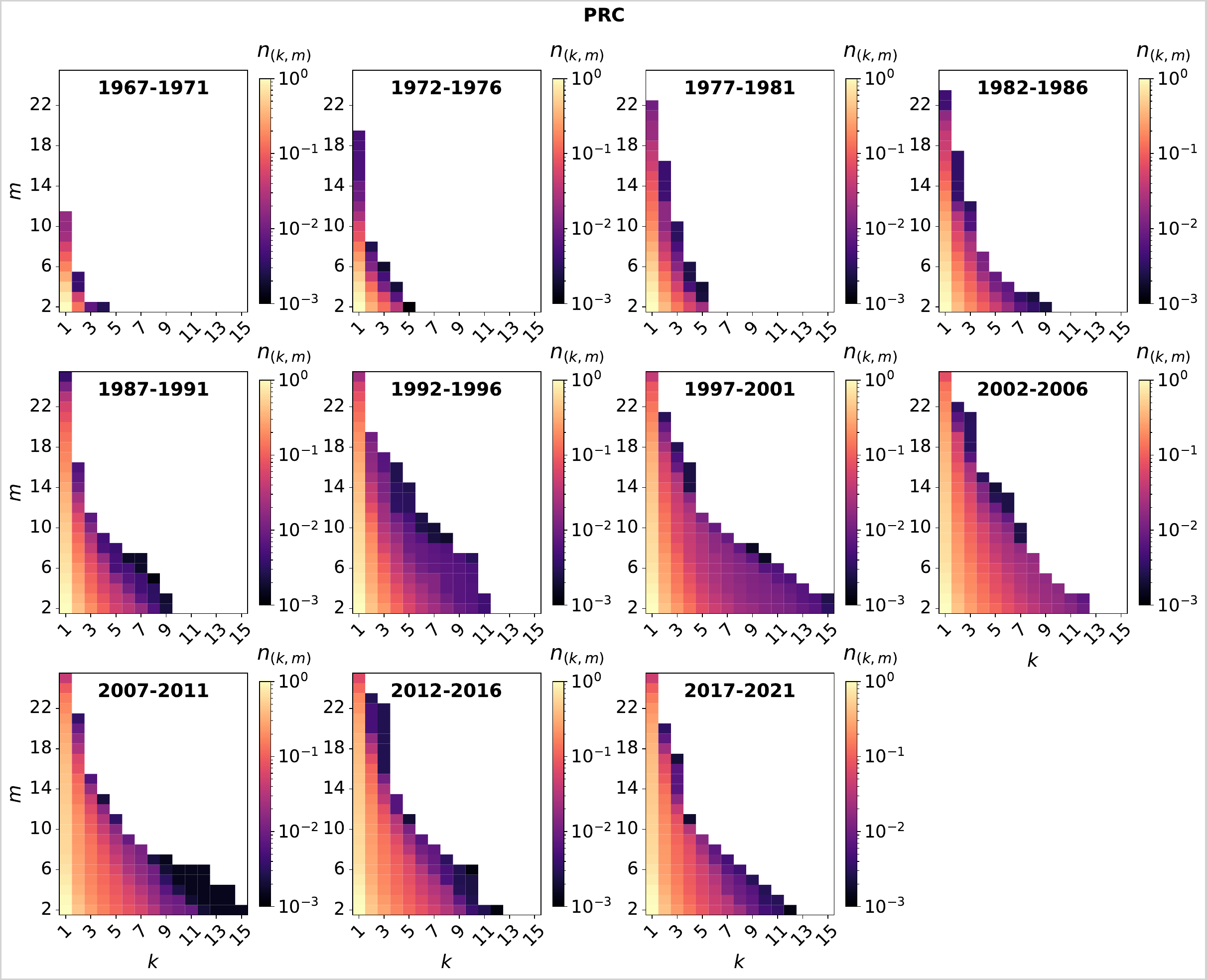} \\ [0.5cm]
    \includegraphics[width=\textwidth]{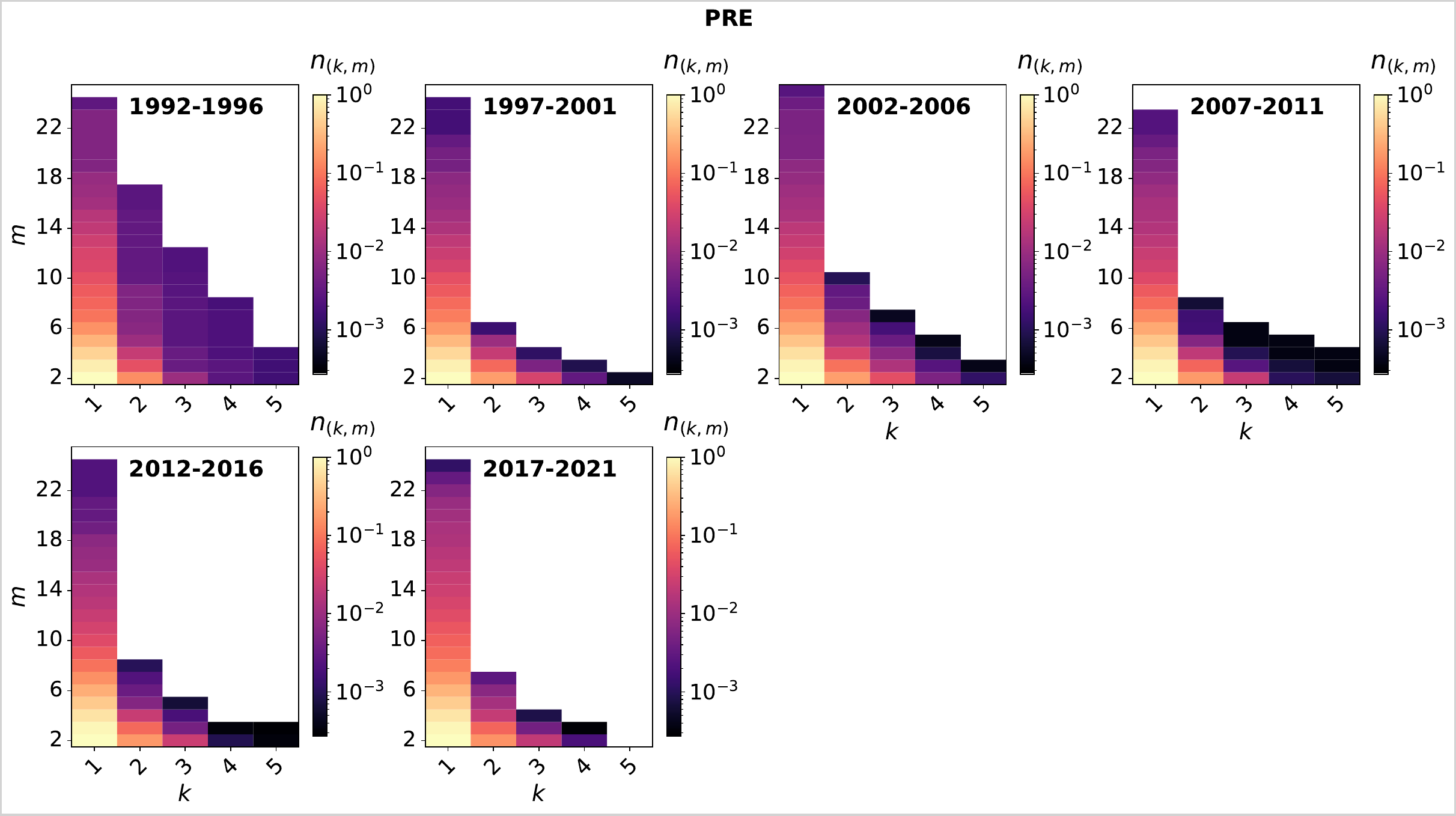}
    \caption{\textbf{Hyper-core structure evolution for specific APS journals - II.} Same as Supplementary Fig. \ref{fig:figure38}, but here we consider the PRC and PRE journals.
    }
    \label{fig:figure39}
\end{figure*}

\begin{figure*}
    \centering
    \includegraphics[width=0.85\textwidth]{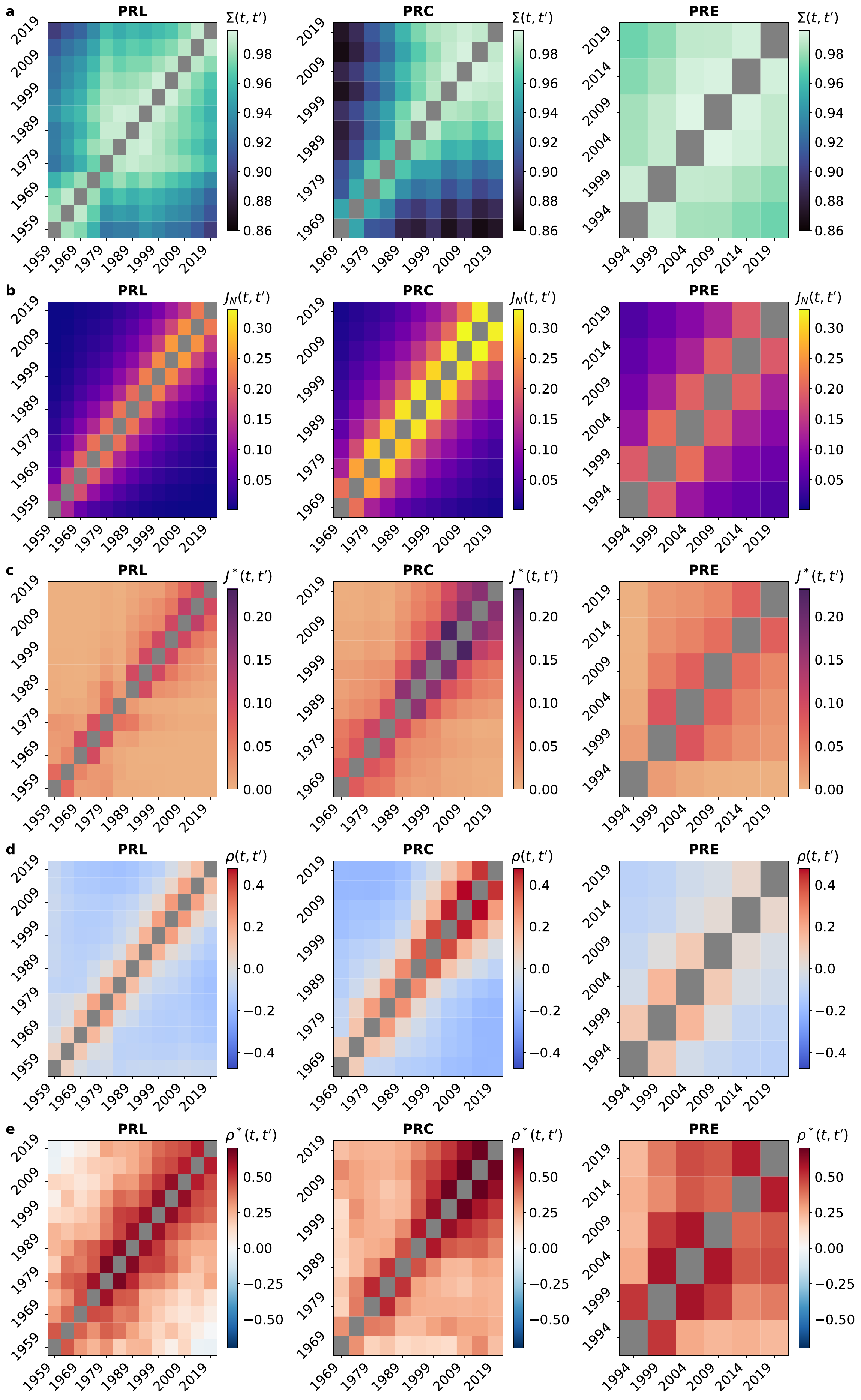}
	\caption{\textbf{Multi-scale stability within specific APS journals.} Same as Supplementary Fig. \ref{fig:figure30}, however here we consider only the empirical case and we build the temporal hypergraphs from the APS data set by considering only the papers published in specific journals separately, considering the snapshot representation with 5-years resolution. Here we consider the PRL, PRC and PRE journals.
	}
    \label{fig:figure40}
\end{figure*}

In this Section we apply the proposed method to specific journals within the APS data set (see Supplementary Table \ref{tab:APS_journals}). We show how the journals share authors over time and we present the statistical properties of some specific journals particularly relevant according to the overall structural characterization of the APS dataset (see main text) or because they clearly identify a scientific community (Supplementary Fig. \ref{fig:figure37}). For three of these journals (i.e., PRL, PRC, PRE) we build the temporal hypergraphs of scientific collaborations generated by considering only the papers published within each of them separately and we apply the proposed characterization method, showing the evolution of the hyper-core structure over time (Supplementary Figs. \ref{fig:figure38}, \ref{fig:figure39}) and the corresponding similarity matrices for the system stability at different topological scales (Supplementary Fig. \ref{fig:figure40}).

\clearpage
\newpage

\clearpage
\newpage

\clearpage
\newpage

\clearpage
\newpage

\section{The effect of temporal resolution}
\label{sez:VIII}

\begin{figure*}[h]
    \centering
    \includegraphics[width=\textwidth]{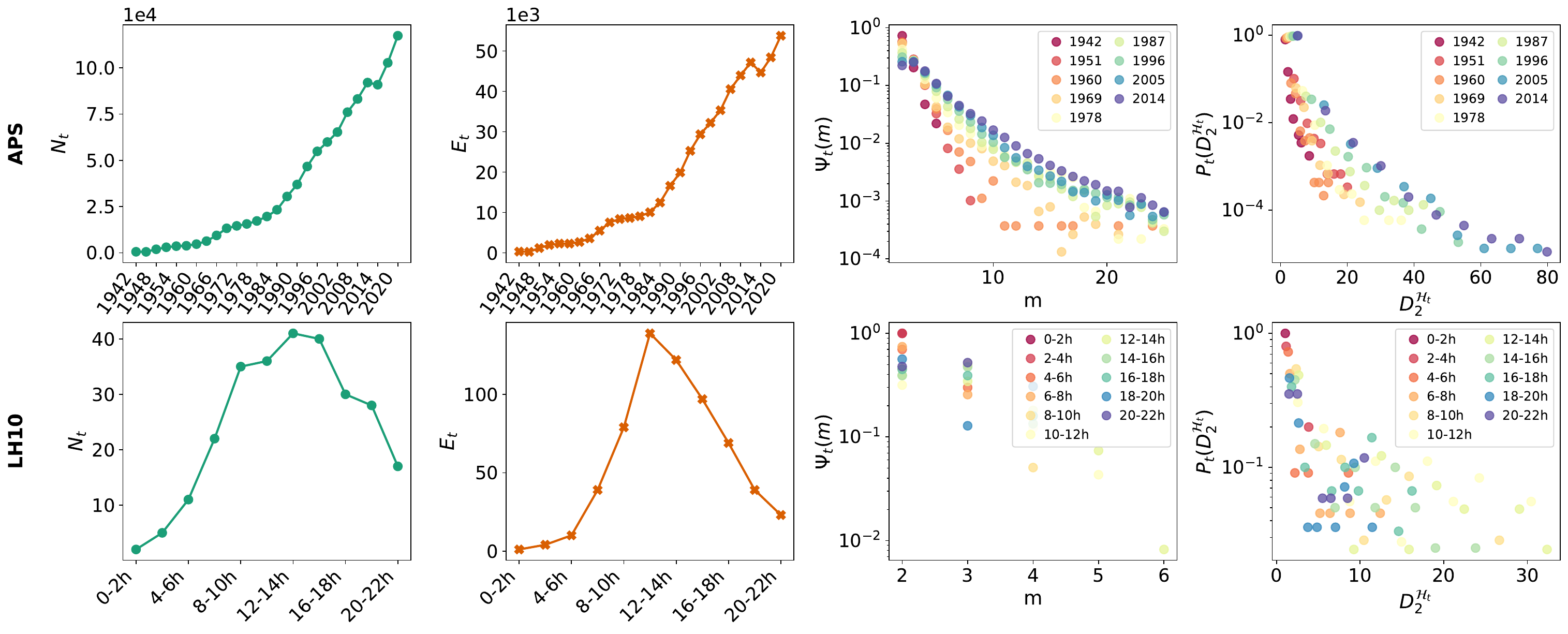}
    \caption{\textbf{Data sets properties evolution for different time resolution.} We consider the APS data set with 3-years resolution (over the period 1941-2021) and the LH10 data set with 2-hours resolution (focusing on the second deployment day). For each data set we show the temporal evolution of: the number of active nodes $N_t$, the number of active hyperedges $E_t$, the distribution of hyperedges size $\Psi_t(m)$ and the distribution of the total degree $P_t(D_2^{\mathcal{H}_t})$ in some time windows, where $D_2^{\mathcal{H}_t}(i)$ is the total number of interactions of arbitrary size in which the node $i$ is involved in the static snapshot $\mathcal{H}_t$.
    }
    \label{fig:figure41}
\end{figure*}

\begin{figure*}[h]
    \centering
    \includegraphics[width=0.8\textwidth]{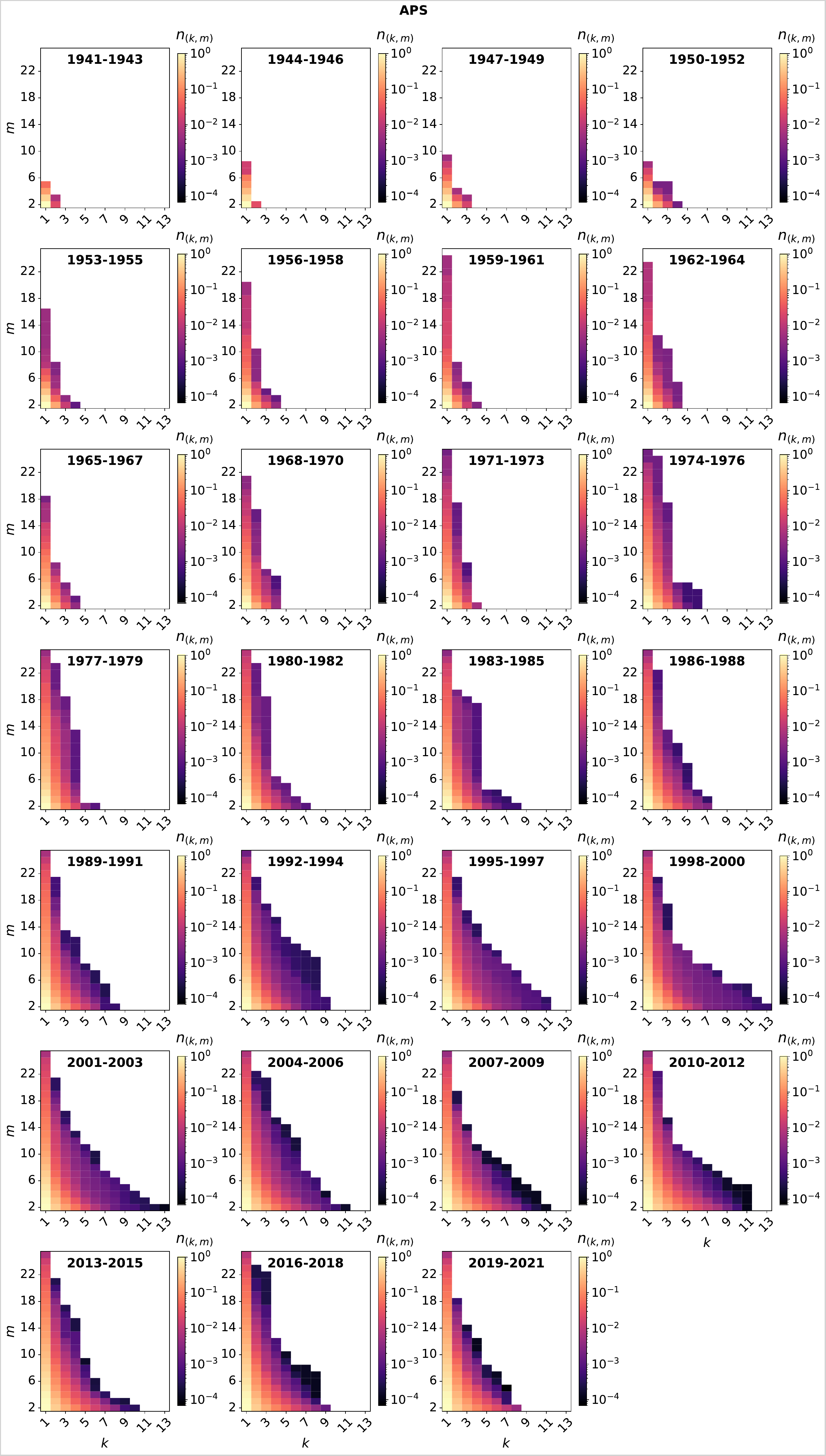}
    \caption{\textbf{Hyper-core structure evolution for different time resolution - I.} We consider the APS data set with 3-years resolution and we show the fraction of nodes $n_{(k,m)}$ in the $(k,m)$-core as a function of $k$ and $m$ for each time window.
    }
    \label{fig:figure42}
\end{figure*}

\begin{figure*}[h]
    \centering
    \includegraphics[width=\textwidth]{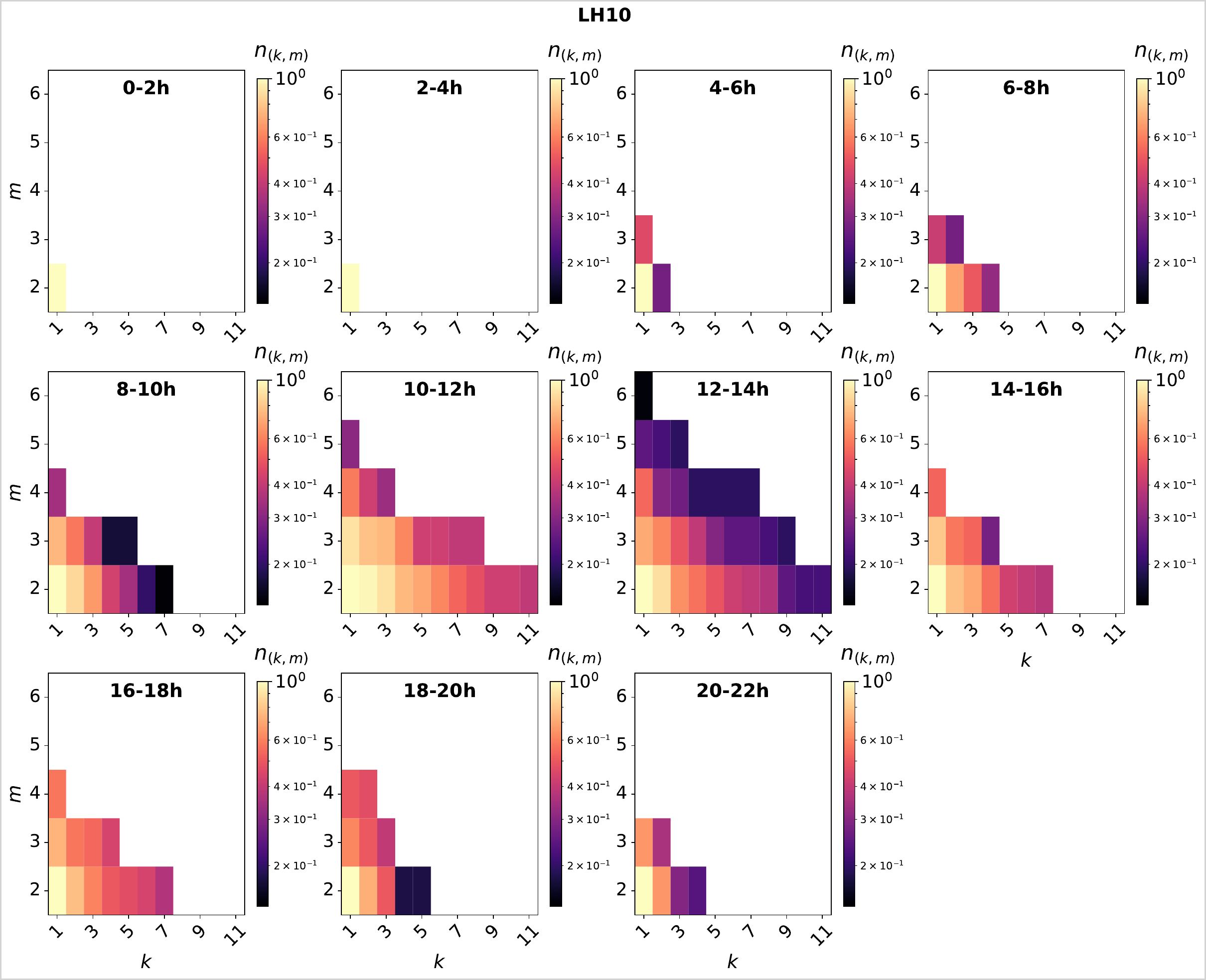}
    \caption{\textbf{Hyper-core structure evolution for different time resolution - II.} Same as in Supplementary Fig. \ref{fig:figure42}, but here we consider the LH10 data set with 2-hours resolution (focusing on the second deployment day).
    }
    \label{fig:figure43}
\end{figure*}

\begin{figure*}[h]
    \centering
    \includegraphics[width=0.95\textwidth]{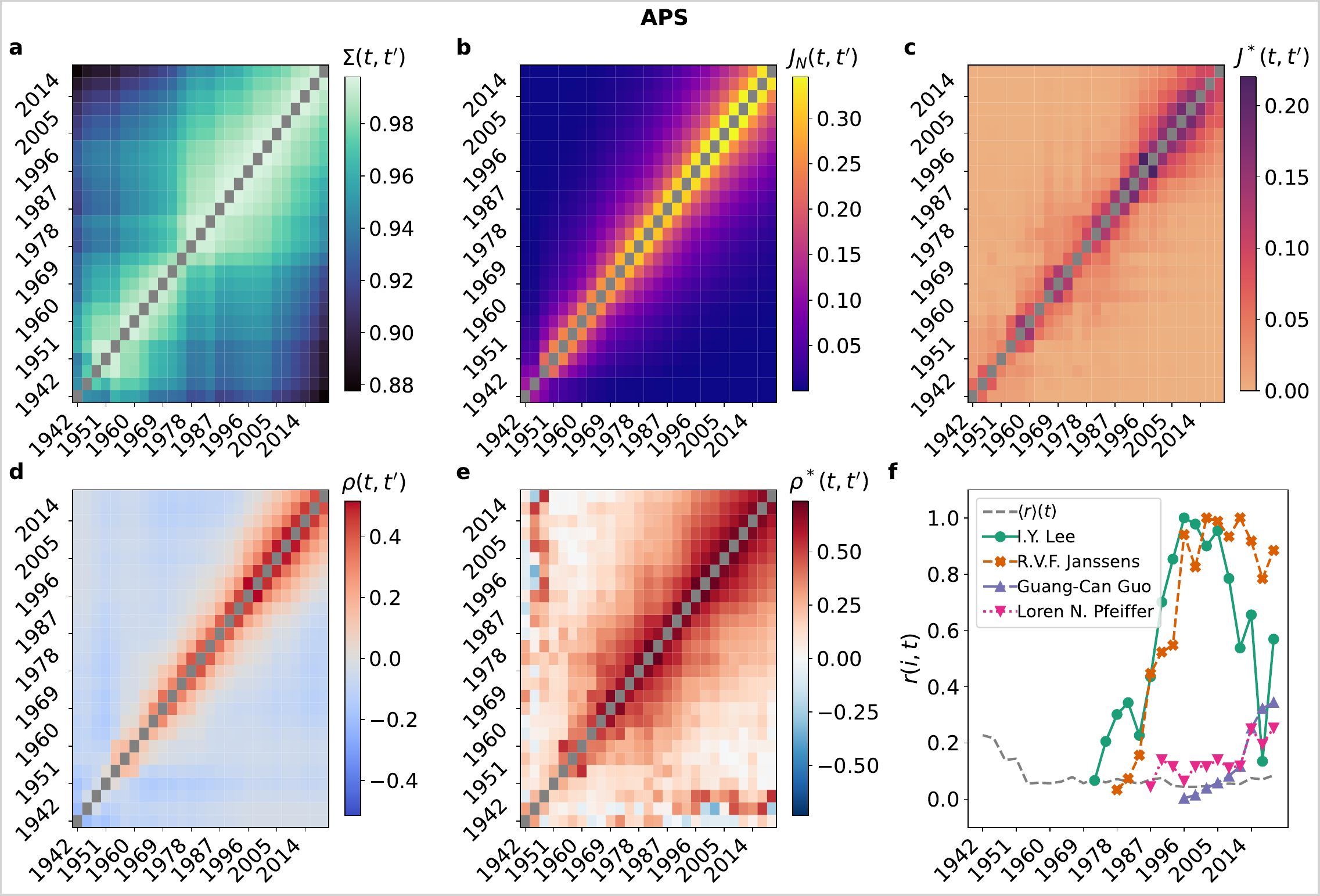} \\ [0.5cm] 
    \includegraphics[width=0.95\textwidth]{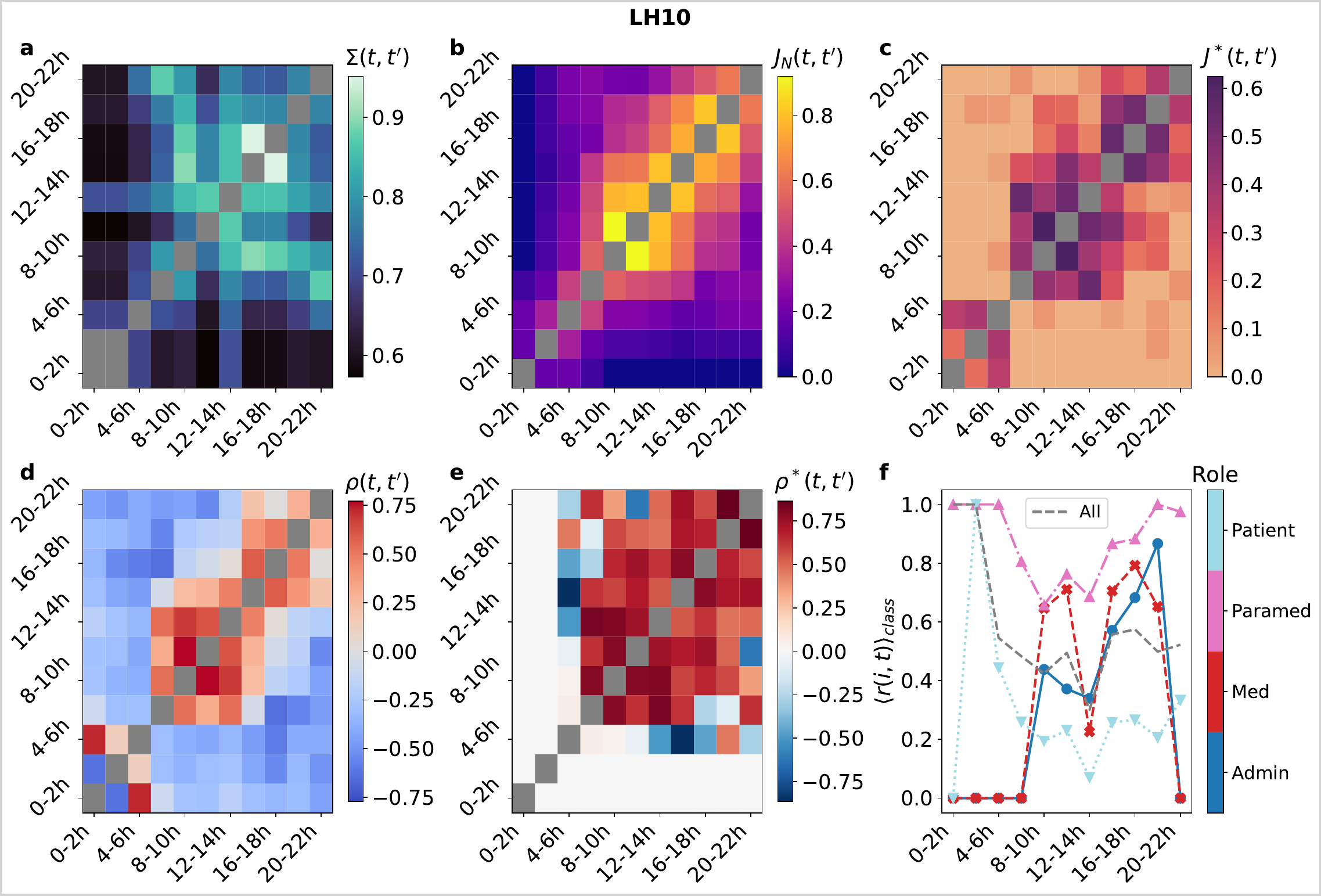}
    \caption{\textbf{Multi-scales stability and hypercoreness evolution for different time resolution.} We consider the APS and LH10 data sets, respectively with 3-years and 2-hours resolution. For each data set we show the similarity matrices: $\Sigma(t,t')$ (panels \textbf{a}), $J_N(t,t')$ (panels \textbf{b}), $J^*(t,t')$ (panels \textbf{c}), $\rho(t,t')$ (panels \textbf{d}) and $\rho^*(t,t')$ (panels \textbf{e}). In panels \textbf{f}, for the APS data set we show the evolution of the hypercoreness $r(i,t)$ for some specific nodes (the same of Supplementary Fig. \ref{fig:figure14}) and the average hypercoreness, while for the LH10 data set we show the evolution of the hypercoreness averaged over all the nodes and over each distinct class.
    }
    \label{fig:figure44}
\end{figure*}

In this Section we replicate the analysis for some data sets by considering a different temporal resolution in the hypergraph aggregation for obtaining the snapshot representation. In particular, we consider the APS data set with 3-years resolution (over the period 1941-2021) and the LH10 data set with 2-hours resolution (focusing on the second deployment day). In both cases, we show the hypergraphs basic properties (Supplementary Fig. \ref{fig:figure41}), the evolution of the hyper-core structure (Supplementary Figs. \ref{fig:figure42}, \ref{fig:figure43}), the corresponding similarity matrices for the system stability at different topological scales and the hypercoreness evolution (Supplementary Fig. \ref{fig:figure44}). Finally, for each time window we show the dominant journal/role that composes each $(k,m)$-core (see Supplementary Figs. \ref{fig:figure45}, \ref{fig:figure46}), considering a dominant journal/role if its frequency exceeds 0.5. Moreover, in Supplementary Fig. \ref{fig:figure47}, for the LH10 data set we show the temporal evolution of the relative frequency of the various node labels within the top positions of the instantaneous hypercoreness ranking, while for the APS data set we report the relative frequency of the various hyperedge labels within the most central hyper-cores, i.e. $(k_{max}^m,m)$-cores $\forall m$.

\clearpage
\newpage

\clearpage
\newpage

\clearpage
\newpage

\clearpage
\newpage

\begin{figure*}[h]
    \centering
    \includegraphics[width=0.79\textwidth]{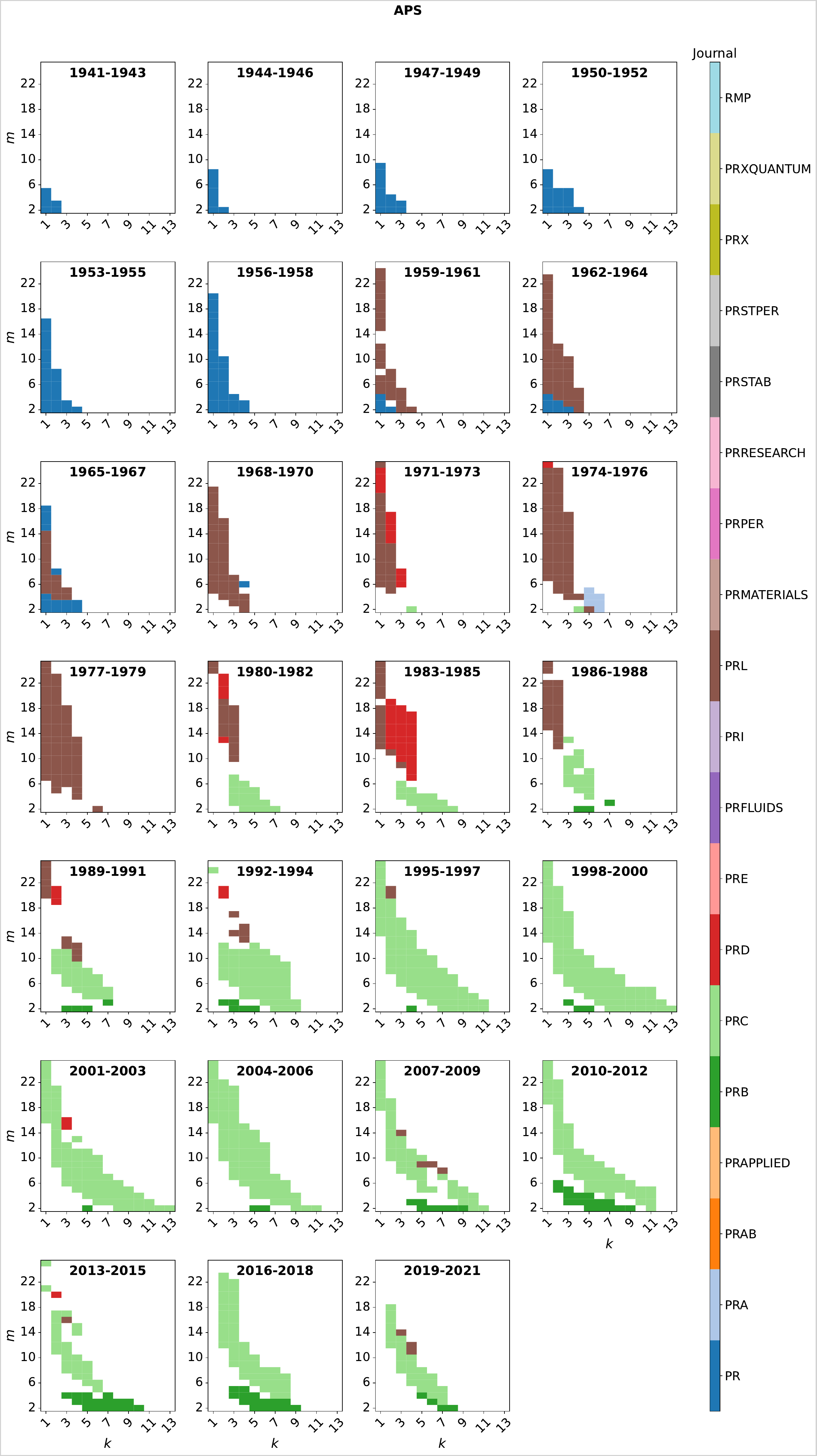}
    \caption{\textbf{Prevalent journal in APS hyper-cores for different time resolution.} For each time window we show the prevalent journal in each $(k,m)$-hypercore: we use a color code for identifying journals and we consider a journal dominant if its frequency is larger than 0.5, white indicates hyper-cores which are empty or where does not exist a dominant journal. We consider the APS data set with 3-years resolution.
    }
    \label{fig:figure45}
\end{figure*}

\clearpage
\newpage

\begin{figure*}[h]
    \centering
    \includegraphics[width=0.83\textwidth]{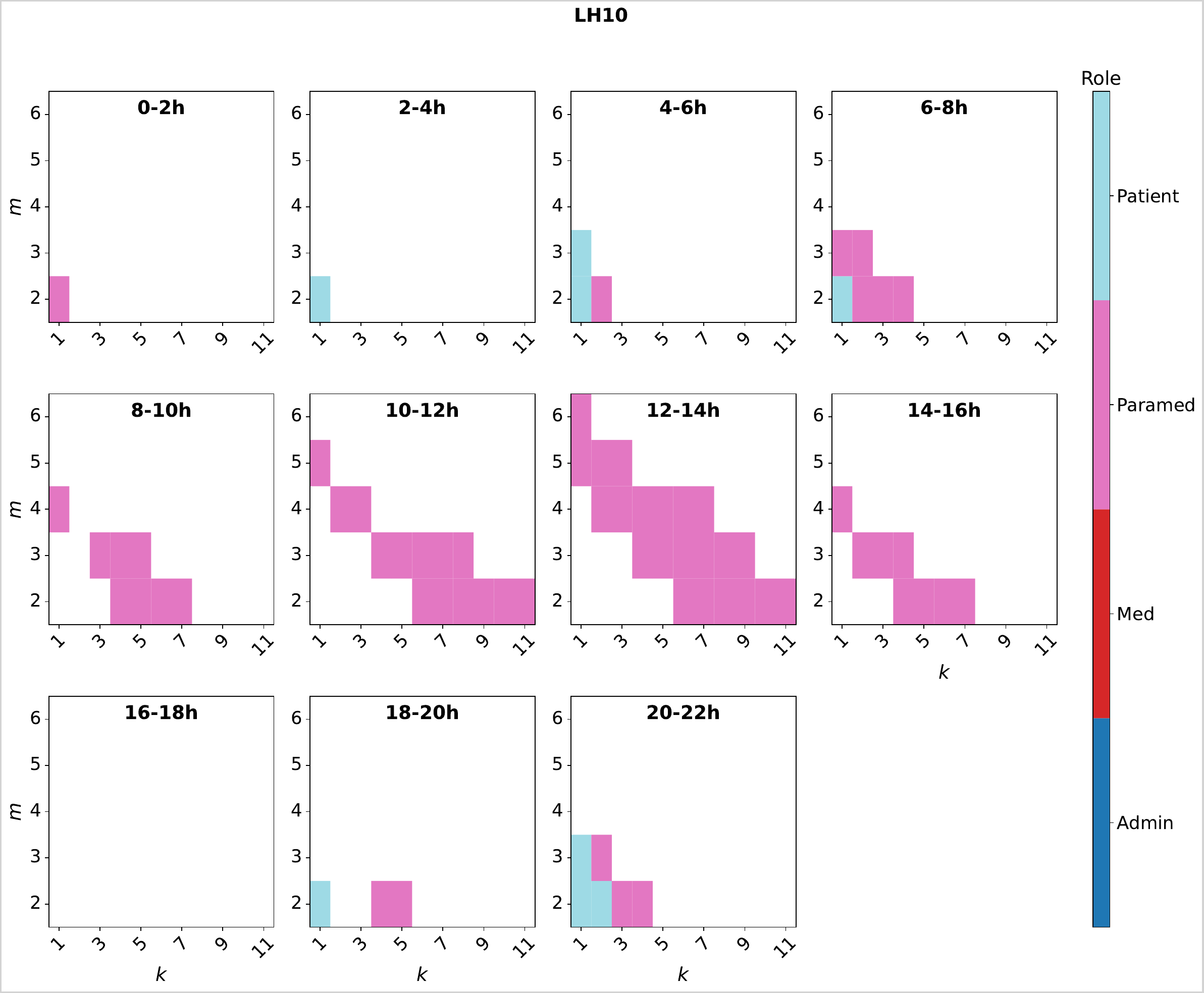}
    \caption{\textbf{Prevalent role in LH10 hyper-cores for different time resolution.} Same as Supplementary Fig. \ref{fig:figure45}, but here we consider the LH10 data set with 2-hours resolution (on the second deployment day) and the prevalent node role in each hyper-core. 
    }
    \label{fig:figure46}
\end{figure*}

\begin{figure*}[h]
    \centering
    \includegraphics[width=0.7\textwidth]{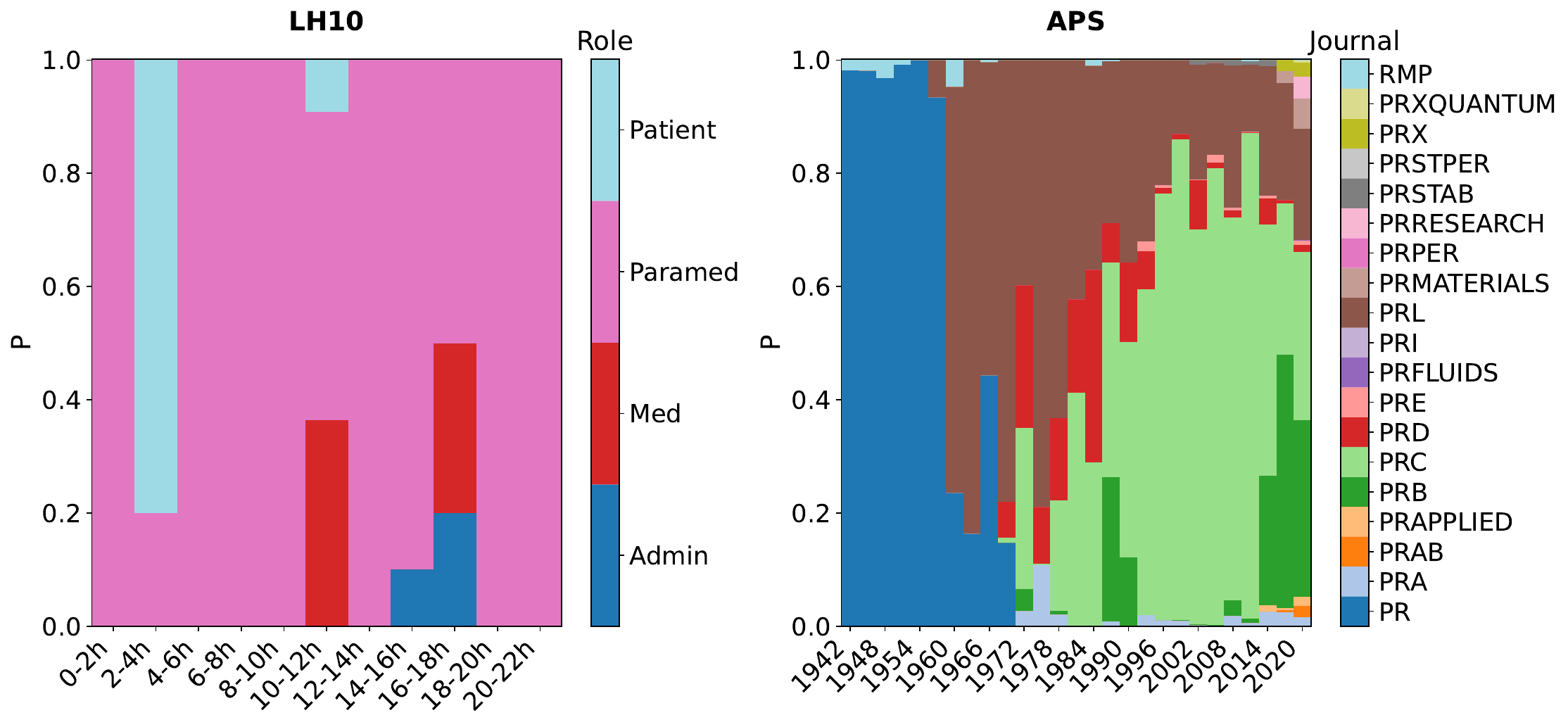}
    \caption{\textbf{Prevalent labels in relevant high-order structures for different time resolution.} For the LH10 data set we show the temporal evolution of the relative frequency $P$ of the various node labels within the top 15\% positions of the instantaneous hypercoreness ranking, through a stacked bar chart; for the APS data set we report the relative frequency $P$ of the various hyperedge labels within the most central hyper-cores.
    }
    \label{fig:figure47}
\end{figure*}

\clearpage
\newpage


%